\documentstyle[psfig,epsf,psfig,12pt]{report}

\def\lesssim{\mathrel{\hbox{\rlap{\hbox{\lower4pt\hbox{$\sim$}}}\hbox{$<$}}}}
\def\gtrsim{\mathrel{\hbox{\rlap{\hbox{\lower4pt\hbox{$\sim$}}}\hbox{$>$}}}}
\newcommand{\sun}{\ifmmode{\odot}\else{$\odot}\fi}
\newcommand{\etal}{et al.}
\newcommand{\msolyr}{\ifmmode{{\rm M}_{\odot}~{\rm 
yr}^{-1}}\else{{M$_{\odot}$~yr}$^{-1}$}\fi}
\newcommand{\msun}{\ifmmode{{\rm M}_\odot}
\else{M$_{\odot}$} \fi}
\newcommand{\lsun}{\ifmmode{{\rm L}_{\odot}}
\else{L$_{\odot}$} \fi}
\newcommand{\rsun}{\ifmmode{{\rm R}_{\odot}}
\else{R$_{\odot}$} \fi}
\newcommand{\teff}{\ifmmode{{\rm T}_{\rm eff}}
\else{T$_{\rm eff}$} \fi}
\newcommand{\mdot}{\ifmmode{\dot{\rm M}} 
\else{$\dot{\rm M}$}\fi} 
\newcommand{\mhbb}{\ifmmode{{\rm M}_{\rm HBB}}\else{M$_{\rm HBB}$}\fi}
\newcommand{\mtd}{\ifmmode{{\rm M}_{\rm 2dr}}\else{M$_{\rm 2dr}$}\fi}
\newcommand{\mcs}{\ifmmode{{\rm M}_{\rm CS}}\else{M$_{\rm CS}$}\fi}
\newcommand{\tip}{\ifmmode{{\rm t}_{\rm ip}}\else{t$_{\rm ip}$}\fi} 
\newcommand{\tp}{\ifmmode{{\rm t}_{\rm p}}\else{t$_{\rm p}$}\fi} 
\newcommand{\np}{\ifmmode{{\rm N}_{\rm p}}\else{N$_{\rm p}$}\fi}
\newcommand{\he}{\ifmmode{^{4}{\rm He}}\else{$^{4}$He} \fi}
\newcommand{\ctw}{\ifmmode{^{12}{\rm C}}\else{$^{12}$C} \fi}
\newcommand{\cth}{\ifmmode{^{13}{\rm C}}\else{$^{13}$C} \fi}
\newcommand{\nit}{\ifmmode{^{14}{\rm N}}\else{$^{14}$N} \fi}
\newcommand{\oxy}{\ifmmode{^{16}{\rm O}}\else{$^{16}$O} \fi}
\newcommand{\nett}{\ifmmode{^{22}{\rm Ne}} \else{$^{22}$Ne} \fi} 
\newcommand{\ls}{\ifmmode{{\rm L}_{\rm s}}\else{L$_{\rm s}$}
 \fi}
\newcommand{\mc}{\ifmmode{{\rm M}_{\rm c}}\else{M$_{\rm c}$} 
\fi}
\newcommand{\mce}{\ifmmode{{\rm M}_{\rm ce}}\else{M$_{\rm ce}$} 
\fi}
\newcommand{\lhemax}{\ifmmode{{\rm L}_{\rm He,max}}\else{L$_{\rm
He,max}$}\fi}
\newcommand{\lhemin}{\ifmmode{{\rm L}_{\rm He,max}}\else{L$_{\rm
He,max}$}\fi}
\newcommand{\ttpagb}{\ifmmode{\tau_{\rm TP-AGB}}\else{$\tau_{\rm
TP-AGB}$}\fi} 
\newcommand{\tbase}{\ifmmode{{\rm T}_{\rm 6,base}}\else{T$_{\rm
6,base}$}\fi} 
\newcommand{\tsix}{\ifmmode{{\rm T}_{6}}\else{T$_6$}\fi}
\newcommand{\rstar}{\ifmmode{{\rm R}_{\star}}\else{R$_{\star}$}\fi}

\newcommand{\aj}{AJ}
\newcommand{\araa}{ARA\&A}
\newcommand{\apj}{ApJ}
\newcommand{\apjl}{ApJ}
\newcommand{\apjs}{ApJS}
\def\ao{\ref@jnl{Appl.~Opt.}}		
\newcommand{\apss}{Ap\&SS}
\newcommand{\aap}{A\&A}
\def\aapr{\ref@jnl{A\&A~Rev.}}		
\newcommand{\aaps}{A\&AS}
\def\azh{\ref@jnl{AZh}}			
\newcommand{\baas}{BAAS}
\def\jrasc{\ref@jnl{JRASC}}		
\newcommand{\memras}{MmRAS}
\newcommand{\mnras}{MNRAS}
\def\pra{\ref@jnl{Phys.~Rev.~A}}	
\def\prb{\ref@jnl{Phys.~Rev.~B}}	
\def\prc{\ref@jnl{Phys.~Rev.~C}}	
\def\prd{\ref@jnl{Phys.~Rev.~D}}	
\def\pre{\ref@jnl{Phys.~Rev.~E}}	
\def\prl{\ref@jnl{Phys.~Rev.~Lett.}}	
\newcommand{\pasp}{PASP}
\def\pasj{\ref@jnl{PASJ}}		
\def\qjras{\ref@jnl{QJRAS}}		
\def\skytel{\ref@jnl{S\&T}}		
\def\solphys{\ref@jnl{Sol.~Phys.}}	
\def\sovast{\ref@jnl{Soviet~Ast.}}	
\def\ssr{\ref@jnl{Space~Sci.~Rev.}}	
\def\zap{\ref@jnl{ZAp}}			
\def\nat{\ref@jnl{Nature}}		
\def\iaucirc{\ref@jnl{IAU~Circ.}}       
\def\aplett{\ref@jnl{Astrophys.~Lett.}} 
\def\apspr{\ref@jnl{Astrophys.~Space~Phys.~Res.}}
\def\bain{\ref@jnl{Bull.~Astron.~Inst.~Netherlands}} 
\def\fcp{\ref@jnl{Fund.~Cosmic~Phys.}}  
\def\gca{\ref@jnl{Geochim.~Cosmochim.~Acta}}   
\def\grl{\ref@jnl{Geophys.~Res.~Lett.}} 
\def\jcp{\ref@jnl{J.~Chem.~Phys.}}	
\def\jgr{\ref@jnl{J.~Geophys.~Res.}}	
\def\jqsrt{\ref@jnl{J.~Quant.~Spec.~Radiat.~Transf.}}
\def\memsai{\ref@jnl{Mem.~Soc.~Astron.~Italiana}}
\def\nphysa{\ref@jnl{Nucl.~Phys.~A}}   
\def\physrep{\ref@jnl{Phys.~Rep.}}   
\def\physscr{\ref@jnl{Phys.~Scr}}   
\def\planss{\ref@jnl{Planet.~Space~Sci.}}   
\def\procspie{\ref@jnl{Proc.~SPIE}}   

\begin{document}

\pagenumbering{roman}
\pagestyle{empty}
\begin{titlepage}
\vspace{2.5in}
~ \\
~ \\
\begin{center}
UNIVERSITY OF OKLAHOMA \\
~ \\
GRADUATE COLLEGE\\
\end{center}
\vspace{1in}
\begin{center}
{\large \bf Abundances in Planetary Nebulae \\ An Autopsy of Low and
Intermediate Mass Stars \\}
~                            \\
~                            \\
~				\\
~               
\end{center}
\begin{center}
  A Dissertation  \\
~ \\
SUBMITTED TO THE GRADUATE FACULTY \\
~ \\
  In partial fulfillment of the requirements \\
  for the degree of         \\
~                             \\
  Doctor of Philosophy       \\
~                             \\

~  \\
~  \\
~                            \\
~ \\
~ \\
~ \\
by \\
~ \\
JAMES F. BUELL \\
~ \\
  Norman,
 Oklahoma \\
        1997 \\
                             
\end{center}
\end{titlepage}
\begin{center}
~ \\
~ \\
~ \\
~ \\
{\large \bf ABUNDANCES IN PLANETARY NEBULAE: \\ AUTOPSIES OF LOW AND
INTERMEDIATE MASS STARS} \\
\end{center}
\vskip 8pt
\begin{center}
A Dissertation APPROVED FOR THE \\
DEPARTMENT OF PHYSICS AND ASTRONOMY
\end{center}
\vskip 160pt
\begin{center}
BY \\ 
\end{center}
\begin{tabbing}
\` \= \underline{\makebox[2.5in]{ }} \\
\`  Richard Henry \\
~ \\
\` \= \underline{\makebox[2.5in]{ }} \\
\`  \> Edward Baron \\
~ \\
\` \= \underline{\makebox[2.5in]{ }} \\
\`  \> David Branch \\
~ \\
\` \= \underline{\makebox[2.5in]{ }} \\
\`  \> Marilyn Ogilvie\\
~ \\
\` \= \underline{\makebox[2.5in]{ }} \\
\`  \> William Romanishin \\

\end{tabbing}
%
%

\setcounter{page}{4}
\pagestyle{plain}

\tableofcontents
\newpage
\listoffigures
\newpage
\listoftables
\newpage

\pagestyle{plain}
\setlength{\baselineskip}{0.8cm}
\setlength{\parindent}{0.6cm}

\pagenumbering{arabic}
\setcounter{page}{1}
\chapter{Introduction}

The planetary nebula (PN) phase is one of the final stages of stellar
evolution for 1-8$\msun$ stars. A planetary nebula is created when an
asymptotic giant branch (AGB) star sheds its hydrogen-rich envelope
via stellar wind, exposing the underlying hot carbon-oxygen (CO)
core. The surface of the remnant star has a surface temperature
T$>$30000K and therefore it emits a significant number of photons with
wavelengths, ($\lambda$), less than 912 Angstroms. These photons with
energy greater than 13.6eV are sufficiently energetic to ionize
hydrogen in the ejecta. The free electrons collisionally excite the
ions in the nebula. The excited atoms then radiatively de-excite
creating emission lines. PNe appear green in a small telescope or
binoculars because of the [OIII]$\lambda$5007.

The energy input from the star is balanced by the energy emitted by
lines from the nebulae. Photons from the star which photoionize the
nebulae are the primary energy source. Emission lines carry this
energy away. This energy balance determines the temperature of the
nebulae and the relative numbers of each ionization stage of each
atomic species.

PN emission lines can be used to determine to reasonable accuracy the
nebular abundances of several atomic species. The density in most PNe
is too low for ions to be collisionally de-excited. Therefore, these
ions must emit one or more photons to get back to the ground state. By
balancing the number of collisional excitations with the number of
radiative deexcitations, the abundance of some ionic species can be
determined, e.g., using the [OIII]$\lambda$5007 line we can get the
abundance of O$^{+2}$. To get the nebular abundance of an atom, an
ionization correction factor is used to account for the unseen ionization
stages. Therefore, the uncertainties in each element vary depending on
how much of the element is in each stage. The procedure for
determining elemental abundances is described in more detail in
elementary textbooks on the subject (e.g. Osterbrock 1989).

Recently, detailed photoionization models have become widely
available, allowing detailed calculations of the nebulae (i.e. CLOUDY).
Such models improve our ability to determine nebular abundances.

The abundances of several important elements can be inferred for
PNe. The list includes hydrogen, helium, carbon, nitrogen, oxygen,
neon, sulfur, and argon. Some of these such as neon and argon can not
be measured AGB phase which preceeds the PN phase. 

The gas that makes up a PN has been processed via stellar
evolution. This processing leads to the difference between the
abundances in PNe and the differences in solar or HII region
abundances. The abundances in the Sun and in HII regions are
respectively believed to reflect the composition of the interstellar
medium (ISM) at 5Gyrs and the current epoch, respectively. The
material making up PNe has been processed by stellar evolution. This
probably leads to enhancements in the abundances of He, C, and N.

\section{The Importance of Stellar Evolution}

The progenitors of PNe are zero-age main sequence (ZAMS) stars in the
mass range 0.8-8.0$\msun$. The upper limit is uncertain and is set by
the minimum mass star capable of igniting carbon-burning in the
core. The lower limit is set by the minimum mass star which has had
sufficient time to reach the PN stage. In fact at least two and
possibly more PNe exist in globular clusters. It is popular to divide
this region into two mass ranges, the low (0.8-4$\msun$) and
intermediate ($>4\msun$) mass ranges.

The progenitors of PNe all appear to go through a number of distinct
and important stages. All of them start on the ZAMS where they burn
hydrogen in the core. When the hydrogen in the core is exhausted the
core contracts but the outer layers expand, causing an increase in
the luminosity and a decrease in the temperature. They continue to
evolve ``redward'' on the HR diagram until hydrogen is reignited in
a shell at which time the stars enter the first giant branch (FGB). By
the time they reach the FGB they have expanded from $\sim\rsun$ to
$\sim100\rsun$. On the FGB, the luminosity of these stars increases
until the helium in the core ignites. The onset of core helium burning
causes the outer layers to contract. This results in an increase in
the surface temperature which causes the stars to evolve ``blueward''
on the HR diagram. When helium is exhausted in the core, the star once
again expands and moves redward on the HR diagram, and when a helium
shell ignites it has entered the asymptotic giant branch (AGB). The
AGB is divided into two distinct stages. When helium burns in a thick
shell and produces the majority of the luminosity this is the
early-AGB (E-AGB) stage. When the helium burning shell narrows and a
hydrogen shell also ignites this is the thermally pulsing asymptotic
giant branch (TP-AGB). The double shell burning mode of the TP-AGB is
unstable and quasi-periodically the luminosity of the helium shell
increases by several orders of magnitude in a few years, a process
which extinguishes the hydrogen burning shell. The sudden increase in
the He shell luminosity is known as a thermal pulse (TP) or as a
helium shell flash. During the TP-AGB stage the mass-loss grows from
10$^{-8}\msolyr$ to 10$^{-4}\msolyr$, a process which ejects the
envelope and allows the PN phase to begin.

During the process of evolution from the main-sequence to the PN
phase, a star can experience a number of possible events in which
material from layers which have been processed by nuclear burning can
be mixed into the surface layers, thereby changing the
composition. Three possible mixing events or ``dredge-ups'' have been
identified (\nocite{it78} Iben and Truran 1978,\nocite{ir83} Iben and
Renzini 1983).

Iben (\nocite{i64}1964, \nocite{i67}1967) showed when a star first
enters the FGB, the 
convective envelope of the star can reach into layers where hydrogen
burning has occurred. The term coined for this event is the ``first
dredge-up''. This increases the abundances of $\he$, $\cth$, and
$\nit$ at the expense of $\ctw$ and $\oxy$. The production of $\cth$
and $\nit$ via the first dredge-up is secondary since these elements
are produced from the carbon that existed at the first
dredge-up. Recent calculations by \nocite{bs97} Boothroyd and Sackmann
(1997) confirm that all stars in the intermediate and low-mass range
encounter the first dredge-up.

The existence of the first dredge-up has been established via
observations of red giant stars. Observations of the red giant stars
confirm both qualitatively and quantitatively the existence and
strength of the first dredge-up (see Sneden 1989 for a review of the
evidence). 

When the star enters the E-AGB, the convective envelope can penetrate
the hydrogen exhausted core, mixing to the surface (Becker and Iben
1979). This process is known as the second dredge-up. Material rich in
helium and nitrogen and totally depleted in hydrogen, carbon, and
oxygen is mixed into the surface layers, increasing the abundances of
$\he$ and $\nit$ at the expense of $\ctw$, $\cth$, and $\oxy$. Becker
and Iben (1979) \nocite{bi79} showed that the second dredge-up occurs
only for progenitors with M$\gtrsim$4.5$\msun$. This limit is
metallicity dependent with the lowest mass model encountering
dredge-up decreasing with decreasing metallicity. Recent calculations
by \nocite{bs97} Boothroyd and Sackmann (1997) confirm the metallicity
dependence and extend it down to very low metallicities.

After each helium shell flash on the thermally pulsing asymptotic
giant branch, the convective envelope can mix helium and carbon rich
material to the surface. During a thermal pulse, a convective helium
burning shell develops (Schwarzschild and H\"{a}rm 1967). The
luminosity of this shell is very high ($\sim10^7\lsun$), and a
significant amount (but not all) of the $\he$ is burned into $\ctw$. At
the end of the thermal pulse, the outer convective envelope can extend
into the ashes of this shell (Iben 1975, 1976, 1977). This process is
known as the third dredge-up. Note that it does not necessarily occur
on every thermal pulse. 

Third dredge-up has been used to explain the presence of carbon
(C-)stars. Some very luminous stars which are most likely on the AGB
have an abundance ratio (by number) C/O$>$1. This is different from
other stars where C/O is generally in the solar ratio
(C/O$\approx1/3$). Schwarzschild and H\"{a}rm realized that the third
dredge-up could explain the existence of C-stars.

One problem that developed was that the stellar evolution theory of
the 1970's (Iben 1975, 1976, 1977) predicted minimum and maximum
luminosities for carbon stars that do not agree with the observed
values. The minimum luminosity carbon star is less luminous than
predicted and so is the maximum luminosity carbon star. Since
luminosity roughly correlates with the mass, the predicted mass range
of carbon stars ranged from $3-8\msun$, whereas the observed mass range
is $\sim1.5-\sim$4.0$\msun$. This is the so called ``carbon-star''
problem.

More recent calculations of the third dredge-up (Iben and Renzini
1982, Lattanzio 1986, Boothroyd and Sackmann 1988abcd) including
improved physics (i.e. better opacities for carbon, semi-convection)
and improved numerical procedures (Straniero \etal\ 1997), find the
third dredge-up in the low-mass range ($1.5-3\msun$). These improved
calculations bring the predictions of the low end of the luminosity
range into agreement.

Scalo (1976) first examined another process which affects the surface
abundances of intermediate mass stars. Between TP-AGB helium shell
flashes, a gigantic convective zone exists between the surface of the
star and the hydrogen exhausted core. For intermediate mass stars, the
base of the convective envelope can get hot enough for the CNO bicycle
to effectively operate. This process, called hot-bottom burning,
effectively converts $\ctw$ to $\nit$ via the CN cycle. It also can
convert some $\oxy$ to $\nit$ via the ON cycle. Renzini and Voli
(1981) showed that this process prevents high luminosity AGB stars
from becoming C-stars, thereby reducing the upper mass limit for
C-star formation.

The recent calculations of Bl\"{o}cker and Sch\"{o}nberner (1991) and
confirmed by Boothroyd and Sackmann (1992), Lattanzio (1992), and
Forrestini and Charbonel (1997) indicate that the luminosity produced
by hot-bottom burning is much larger than previously thought. These
studies have found that up to 50\% of the luminosity of the
intermediate mass star between thermal pulses can be produced by this
envelope burning. This effect has significant effects on the lifetime
and nucleosynthesis of intermediate mass stars.

Planetary Nebulae provide an excellent observational test of the
dredge-up theory as well as hot-bottom burning. They immediately
follow the TP-AGB phase, so the PNe should retain the final surface
abundances. The elements most affected by the three dredge-ups and
hot-bottom burning are hydrogen, helium, carbon, nitrogen, and oxygen;
all of which can be measured in PNe. Also, PNe span the entire range
of mass where the three dredge-ups and hot-bottom burning occur
allowing us to examine objects where the different processes have
affected the surface abundances by different relative effects.

Qualitative studies of the abundances in planetary nebulae have
confirmed the general picture outlined above. These studies show that
they differ significantly from those in HII regions and the Sun. The
difference in abundances reflects the processing by stellar evolution
of the PN gas. Recent studies confirm that in some PNe the abundance
ratio N/O is 2-4 times higher than that found in the Sun or in HII
regions. The same studies indicate the PN helium abundance can be
10-40\% higher than that found in HII regions or the Sun. The C/O
abundance ratio is enhanced in some PNe (Rola and Stasinska 1993,
Henry and Kwitter 1995, 1996, 1997). The C/O ratio can be greater than
1. The enhancements in the abundances of helium, carbon, and nitrogen
are all expected in the picture of stellar evolution outlined above.

Recent quantitative studies also confirm a general pattern. Using
trends in TP-AGB models, Renzini and Voli (1981), Groenewegen and
deJong (1993, 1994), and Marigo \etal\ (1996) created models to make
synthetic TP-AGB populations. Each of these studies examined the
evolution of stars with masses between 1 and 8 $\msun$ incorporating
the three dredge-ups and hot-bottom burning to PNe models. All found
that nitrogen is enhanced in intermediate mass stars via hot-bottom
burning and that carbon is enhanced in low-mass stars via third
dredge-up. These studies also provide {\em qualitative} agreement with
the trends in the PN data.


\section{The Importance of Population}

The mass range of PNe progenitors means that PNe will represent a
range of populations. PNe are known to be members of the bulge, disk,
and halo. Using kinematics, studies have established the differences
between older and younger disk PNe (Acker 1980, Maciel and Dutra
1992). Bulge and halo PNe are also distinguishable via their
kinematics. 

Abundance studies have established differences between different
populations, PNe nearest the Galactic Plane typically have the highest
average N/O and O/H. Both these quantities generally decrease with
distance from the plane.

Several halo PNe have been discovered which give us a unique window
into the advanced stages of very low-mass stellar evolution. There is
a PN in globular cluster M15 (K648) for which we can reliably infer
most of the stellar parameters, e.g., mass, age, luminosity. The
wealth of information on K648 will strongly constrain the parameters
used to model it. By using this PN we hope to establish a model(s) for
the origin of halo PNe.

\section{Thesis Goals}
\begin{enumerate}
\item{Establish the differences in chemistry between PNe of different
stellar populations, in particular the differences between bulge and
the disk.}
\item{Create a new and more detailed model for computing synthetic
TP-AGB populations and the resulting PNe. This model incorporates an
improved treatment of hot-bottom burning based on the latest
calculations. This model will also incorporate many updated
parameters and physics.}
\item{Explore the range of possible PNe parameter space. In
particular, we will explore models of other than solar metallicity to
improve agreement between model results and PN data.}
\item{By incorporating new results, we hope to improve the
quantitative agreement between models and data.}
\item{Establish models for the TP-AGB evolution of halo PNe.}
\end{enumerate}
 
\newpage
\chapter{Multivariate Data Analysis}

Planetary nebulae (PNe) are a heterogeneous set of objects, believed
to have zero-age main sequence (ZAMS) masses between 0.8 to 8.0 $\msun$
and pre-PN lifetimes of $\sim$0.03-10 Gyrs, implying a variety
of nucleosynthetic histories, membership in all but the youngest
stellar populations, and a significant range of initial metallicities.
Thus, a good, objective classification scheme which effectively
separates the entire PN population into useful subsets according to
empirical data is extremely useful when one desires to understand the
relationship of these empirical properties to the underlying
characteristics of PNe.

A number of classification schemes based on a variety of criteria are
currently in use: the Peimbert scheme (Peimbert 1978, 
Fa\'{u}ndez-Abans and Maciel 1987) based on the He and N abundances
and the kinematic properties of the nebula population; the
classification system of Balick (1987) based on morphology; and the
scheme of Amnuel, Guseinov, and Rustanov (1989) which relies on a
number of characteristics related to the masses of the nebula and
central star. These systems have recently been employed as bases of
comparison while examining correlations between chemical abundance
ratios (Amnuel 1993, Clegg 1991, Henry 1989, 1990, Perinotto 1991,
Kingsburgh and Barlow 1994), radial abundance gradients (Amnuel 1993
, Maciel and K\"{o}ppen 1994, 
Pasquali and Perinotto 1993
), vertical abundance gradients (
Fa\'{u}ndez-Abans and Maciel 1988
), the
kinematics of PNe
(Dutra and Maciel 1990, Corradi and Schwarz 1995)
, correlations between
abundance ratios and morphology (Corradi and Schwarz 1995), and correlations
between central star parameters and morphology 
(Stanghellini, Corradi, and Schwarz  1993). In addition, all of these
schemes derive from a parameter space whose properties have not been
precisely determined.  

Components of the bulge and disk are expected to be different, as the
disk should contain a mix of old and young stars (0-10Gyrs old),
whereas the bulge is expected to contain older stars ($>9$Gyr
old). Comparison of disk and bulge PNe has revealed that the average
abundances of heavy elements in bulge PNe excluding He and N is $\sim
25$\% less than solar values (Ratag 1991). The same study also found
a paucity of central star masses above 0.65$\msun$ . PNe with enriched
nitrogen and helium have been found in the galactic bulge (Ratag 1991,
Webster 1988). By comparing bulge and disk PNe with multivariate data
analysis we hope to find a simple method of distinguishing the two.

The purpose of this study is to use multivariate analysis techniques
to determine: 1) the effective dimensionality of PN parameter space
and the specific parameters that represent each dimension; 2) the most
appropriate classification system, given all of the data now extant;
3) the differences and similarities between bulge and disk PNe;
and 4) a classification scheme for the bulge PNe.  Multivariate data
analysis (MVDA) allows a formal, objective, and detailed examination
of an observational data set consisting of many objects, each of which
is described by numerous parameters. In this chapter we specifically
employ principal components analysis to probe the dimensionality of
planetary nebula parameter space, while the taxonomy of planetary
nebulae is explored using cluster analysis. The parameter space we
investigate extends over numerous abundance ratios as well as
kino-spatial properties of both disk and bulge PNe in the galaxy. 


\section{The Sample}
Our sample of PNe comprise objects of the galactic disk and bulge 
compiled in Henry (1989b) and Henry (1990).  The nine parameters
considered in our study include six abundance ratios He/H, N/H, O/H,
Ne/H, N/O, and Ne/O; and three kino-spatial parameters $R$, $Z$, and
$V_{LSR}$, i.e. the galactocentric distance, the height above the
galactic plane, and the radial velocity, respectively. The abundance
information was taken 
from Henry (1989b) and Henry (1990). Values for $R$ and $Z$ for each PN, both in kpc,
were calulated using the relations 
\begin{equation}
R=[R_0^2+d^2cos^2(b)-2R_0 d \cdot cos(b)cos(l)]^{1/2},
\end{equation}
\begin{equation}
Z=d \cdot sin(b)
\end{equation}
\noindent
where {\it d} is the distance from the sun from Cahn, Kaler, and
Stanghellini (1991), {\it l} and {\it b} are
galactic latitude and  longitude, respectively, from 
Perek and Kohoutek (1967)
or
Blackwell and Burton (1981), and $R_0$ is the sun's galactocentric
distance, which is assumed to be 8.5kpc.  Values for $V_{LSR}$ were taken
from Schneider et al. (1983).  Objects in Henry (1989b) and Henry (1990) for which data
for any of the study parameters were unavailable were excluded from
our analysis altogether.  PNe NGC~2022, M2-6,
M4-3, M2-10, H1-18, H2-18, M3-15, and NGC~6804 were excluded because
of lack of complete data.  Also, because of the
large uncertainty in their galactocentric distances, the halo PNe were
not included in our sample. 

Objects within $20^{\circ}$ of the galactic center and having
$\vert V_{LSR}\vert\geq 25 km s^{-1}$ were designated bulge objects. 
The resulting list of 17 bulge objects are all contained in the lists
compiled by Ratag (1991) and Webster (1988). Note, some bulge objects
with low velocities maybe misclassified as disk objects, but this
should not be a serious problem. 

Our final list of 76 PNe is presented in table \ref{pnsamp}, where,
for ease of analysis later on, objects are grouped according to the
cluster analysis results.  For each PN, column 1 provides the object's
most common name, columns 2 and 3 list our calculated values for R and
Z in kpc, column 4 refers to cluster analysis results which will be
discussed below, and the designation of a PN as either a bulge (B) or
a disk (D) object is indicated in column 5.

\begin{table}
\begin{tabular}{rrrrr||rrrrr}
PN&R&Z&CL&Pop.&PN&R&Z&CL&Pop.\\ \hline
Hub4&6.4&0.11&1a&B&M1-80&11.4&0.22&2b&D\\
N6620&0.9&0.94&1a&B&N7354&9.0&0.05&2b&D\\
Hub6&6.8&0.05&1a&B&Hu1-1&13.1&0.79&2b&D\\
N6439&4.6&0.42&1a&B&I1747&10.6&0.07&2b&D\\
N6778&6.2&0.36&1a&B&I2003&11.7&3.32&2b&D\\
N6781&7.5&0.08&1a&D&N2346&9.7&0.09&2b&D\\
N6803&6.8&0.22&1a&D&N2452C&10.1&0.05&2b&D\\
Me1-1&6.8&0.24&1a&D&N2452S&10.1&0.05&2b&D\\
M1-75&8.0&0.00&1a&D&I4634&5.8&0.59&3a&D\\
N6894&8.1&0.08&1a&D&N6629&6.6&0.17&3a&D\\
N7026&8.7&0.01&1a&D&N6210&7.4&1.22&3a&D\\
N650&9.0&0.13&1a&D&N6826&8.5&0.35&3a&D\\
N2371W&9.9&0.51&1a&D&I5117&8.6&0.12&3a&D\\
N2371E&9.9&0.51&1a&D&N7662&8.9&0.36&3a&D\\
N2452N&10.1&0.05&1a&D&M1-1&14.6&2.28&3a&D\\
Hub5&7.3&0.02&1a&B&M1-4&11.1&0.12&3a&D\\
M1-42&3.1&0.45&1b&B&M2-2&12.4&0.32&3a&D\\
M1-35&4.5&0.16&1b&B&I351&13.6&1.44&3a&D\\
Me2-2&10.6&0.76&1b&D&J320&14.3&1.86&3a&D\\
I4776&4.7&0.91&2a&D&J900&11.2&0.13&3a&D\\
I4673&5.3&0.14&2a&D&N1535&10.1&1.50&3a&D\\
N6309&6.1&0.64&2a&D&I2165&10.1&0.43&3a&D\\
N6578&6.3&0.07&2a&D&N3242&8.7&0.58&3a&D\\
N6818&6.9&0.58&2a&D&Ha2-1&4.9&0.28&3a&D\\
N6751&6.4&0.27&2a&D&I4593&6.4&2.09&3c&D\\
N6807&5.9&0.61&2a&D&H1-23&3.8&0.15&3c&D\\
M1-74&6.8&0.29&2a&D&N6833&9.2&0.95&3b&D\\
N6891&7.1&0.67&2a&D&K3-67&11.8&0.38&3b&D\\
N6879&7.6&1.15&2a&D&K3-68&14.4&0.26&3b&D\\
N6905&7.8&0.28&2a&D&M2-21&2.4&0.27&4a&B\\
N6881&8.2&0.09&2a&D&M2-33&1.8&0.74&4a&B\\
N6884&8.5&0.26&2a&D&M2-23&4.6&0.19&4a&B\\
I418&9.0&0.25&2a&D&M3-20&3.8&0.18&4a&B\\
N6369&7.8&0.07&2b&B&M2-30&0.5&0.61&4a&B\\
N6790&7.4&0.17&2b&D&I4732&3.7&0.58&4a&B\\
Hu1-2&8.5&0.23&2b&D&N6567&6.2&0.03&4a&B\\
N6543&8.6&0.50&2b&D&I4846&5.4&0.64&4a&D\\
I5217&10.4&0.43&2b&D&CN2-1&4.6&0.30&4b&B\\ \hline
\end{tabular}
\caption{Objects used in Multivariate Data Analysis}
\label{pnsamp}
\end{table}

\section{Multivariate Data Analysis}
Given a database comprising numerous parameters for a large sample of
objects, principal components analysis (PCA) enables one to determine
the minimum dimensionality of the parameter space required to adequately
characterize the sample. On the other hand, cluster analysis (CA)
objectively groups objects according to their location in parameter
space. Here we present only short discussions of the principal
components and cluster analysis techniques. For more detailed
information the reader is referred to one of many texts on the
subject, e.g., Kendall (1975), Murtagh and Heck (1987; MH87), and
Rummel (1970). An excellent practical astronomical example of
principal components analysis can  be found in Whitmore
(1984). Throughout our analysis we employed modified versions of the
computer programs found at the end of chapters 2 and 3 of MH87. 

\subsection{Principal Components Analysis}
Given a set of N objects with {\it n} observational parameters, each
object can be represented in $n$-space by a set of coordinates
$x_{ij'}$ ($1 \le i \le N$; $1 \le j' \le n$), where the orthogonal
axes (${\bf e}'_1,...,{\bf e}'_n$) correspond to the $n$ observational
parameters. The goal of PCA is to find a set of {\it m} (${\it
m}\ll{\it n}$) parameters which are sufficient to describe the
objects, i.e. the original coordinate system is rotated to a new
coordinate system (${\bf e}_{1},...,{\bf e}_{n}$) such that a subspace
of dimension $m$ (${\bf e}_{1},...,{\bf e}_{m}$) is sufficient to
describe the data set. 

To proceed, the data are standardized to values $y_{ij'}$, where 
\begin{eqnarray}
y_{ij'}=\frac{x_{ij'}-\overline{x_{j'}}}{\sqrt{n}\sigma_{j'}}.
\end{eqnarray}
\noindent
In this expression $\overline{x_{j'}}$ is the average value for parameter
$j'$, while $\sigma_{j'}$ is the standard deviation of parameter
$j'$ for all $N$ points. As a result, $\sum_{i=1}^{N}y_{i,j'}=0$,
$\sigma_{j'}=1$, and each $y_{i,j'}$ is unitless. The next step is to
rotate the n-dimensional coordinate system so that the $y_{i,j'}$'s
are transformed into $y_{i,j}$'s with axes ${\bf e}_1,...,{\bf
e}_n$. The direction of each axis in the new coordinate system is
determined sequentially, i.e. ${\bf e}_1$ is determined first,
followed by ${\bf e}_2$, etc. The variance, $\lambda_j$, is maximized
for each axis, where the variance of the $j$th axis is defined as 
\begin{eqnarray}
\lambda _{j}=\sum_{i=1}^{N}y_{ij}^{2};
\end{eqnarray}
\noindent
and the $j$th axis must be perpendicular to all the axes determined
before it, i.e. the third axis (${\bf e}_{3}$) must be
perpendicular to the first (${\bf e}_1$) and second (${\bf e}_2$)
axes. This puts a constraint on $\lambda_j$ such that 
\begin{eqnarray} 
\lambda _{1}\geq\lambda _{2}\geq \lambda_{3}\geq ...\geq \lambda _{n}. 
\end{eqnarray}
The above procedure of maximizing the $\lambda_j$ is equivalent to
finding the eigenvalues ($\lambda_j$'s) and the eigenvectors (${\bf
e}_k$'s) of the $n\times n$ correlation matrix.

The next step in the analysis is to determine the dimensionality $m$
($m<n$) of the subspace which can best describe the data. This stage
of the analysis is more subjective than the previous ones, as there is
no standard way to do it. For instance, it is possible to use the three
criteria set forward by Gutman (1954),
one of which states that any
principal component with an eigenvalue greater than one is
significant. However, this criterion can only be considered as a rule 
of thumb. 
In this work, the dimensionality is determined by matching the
residual to the average observational error. There are two types of
residuals, individual and average. Using the $n$ dimensional rotated
coordinate system, it is possible to reconstruct for each object the
values of the observational parameters. For an $m$ dimensional
subspace, it is also possible to reconstruct a value for each
observational parameter for each object. However there will be some
difference between this reconstructed value and the actual value of
the observation for the individual object. The difference between the
reconstructed value for each parameter and the actual value is the
residual for the individual object. The average residual is the
average of all the individual residuals for a particular observational
parameter. We compute the average residual, $r_{j'}$ using the formula
in table~4 of 
Brosche (1973).
The average residual $r_{j'}$ is a measure of how well, on average, we
can reconstruct the value of the {\em j'th} observational parameter
using the $m$ dimensional subspace.


\subsection{Cluster Analysis}
The goal of cluster analysis is to group objectively a set of objects
according to distances of separation in n-dimensional space.  Cluster
analysis is an iterative procedure which starts with a set of $N$ data
points, each described by $n$ parameters, and identifies the pair of
points with the smallest dissimilarity, subsequently agglomerating
this pair into a single composite point. The new composite point and
the other points now form a new set of $N$-1 data points and the least
dissimilar pair of points is identified and agglomerated. This
procedure is repeated until the set has been agglomerated into a
single point. Then, for the results of our cluster analysis to be
useful one looks at the agglomeration process in reverse, starting
with the single composite point, to find a functional set of points or 
groups. 

Dissimilarity is a mathematical measure of the differences between
each pair of points in $n$ dimensional space. There are several types
of dissimilarity, Euclidean distance being the most familiar type. Out
of several possible methods for performing cluster analysis, we have
chosen Ward's method (1963), which is distinguished by the type of
dissimilarity employed as well as the method for determining new
composite points. In this method the cluster center {\bf g} is the
composite point consisting of $l$ individual points, where {\bf g}
represents the center of gravity of the $l$ points.  Thus, if $x_{ij}$
is the $j$th component of the $i$th point, then the $j$th component of
the cluster center is  
\begin{eqnarray}
g_j=\frac{1}{l}\sum_{i=1}^{l}x_{ij},
\end{eqnarray}
\noindent
where the sum is over all points comprising the composite. The
average variance of each cluster is then defined as
\begin{eqnarray}
Variance=\frac{1}{l}\sum_{j=1}^{n}(\sum_{i=1}^{l}(x_{ij}^2-g_{j}^{2}))
\end{eqnarray}
\noindent
where $m$ is the dimension of the parameter space. In Ward's method
the dissimilarity between a pair of points is defined as the increase
in the average variance of the potential cluster which is introduced
by adding the new point. 

\section{Results and Discussion}
\subsection{Results of Principal Components Analysis}

To examine the effect of population differences on our PCA results,
our analysis was performed separately on bulge (B) and disk (D)
subsamples as well as the combined (C) sample\footnote{We note that
when placed in the He-N/O plane the disk objects I4593 and H1-23
appear anomalous relative to the remainder of disk PNe.  Therefore, we
have ignored these two objects in our analysis.} All runs were made
with the following parameters: log(He/H)+12, log(N/H)+12, log(O/H)+12,
log(Ne/H)+12, log(N/O), log(Ne/O), and $V_{LSR}$ designated from now
on as He, N, O, Ne, N/O, Ne/O, and $V_{LSR}$ respectively. The
parameters R and Z were included in the D and C runs but omitted from
the B run. 



The main PCA results are reported in Table \ref{evs}.  Letters
in column 1 indicate the set of objects, while the parameters
are listed in column 2. Values for residuals for dimensionalities
1-5 are given in columns 3-7. Estimated observational uncertainties
and the eigenvalues appear in columns 8 and 9, respectively. 

\begin{table}
\begin{tabular}{rcrrrrrrr}
Set&Var&\multicolumn{5}{c}{{\bf Residuals}$^{\rm a}$}&OU$^{\rm
b}$&EV$^{\rm c}$ \\  
& &m=1 &m=2&m=3&m=4&m=5&&\\ \hline 
B&He&16\%&11\%&10\%& & &10\%&$\lambda_1=3.12$\\ 
&N$^{\rm d}$&0.26&0.16&0.16& &
&0.15&$\lambda_2=2.22$\\ 
&O$^{\rm d}$&0.18&0.10&0.07& &
&0.15&$\lambda_3=0.77$\\ 
&Ne$^{\rm d}$ &0.20&0.06&0.00& &
&0.10&$\lambda_4=0.61$\\ 
&N/O$^{\rm d}$ &0.38&0.13&0.12& &
&0.20&$\lambda_5=0.27$\\ 
&Ne/O$^{\rm d}$&0.08&0.07&0.07& &
&0.20& \\ 
&$V_{LSR}$$^{\rm e}$ &62.9&51.0&9.3& & &10.0 
&\\ \hline
D&He &11\%&10\%&9\%&9\%&9\%&10\%&$\lambda_1=3.23$\\
&N$^{\rm d}$ &0.19&0.17&0.12&0.11&0.10&0.15&
$\lambda_2=1.87$\\ 
&O$^{\rm d}$ &0.21&0.09&0.06&0.03&0.00&0.15&
$\lambda_3=1.28$\\
&Ne$^{\rm d}$ &0.18&0.09&0.09&0.05&0.04
&0.10&$\lambda_4=1.04$\\
&N/O$^{\rm d}$
&0.30&0.11&0.10&0.10&0.10&0.21&$\lambda_5=0.69$\\  
&Ne/O$^{\rm d}$
&0.10&0.10&0.07&0.07&0.04&0.20&$\lambda_6=0.53$\\ 
&$V_{LSR}$$^{\rm e}$&33.5&32.3&16.3&16.2&9.6&10.0
&$\lambda_7=0.36$\\
&R$^{\rm f}$ &2.2&2.0&2.0&1.3&0.8 &&\\
&Z$^{\rm f}$ &0.52&0.52&0.52&0.31&0.22
&&\\ \hline
C&He&12\%&10\%&10\%&10\%&10\%&10\%&$\lambda_1=3.24$
\\
&N$^{\rm d}$&0.20&0.18&0.18&0.18&0.09&0.20&
$\lambda_2=1.79$\\
&O$^{\rm d}$&0.21&0.06&0.06&0.06&0.02&0.20&
$\lambda_3=1.21$\\
&Ne$^{\rm d}$&0.20&0.06&0.03&0.02&0.02&
0.10&$\lambda_4=0.93$\\
&N/O$^{\rm d}$&0.31&0.14&0.13&0.13&0.09&0.31&$\lambda_5=0.89$\\
&Ne/O$^{\rm d}$&0.10&0.10&0.09&0.08&0.04&0.21&$\lambda_6=0.58$\\
&$V_{LSR}$$^{\rm e}$&46.1&46.1&31.7&18.0&16.3&10.0 &
$\lambda_7=0.45$\\
&R$^{\rm f}$ &2.8&2.8&2.0&2.0&1.3 &&\\ 
&Z$^{\rm f}$&0.49&0.49&0.46&0.21&0.21&&\\ \hline
\end{tabular}
\caption{Residuals and Eigenvalues}
\label{evs}
a. $m$ is the number of dimensions being considered.\\
b. Observational Uncertainty\\
c. Eigenvalue for each dimension\\
d. Residuals and observational uncertainty in dex\\
e. Residuals and observational uncertainty in km/s\\
f. Residuals and observational uncertainty in kpc
\end{table}

To illustrate the use of Table \ref{evs}, consider the case of 
the bulge (B) objects. First, the feasibility of using 
a one-dimensional parameter space (m=1) for representing these
objects is tested.  Note, except for Ne/O, none of the residuals 
is less than its associated observational uncertainty.  Therefore, the
bulge PNe cannot be represented satisfactorily by a one dimensional 
parameter space. On the other hand, for two-dimensional space (m=2),
all of the residuals with the exception of the one associated with
$V_{LSR}$, are less than or approximately equal to their
uncertainties.  Thus, the two dimensional space satisfactorily describes
the base set of chemical parameters: He, N, O, Ne, N/O, and Ne/O.  For
the three dimensional space (m=3), all residuals are now less than or
equal to their associated uncertainties, and thus three dimensional
parameter space also adequately represents the set of bulge PNe. 

Using the same method of comparing residuals to observational
uncertainties, the D and C sets are shown to require five
dimensions, with the base set of chemical parameters being
described by the first two dimensions and $V_{LSR}$, R, and Z by
the last three.

\begin{figure}
\centerline{\hbox{\psfig{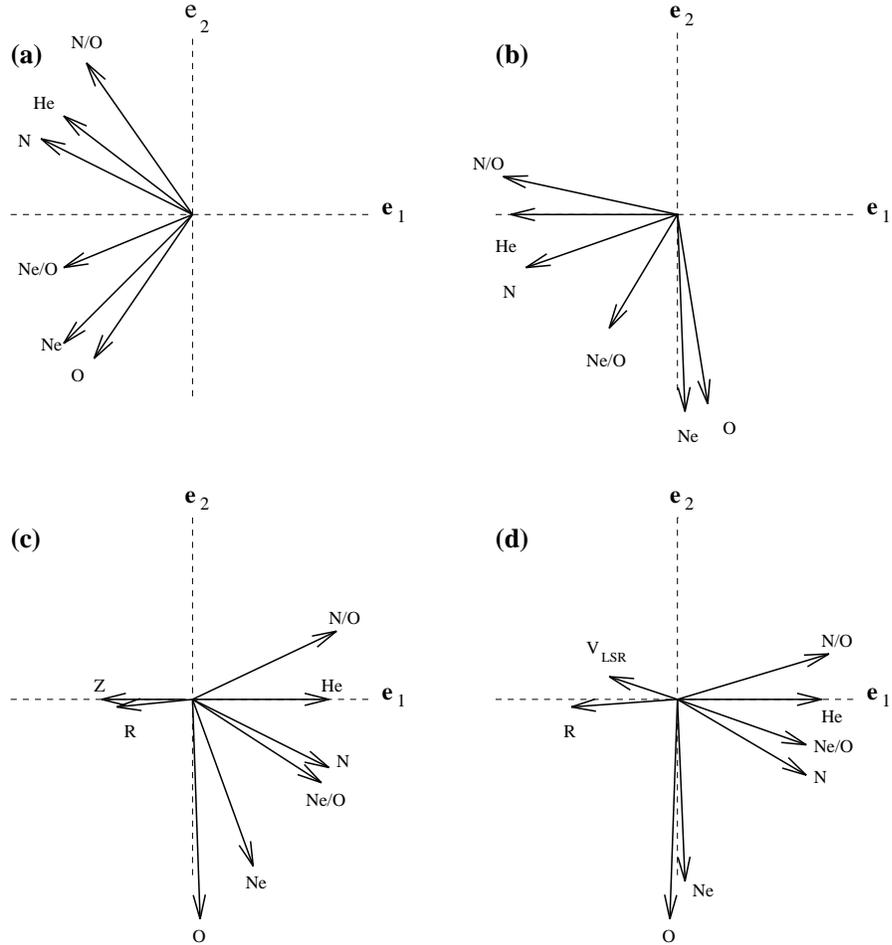}}}
\caption{The correlation vector diagrams generated by our PCA
runs. The diagrams are labeled as follows: (a) The unrotated diagram
for the bulge PNe subset, (b) the rotated diagram for the bulge PNe,
(c) the rotated diagram for the disk PNe subset, and (d) the rotated
diagram for our bulge and disk PNe set. In each diagram the distance
from the origin to the end of the dotted line is 0.6. It should be
noted that it was impossible to get the angles between the vectors
exactly correct due to limitations of our plotting package.}
\label{fig:cvs2s5.eps}
\end{figure}

A correlation vector (CV) diagram is a graphical method for
representing the relationship between the parameters. The CV diagram
is created by projecting unit vectors lying along the individual
parameter axes, i.e. He, N, O, etc., into the space defined by the 
eigenvectors determined by the PCA.  Figure \ref{fig:cvs2s5.eps}a is the CV
diagram for the B set. Our CV diagrams are shown in two dimensions
only, since higher dimensional diagrams become visually confusing. 
The cosine of the angle between any two vectors is approximately 
equal to the correlation coefficient\footnote{Correlation coefficients
were computed as part of the PCA analysis but are not tabulated here.}
of the corresponding quantities. For example, the correlation between N 
and N/O is 0.90, implying an angle of $26^{\circ}$, which is very
close to the actual angle of $29^{\circ}$ between the vectors. 

\begin{table}
\begin{center}
\begin{tabular}{crrr}
Var&{\bf e}$_1$&{\bf e}$_2$&{\bf e}$_3$\\ \hline
He&-0.5626&0.0000&0.0045\\
N&-0.5032&-0.1720&0.1592\\
O&0.1289&-0.6315&0.0000\\
Ne&-0.0286&-0.6399&-0.0433\\
N/O&-0.5837&0.1125&0.1663\\
$V_{LSR}$&-0.1616&0.0642&-0.9656\\
Ne/O&-0.2158&-0.3812&-0.1128\\
\end{tabular}
\end{center}
Each vector, i.e. He, is a unit vector in the 7
dimensional space, its direction lies in the direction of the original
axis. The numbers in columns 2-4 are the components of each parameter
in the new space. The vectors shown are the projections of the
original axes (i.e. He, N, etc.) into the first two new axes.
\caption{Components of Vectors in the Rotated Bulge
Parameter Space}
\label{b2short}
\end{table}

\begin{table}
\begin{center}
\begin{tabular}{crrrrr}
Var&{\bf e}$_1$&{\bf e}$_2$&{\bf e}$_3$&{\bf e}$_4$&{\bf e}$_5$\\ \hline
R&-0.2512&-0.0085&0.6871&-0.4782&-0.0848\\
Z&-0.2975&-0.0009&0.1974&-0.2099&0.7484\\
$V_{LSR}$&0.0549&-0.0050&0.3511&0.7409&0.3494\\
He&0.4489&0.0000&0.0503&-0.3045&-0.0612\\
N&0.4414&-0.2184&0.3535&0.0520&-0.0552\\
O&0.0123&-0.7231&0.0000&0.0000&0.0000\\
Ne&0.2118&-0.5543&-0.1284&-0.1195&0.2314\\
N/O&0.4687&0.2276&0.3816&0.0550&-0.0604\\ 
Ne/O&0.4294&0.2651&-0.2755&-0.2561&0.4966\\ \hline
\end{tabular}
\end{center}
\caption{Components of Vectors in the Rotated Disk
Parameter Space}
\label{d2short}
Same as table \ref{b2short} except that the first 5
components of each vector are presented.
\end{table}

\begin{table}
\begin{center}
\begin{tabular}{crrrrr}
Var&{\bf e}$_1$&{\bf e}$_2$&{\bf e}$_3$&{\bf e}$_4$&{\bf e}$_5$\\ \hline
R&-0.2583&-0.0287&-0.5896&-0.4868&-0.1639\\
Z&-0.2194&-0.0858&0.1075&-0.3222&-0.7632\\
$V_{LSR}$&0.1252&-0.0599&0.0377&0.6364&-0.5926\\ 
He&0.4833&0.0000&-0.0420&-0.1681&0.0362\\
N&0.4290&-0.2579&-0.3757&0.0099&-0.1030\\
O&-0.0874&-0.7293&0.0000&0.0000&0.0000\\
Ne&0.1092&-0.5917&0.2365&-0.1934&0.0510\\
N/O&0.5049&0.1403&-0.3987&0.0944&-0.1090\\
Ne/O&0.4211&0.1416&0.5306&-0.4340&-0.1145\\ \hline
\end{tabular}
\end{center}
\caption{Components of Vectors in the Rotated Bulge and
Disk Parameter Space}
\label{a1short}
Same as table \ref{b2short} except that the first 5
components of each vector are presented.
\end{table}

For greater clarity, it is useful to rotate the coordinate system
orthogonally so that the eigenvectors ${\bf e}_1$, ${\bf e}_2$, and
${\bf e}_3$ are aligned as closely as possible with the parameters.
Such a rotation does not change the relationship between the vectors
but allows easier inspection of the relationships between them. For
each set, the original orientation of the parameter vectors has been
rotated so that the He and O vectors now have no ${\bf e}_2$ and ${\bf
e}_3$ components, respectively.  The resulting CV diagrams for B, D,
and C are respectively shown in figures 1b,c,d, while the vector
components are given in Tables \ref{b2short}, \ref{d2short},
\ref{a1short}. If we consider only the first two eigenvectors in each
set, i.e. ${\bf e}_1$ and ${\bf e}_2$, the He, N, O, Ne, and N/O
vectors show a clear pattern: the vectors He, N, and N/O are all
roughly correlated with each other, while the same is true of the
vectors O and Ne.  At the same time, these two vector sets are
uncorrelated with each other.  This pattern is evident from the
general orthogonal orientation of these two vector groups in the CV
diagrams as well as the component values in Tables \ref{b2short},
\ref{d2short}, \ref{a1short}. Finally, the same diagrams and tables
indicate clearly that the kino-spatial parameters $V_{LSR}$, R, and Z
are orthogonal to each other as well as to the chemical parameters.

Studies of PNe, both observational 
(Henry 1990, Clegg 1990, Kingsburgh and Barlow 1994, Perinotto 1991 )
and theoretical (Iben and Renzini 1983, Renzini and Voli 1981),
suggest He, N, and N/O are related to the nucleosynthesis which
occurs in the progenitor star and O and Ne are related to the
progenitor metallicity. Thus, considering the parameters associated
with each eigenvector in the CV diagrams, ${\bf e_1}$ is related to
nucleosynthesis during progenitor star evolution, while ${\bf e_2}$ is
related to progenitor metallicity. At the same time, the
spatio-kinematic quantities do not correlate with one another nor with
the abundance parameters, although it is important to remember that
since our study is confined to linear correlations, possible higher
order relationships between the kinematic and chemical quantities may
be masked.  Of particular interest is the absence of projection of
these three vectors along the ${\bf e}_2$ eigenvector, i.e. the one
closely associated with progenitor metallicity, as seen in Tables
\ref{b2short},\ref{d2short},\ref{a1short}.  Our results would seem to
indicate that for the bulge and disk, kino-spatial properties and
progenitor metallicity are uncorrelated. 

\subsection{Results of Cluster Analysis}

It is possible to classify the PNe according to their location in the
subspace derived by PCA. To locate the natural groups in this
subspace we perform cluster analysis on the sample using the seven
variables: R, $V_{LSR}$, He, N, O, Ne, and N/O.  The five abundance
parameters were chosen because they correlate with the first two
principal components. R and $V_{LSR}$ have been included to
account for the kinematic differences between the bulge and disk
objects. The height above the galactic plane (Z) has been excluded,
since it is meaningless for bulge objects. Finally, the ratio Ne/O has
been excluded since it is unclear if differences in this parameter
among PNe are real or simply due to observational scatter. 

The inclusion of R as a parameter may seem questionable since the
determination of R depends on the distance derived to the PNe. The
distances to galactic PNe are poorly known (see Terzian 1993 for a
recent review). However, R turns out to be an important classification
parameter only between objects with large differences in
galactocentric radii, i.e. between the bulge and disk
objects. Therefore, it is probable the uncertainties in R will not
have a large impact on the results. 

\begin{figure}
\centerline{\hbox{\psfig{figure=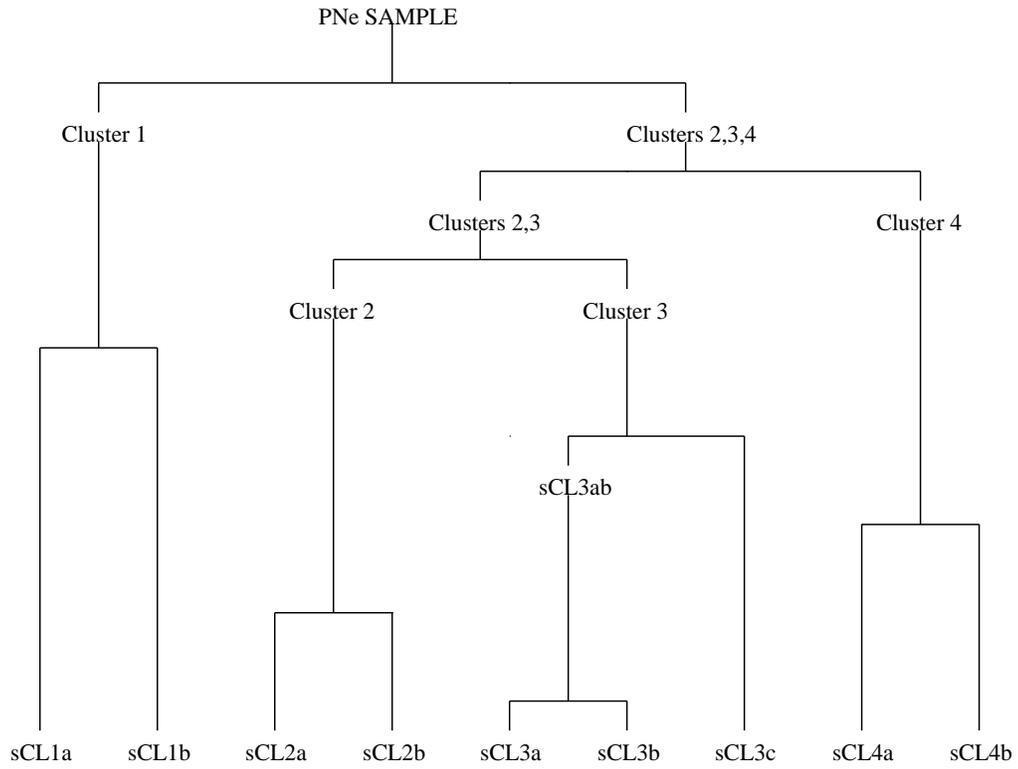,height=4truein}}}
\caption{The dendridic diagram generated by our cluster analysis using
the parameters R, $V_{LSR}$, He, N, O, Ne, and N/O. The branches where
each of the clusters and subclusters are located are labeled.}
\label{fig:dendridic2.eps}
\end{figure}

Starting at the top and moving downward, the dendridic diagram in
figure \ref{fig:dendridic2.eps} shows the reverse of the agglomeration process,
i.e., it shows how CA divided the PNe sample into subsets. We decided
to first analyze the first four subgroups which we label as clusters
1, 2, 3, and 4. Each of these groups contains a reasonable number
($\geq 9$) objects which are hopefully fairly homogeneous. The four
clusters were divided further into 9 subclusters, labeled on figure
\ref{fig:dendridic2.eps} as sCL. The relative vertical distances of connection
points from the bottom indicate the order in which the points were
combined, e.g., the first two points to be joined were 3a and 3b
(these two groups of PNe are the most similar), followed by the
joining of 2a and 2b, and so on. The cluster and subcluster membership
of each PN is listed in column 4 of table \ref{pnsamp}. 

\begin{table}
\begin{center}
\begin{tabular}{lccccccc}\hline
Cluster&Type I&Type IIa&Type IIb&Total
&$V_{LSR} >60$
&Disk&Bulge\\ \hline
1a&11&5&0&16&3&10&6\\
1b&3&0&0&3&3&1&2\\
2a&1&6&7&14&0&14&0\\
2b&3&8&2&13&3&12&1\\
3a&0&0&16&16&0&16&0\\
3b&2&0&1&3&1&1&2\\
3c&2&0&0&2&2&1&1\\
4a&0&0&8&8&8&1&7\\
4b&0&1&0&1&1&0&1\\ \hline
\end{tabular}
\end{center}
\caption{Shows the breakdown of each subcluster into
Peimbert types, bulge and disk members, and into objects with $V_{LSR}
> 60$ km/s. Only composition criteria are used in the Peimbert scheme classification.}
\label{clnumtable}
a  The total number of PNe in each subcluster\\
b  Number of PNe with $V_{LSR}> 60$km/s\\
\end{table}

The parameter averages for each cluster and subcluster are
presented in table \ref{clavg}. The average of the chemical parameters
and the radial velocity $V_{LSR}$ decrease along the sequence of 
clusters: 1, 2, and 3, whereas the parameters R and Z increase along the
same sequence. Cluster 4 appears distinct from the others as it has
the smallest average R and the largest average $V_{LSR}$. However, the
average chemical abundances of cluster 4 resemble those of clusters 2
and 3. 

\begin{table*}
\begin{center}  
\begin{tabular}{crrrrrrrrr}
CL&R$^{\rm a}$&Z$^{\rm a}$&$V_{LSR}$$^{\rm b}$&He/H$^{\rm
c}$&N/H$^{\rm c}$&O/H$^{\rm c}$&Ne/H$^{\rm c}$&N/O$^{\rm
d}$&Ne/O$^{\rm d}$\\ \hline
Solar$^{\rm e}$&----&----&----&10.99&8.05&8.93&8.09&-0.88&-0.84\\ \hline
\multicolumn{10}{l}{Clusters}\\ \hline
1 all&7.11&0.27&47.69&11.12&8.67&8.80&8.14&0.02&-0.66\\
2 all&8.34&0.45&35.79&11.03&8.10&8.77&7.96&-0.58&-0.77\\
3 all&9.36&0.74&31.43&11.00&7.64&8.51&7.71&-0.73&-0.79\\
4 all&3.66&0.39&156.6&11.00&8.07&8.73&7.97&-0.71&-0.79\\ \hline
\multicolumn{10}{l}{Subclusters}\\ \hline
1a&7.3&0.23&36.8&11.11&8.64&8.85&8.19&-0.18&-0.65\\
1b&6.1&0.46&105.8&11.19&8.81&8.33&7.59&0.48&-0.74\\
2a&6.9&0.44&24.6&11.03&8.01&8.86&8.03&-0.78&-0.76\\
2b&9.9&0.46&47.9&11.03&8.19&8.64&7.86&-0.43&-0.78\\
3a&9.8&0.73&20.6&10.98&7.62&8.55&7.76&-0.91&-0.79\\
3b&11.8&0.53&76.8&10.95&7.90&8.07&7.34&-0.21&-0.76\\
3c&5.1&1.12&50.4&11.16&7.02&8.58&7.71&-1.47&-0.87\\
4a&3.5&0.41&143.4&10.99&7.80&8.58&7.78&-0.74&-0.80\\
4b&4.6&0.30&262.2&11.04&8.74&9.24&8.56&-0.50&-0.68\\ \hline
\end{tabular}
\end{center}
\caption{Shown in this table are the parameter averages for each PNe
Cluster.} 
\label{clavg}
a.  in units of kpc\\
b.  in units $kms^{-1}$\\
c.  in form log(X/H)+12\\
d.  in form log(X/O)\\
e.  Solar abundances calculated from Anders and Grevasse (1989) 
\end{table*}

\begin{figure}
\centerline{\hbox{\psfig{figure=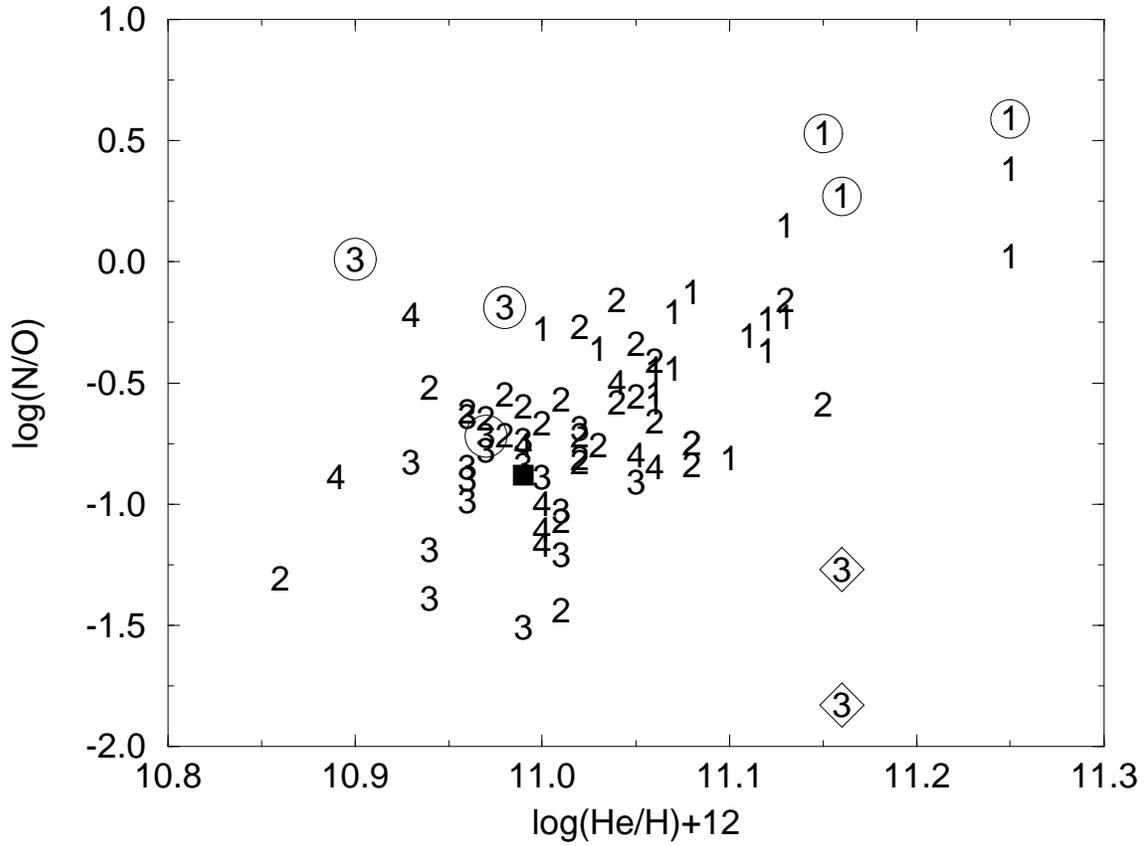,height=5truein}}}
\caption{Cluster comparison diagram for log(He/H)+12 versus
log(N/O). Cluster membership of each PN is indicated by a
number. Members of subcluster 1b are indicated by circled 1's, members
of 3b by circled 3's and members of 3c by a 3 enclosed by a
diamond. The solar values are indicated by the solid square.}
\label{fig:helno.eps}
\end{figure}

\begin{figure}
\centerline{\hbox{\psfig{figure=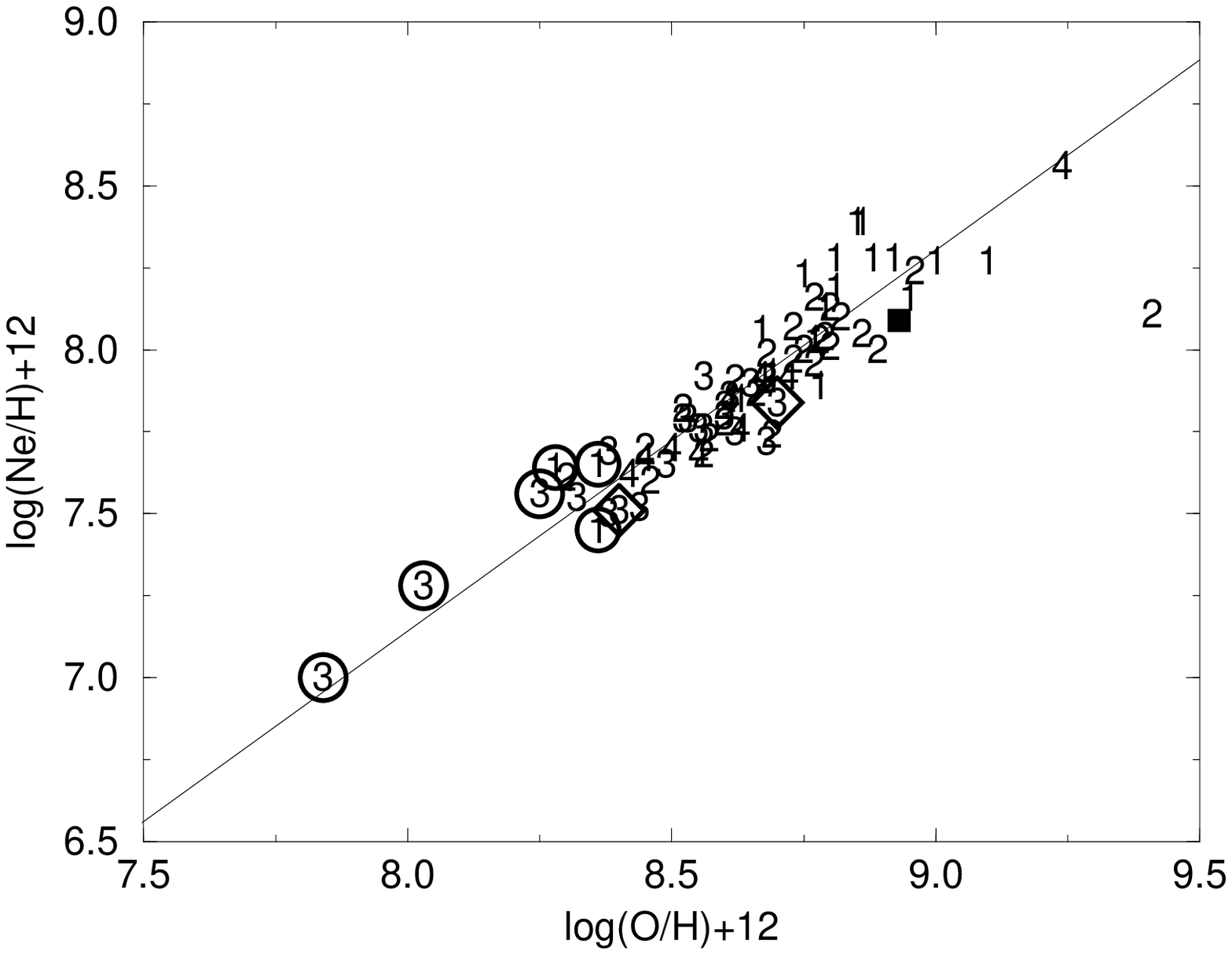,height=5truein}}}
\caption{Cluster comparison diagram for log(Ne/H)+12 versus
log(O/H)+12. The symbols have the same meaning as in figure
\ref{fig:helno.eps}.} 
\label{fig:ne-o.eps}
\end{figure}

\begin{figure}
\centerline{\hbox{\psfig{figure=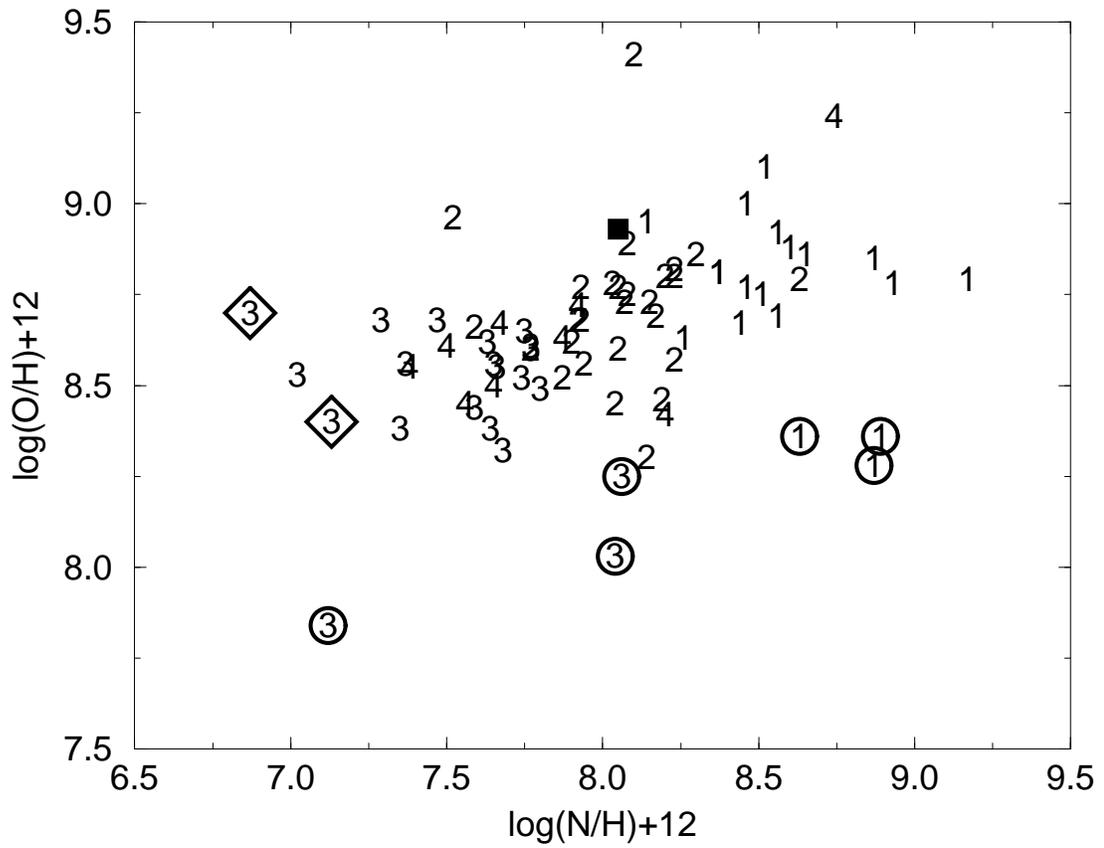,height=5truein}}}
\caption{Cluster comparison diagram for log(O/H)+12 versus
log(N/H)+12. The symbols have the same meaning as in figure
\ref{fig:helno.eps}.} 
\label{fig:on.eps}
\end{figure}

Figures \ref{fig:helno.eps}, \ref{fig:ne-o.eps}, and \ref{fig:on.eps}
are used to explore the ranges of the chemical parameters for each
cluster and subcluster. Figures \ref{fig:helno.eps} and
\ref{fig:on.eps} show a clear separation of clusters 1, 2, and 3 on
the He-N/O plane and in the abundance of N, i.e. it is possible to
draw lines on these figures which separate the clusters\footnote{The
most straightforward way to put other objects into these categories by
using the nitrogen abundances. Cluster 1 objects generally have
$log{N/H}+12\geq 8.3$, cluster 2 objects have $8.3\leq log{N/H}+12
\geq 7.8$, and cluster 3 objects have $log{N/H}\leq 7.8$.}. The ratios
He/H, N/H, and N/O increase along the sequence cluster 3, 2, and
1. There is no clear line of separation between clusters 1, 2, and 3
that can be seen on figures \ref{fig:ne-o.eps} and \ref{fig:on.eps},
although the average O/H and Ne/H also increases along the sequence
cluster 3, 2, 1. At the same time, cluster 4, while overlapping
clusters 2 and 3 in abundance patterns, is significantly separated
from these clusters in parameter space because of systematic
differences in velocity and galactocentric distances.  

The bulge PNe in the sample are almost exclusively confined to
clusters 1 and 4. Cluster 1 contains a mix of bulge and disk PNe with
similar abundances and it is very striking how similar the bulge and
disk PNe appear to be in abundance space. To test how well the bulge and
disk PNe could be separated by using only abundances as classification
parameters a trial cluster analysis was carried out using only the
abundance parameters resulting in the bulge and disk PNe being
completely mixed together in 3 clusters that roughly correspond to
clusters 1, 2, and 3. {\em The conclusion is that it is impossible to
distinguish between individual bulge and disk PNe based on abundances
alone.} 

We now turn our attention to the subclusters arising in
figures \ref{fig:helno.eps}, \ref{fig:ne-o.eps}, and
\ref{fig:on.eps}. The importance of the subclusters is that each may
contain objects which require explanation or may contain objects with
significant errors in the parameter determination. In clusters 1, 3,
and 4, the subcluster containing the most objects is designated
``a''. The ``a'' subclusters of clusters 1, 3, and 4 each contains
75\% or more of the total number of cluster objects, whereas
subclusters other than ``a'' each contain less than four objects.  The
subclusters of clusters 1 and 3 have been distinguished in figures
3a-c, the three 1b members are indicated by circled ones, the three 3b
members by circled threes and the two 3c members by threes enclosed by
a diamond. 

In figs. \ref{fig:helno.eps} and \ref{fig:ne-o.eps}, the 1b objects have He/H
and N/O values that are generally higher than the 1a objects, but the
1b objects have O/H and Ne/H that are $0.4dex$ less than the closest 1a
object.  This difference is larger than the average observational 
uncertainties, however this does not rule out the possibility of
observational error. If this gap is real then the 1a and 1b PNe
respectively represent oxygen-rich and oxygen-poor cluster 1 objects. 

Subcluster 4b contains a single bulge PN, CN2-1, which is high in
O/H and Ne/H, while subcluster 4a contains bulge planetaries with
much lower abundances.

The subclusters of cluster 3 are widely separated in parameter space.
The PNe in cluster 3c are I4593 and H1-23 which appear anomalous as
discussed in the PCA section. The PNe of cluster 3a have
$V_{LSR}<40kms^{-1}$ and $log(O/H)+12>8.55$ and those of cluster 3b
have $V_{LSR}>50kms^{-1}$ and $log(O/H)+12<8.55$. The low abundances
and large velocities of the 3b objects suggests that these objects are
thick disk objects and the abundances and velocities of the 3a objects
suggest a thin disk origin.

\section{Discussion}
The results of our cluster analysis confirm the expected, the primary
classification criterion for PNe is the nitrogen and helium
content. Spatio-kinematic criteria determine the population of each
PN. There also appears to be some use in looking at the oxygen and
neon content of each nebula, since the 1b PNe appear to be a class of
oxygen poor nebulae with high nitrogen and helium content.

What accounts for the differences among the clusters produced in the
analysis? The models of Becker and Iben (1980) and Renzini and Voli
(1981) suggest that as the mass of the progenitor increases, the
helium and nitrogen abundances in the resulting PN increase. Studies
comparing N/O to core mass indicate that a correlation exists between
these quantities 
(Kaler and Jacoby 1989, Kaler and Jacoby 1990, Stasi\'{n}ska and
Tylenda 1990) for disk PNe. If we make the reasonable assumption that
as core mass increases the progenitor mass increases, which is
supported by the initial-final mass relation of 
Weidemann and Koester (1983) then we infer that since nitrogen increases
along the sequence from cluster 3 to cluster 1, the progenitor mass
also increases. 

\begin{figure}
\centerline{\hbox{\psfig{figure=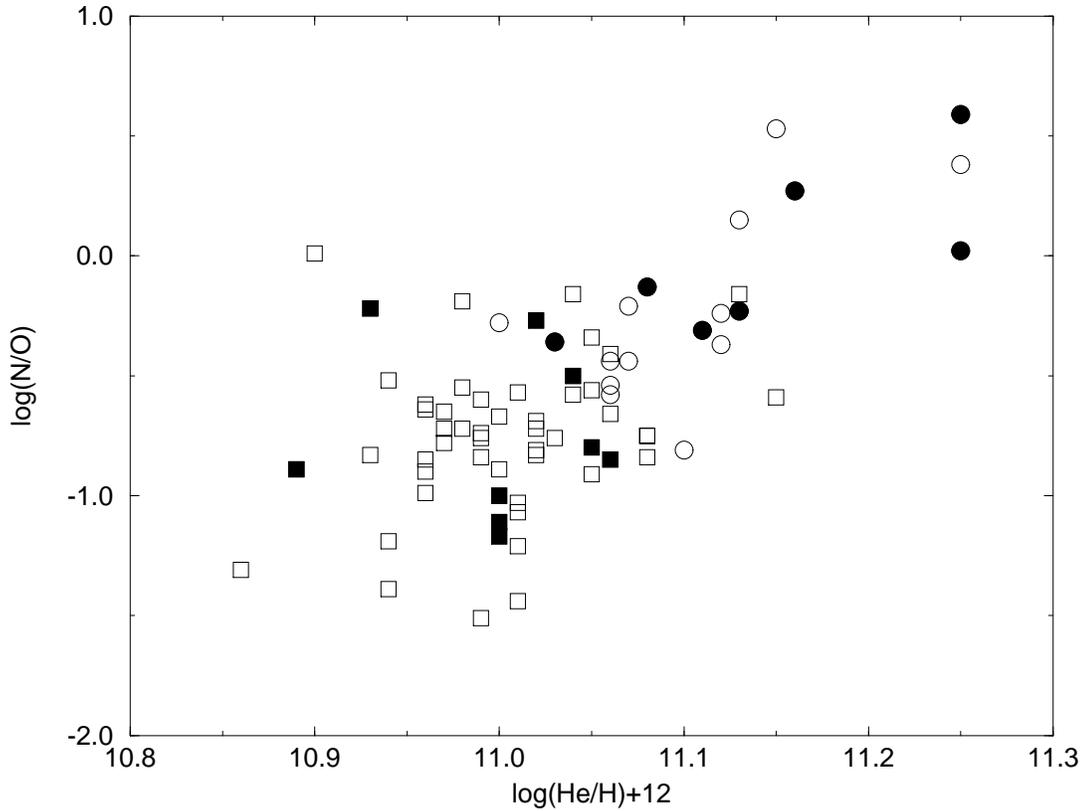,height=5truein,angle=270}}}
\caption{log(He/H)+12 versus log(N/O) for bulge and disk PNe. The
filled symbols represent the bulge objects and the open symbols the
disk objects. The circles represent the cluster 1 objects and the
squares the non-cluster 1 objects. The two PNe of cluster 3c have been
removed.}
\label{fig:henobd.eps}
\end{figure}

This picture works for the disk objects. However, eight of nineteen
PNe in cluster 1 are classified as bulge objects. Does this trend
hold for them as well?  In Fig.~\ref{fig:henobd.eps} we have plotted log(N/O)
versus log(He/H)+12, where the bulge objects are represented by filled
symbols and the disk objects by open symbols. Cluster~1 objects have
been distinguished by circles, and members of other clusters by squares.
It can be seen that the bulge and disk objects are completely mixed,
{\it suggesting that there is no difference between disk and bulge PNe
in terms of their nucleosynthesis}. 

The apparent similarity in He/H and N/O (nucleosynthesis) patterns of
bulge and disk PNe implies that their progenitors represent the same
mass range. However, studies of bulge stars by Terndrup (1988) and van
der Veen and Habing (1990) suggest that the upper mass limit for bulge
stars is $\approx 2 M_{\odot}$, whereas according to theory
(Renzini and Voli 1981), to account for the ratios of He/H and N/O
seen in our sample, main sequence progenitors of $5-8M_{\odot}$ are
needed, which is in disagreement with the $2 M_{\odot}$ upper mass
limit for bulge stars. We suggest the following possibilities to
account for this discrepancy: 
\begin{enumerate} 
\item An unidentified population of intermediate mass stars exists in
the bulge. This has also been suggested by Webster (1988) on the basis
of her study of bulge PNe. The number of Mira variables with
$M_{bol}<-5.0$ has been estimated at 100 
(Whitelock 1992). These luminous Miras are believed to be the progeny
of main sequence stars with $M>3M_{\odot}$, suggesting that the number
of progenitors in the right mass range is small. It has been suggested
that these luminous variables are the progeny of binary mergers in the
bulge 
(Renzini 1994).
\item The progenitors of the cluster 1 bulge PNe are low mass stars
that formed with He/H and N/O similar to that found in disk PNe. This
scenario appears to be a better explanation for the lack of a
correlation between N/O and core mass.
\item The progenitors of the cluster 1 bulge PNe are binary stars,
where the companion is now an unseen white dwarf, but in the past was
an AGB star with an envelope enriched in He and N which was transferred
to the PN progenitor. Recently 
McWilliam and Rich 1994 found a Li rich star in a small sample (12
stars) of galactic bulge K giants, which suggests that this binary
scenario may be reasonably common in the galactic bulge, since the
most massive intermediate mass stars may become enriched in Li as a
result of hot bottom burning 
(Sackmann and Boothroyd 1992). Lithium rich AGB stars have also been
observed in the Magellanic Clouds 
(Smith and Lambert 1989, Smith and Lambert 1990)
and in the galaxy
(see, e.g. Abia \etal\ 1991). 
\item The cluster 1 bulge planetary nebulae may really be disk
planetaries at the same distance as the bulge (Jacoby 1996). It seems
possible that high mass objects near the bulge could look
kinematically similar to bulge objects. Kinematic studies of these
objects might reveal this possibility.
\end{enumerate}

\begin{figure}
\centerline{\hbox{\psfig{figure=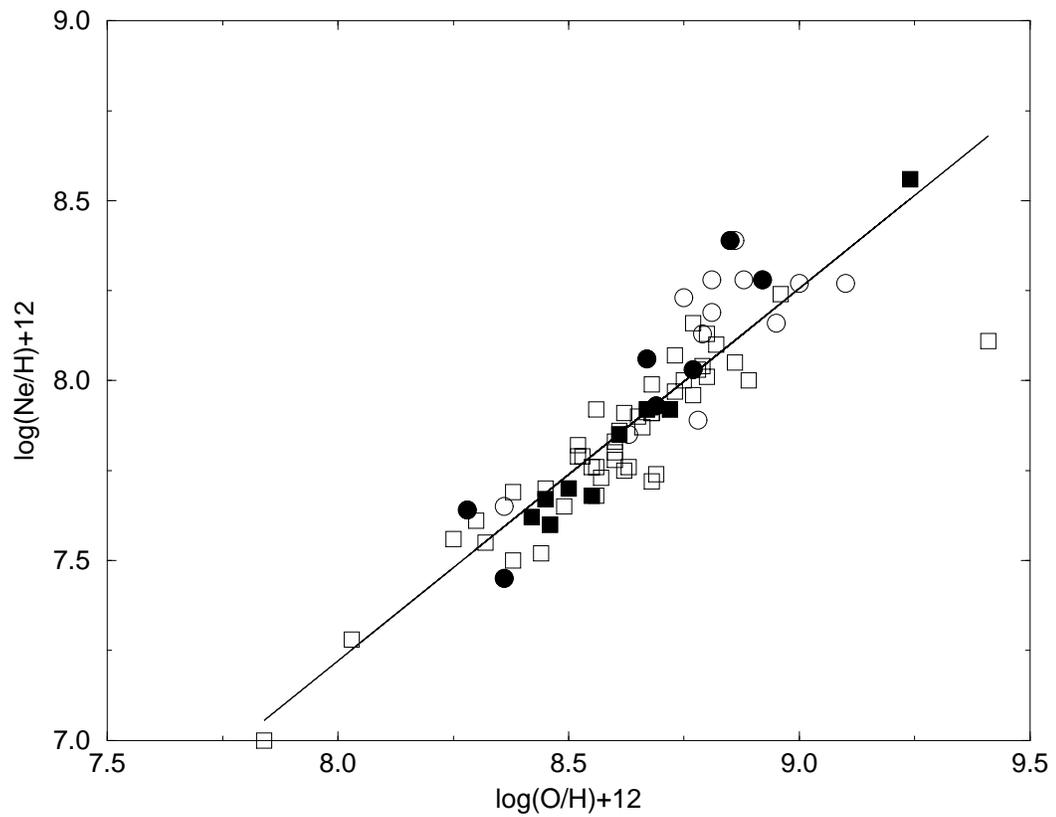,height=5truein,angle=270}}}
\caption{log(Ne/H)+12 versus log(O/H)+12 for bulge and disk PNe. The
symbols have the same meaning as in the previous graph. The line is a
least squares fit to the data.}
\label{fig:neobd.eps}
\end{figure}

Is there a difference in progenitor metallicity between the bulge and
disk PNe? Figure \ref{fig:neobd.eps} is a plot of Ne vs. O, where the
symbol notation is the same as in fig \ref{fig:henobd.eps}. We assume
that Ne and O trace the progenitor metallicity. The line is a least
squares fit to the data, the equation being given by; 
\begin{eqnarray}
log(Ne/H)+12=(1.034\pm 0.052)(log(O/H)+12)+(-1.057\pm 0.454).
\end{eqnarray}
\noindent
The bulge PNe clearly fall along the same line as the disk PNe,
implying that bulge objects have the same constant Ne/O ratio found in
disk PNe. This evidence suggests that the rates of enhancement of the
ISM by Ne and O relative to each other are the same in the bulge and
disk. 

We now draw attention to the obvious fact that the natural separation
of PNe in our sample into four distinct clusters forms the statistical
equivalent of the well-known Peimbert classification scheme (Peimbert
1978). Table 4 shows a comparison of Peimbert type with our cluster
classification.  For each of our subclusters in column~1, columns 2-5
give the number of objects in that group which belong to the Peimbert
types indicated by the headings of those columns. Clearly cluster 1
corresponds to the type I PNe, cluster 2 contains PNe with
intermediate nitrogen abundances similar to the type IIa PNe, cluster
3 contains PNe with the lowest nitrogen abundances which is similar
to the type IIb and III PNe, and finally the cluster 4 PNe correspond
to solar metallicity type V or bulge PNe. The cluster 3 subgroups
further bifurcate into type IIb and type III PNe. There also appears
to be a bifurcation of the type I PNe. In our scheme the cluster 1a
PNe would correspond to an oxygen rich type I and the 1b PNe would
correspond to an oxygen poor type I PNe. It seems that the Peimbert
scheme is also applicable to the bulge since these planetaries are
similar to the disk PNe in composition.

\section{Conclusions}
A multivariate data analysis of galactic planetary nebulae has been
carried out with the goal of determining the effective dimensionality
of PN parameter space as well as the importance of the numerous
observational parameters in classifying PNe.  The parameters employed
for this study are the abundance ratios He/H, N/H, O/H, Ne/H, N/O,
Ne/O, and the spatial and kinematic parameters R, Z, and $V_{LSR}$. Our
main results are: 

\begin{enumerate}
\item Planetary nebula abundance parameter space is two dimensional,
i.e. the data can be described by two principal components.  The first
and most important one is related to the products of stellar
nucleosynthesis during the evolution of PN progenitors.  The second
component is related to progenitor metallicity. 

\item The kinematic properties are not linearly correlated with any
other parameters.

\item The disk PNe separate into three clusters, each having a
distinct location in the nucleosynthesis-metallicity plane.  A fourth
cluster comprises many of the bulge PNe. The bulge PNe only separate
from the disk PNe because of the inclusion of the parameters R and
$V_{LSR}$. 

\item We have identified some unusual type I PNe with extremely low
oxygen abundances. However, it is not clear if these are real.

\item Bulge and disk PNe show the same distributions along the
nucleosynthesis and metallicity components. 

\end{enumerate}

This analysis is a first attempt to characterize planetary nebula
parameter space in an objective way. The future addition of parameters such as
carbon abundance, nebular expansion velocity, morphology, binarity,
and central star mass further improve our understanding of these objects.

\newpage
\chapter{Thermally Pulsing AGB Models}
\label{modelchap}

Thermally pulsing asymptotic giant branch (TP-AGB) stars
experience two distinct and repeating phases: the helium shell flash
or thermal pulse (TP) and the time between pulses or the interpulse
phase (IP). The duration of the thermal pulse ($\tp$) is short
compared to the duration of the interpulse phase ($\tip$) with $\tp
/\tip\approx 0.01$. During the TP-AGB, these stars spend most of the
time in the interpulse phase, burning both He and H burn quiescently
in thin shells. During the interpulse phase, most of the luminosity 
is produced by the hydrogen burning shell. However, the helium shell
in this configuration is unstable, and it will eventually result in a
helium shell flash. Thermal pulses occur quasi-periodically, resulting
in the luminosity of the helium burning shell increasing by several
orders of magnitude. The increased luminosity causes the outer layers
of the star to expand, including the hydrogen burning shell. The
luminosity generated by hydrogen burning during the shell flash
decreases to essentially zero due to the expansion. After the thermal
pulse the star resumes its interpulse configuration. This cycle
repeats until the envelope is lost via mass-loss. 

The two distinctive phases of the TP-AGB have important consequences
for the surface abundances of the star. At the end of a thermal pulse
it is possible for the convective envelope to penetrate into regions
where He has been partially burned into $\ctw$ and minor amounts of
$\oxy$, mixing these products to the stellar surface. This process is
called the third dredge-up and is believed to be the process by which
carbon stars are produced. In the more massive TP-AGB stars it is
possible for the base of the convective envelope to get hot enough to
burn $\ctw$ to $\cth$ and $\nit$ and $\oxy$ to $\nit$ in a process
known as hot-bottom burning. This process is believed to produce the
nitrogen rich PNe. Hot bottom burning can take the carbon dredged-up
into the envelope and process it into nitrogen. This is one of the
possible solutions to the so-called carbon star mystery.

The third dredge-up occurs at the end of a thermal pulse and mixes
helium and carbon rich material into the stars outer layers. During
the thermal pulse a convective shell appears in the helium burning
zone. This convection zone spans the mass range from the base of the
interpulse He burning shell to just below the position of the
convective envelope during the preceding interpulse phase. In this
zone these important helium burning reactions take place:
\begin{itemize}
\item{$\he +\he\longrightarrow ^8{\rm Be}+\gamma$}
\item{$^8{\rm Be}+\he\longrightarrow\ctw +\gamma$}
\item{$\ctw +\he\longrightarrow\oxy +\gamma$}
\end{itemize}
with the first two being the most important. The third reaction does
not occur very often. Therefore, the convective shell at the end of
the pulse is composed primarily of helium and carbon. During the
transition period between the thermal pulse and the resumption of the
interpulse phase the convective shell disappears and the convective
envelope can penetrate into this region. The mixing of carbon and
helium rich material to the surface layers of the star at the end of a
thermal pulse is known as the third dredge-up to distinguish it from
the first two dredge-ups which we will describe later.


\section{Models}
\subsection{The Structure of the Envelope}
Computing the structure of a TP-AGB star during the thermal pulse is a
difficult task. To follow the thermal pulse from beginning to end
requires the computation of a new stellar model with several hundred
mass zones over several hundred time steps for each thermal pulse. This
process is obviously very time consuming. In addition, these
computations are beset by numerical difficulties, in particular the
difficulty getting these models to converge (e.g. Frost and Lattanzio
1996). It is also difficult to get the third dredge-up to occur
without the presence of some form of extra mixing, i.e. overshoot,
semi-convection, etc. Most of these difficulties can be traced to the
rapid changes of the stars structure.

Calculating the structure during the interpulse phase is a far less
demanding task than doing the same during the thermal pulsing
phase. During the interpulse phase the structure changes slowly, so at
most only a few models are needed. It is possible to parameterize the
effects of the thermal pulses on the star and to partially
parameterize the interpulse phase. This technique is known as synthetic
modeling and was pioneered by Iben and Truran (1978). The thermal
pulses are parameterized so as to calculate the amount of dredge-up
and the change in the abundances of all the regions of the star. The
parameterization of the interpulse phase is limited to calculating the
luminosity as a function of the stellar parameters. The evolution of
the envelope of the star can then be calculated using standard
techniques.  

Our computer code, XYCNO, calculates synthetic TPAGB models. XYCNO is
a significantly updated version of a code kindly supplied to us by
Dr. Renzini. The changes in the star due to the thermal pulses are
parameterized, allowing rapid computation of this phase. The changes
in the star (chemical and physical) during the interpulse phase are
done by first computing a model of the envelope. The code 
computes the envelope structure by taking as an input the interpulse
surface luminosity, $\ls$, guessing $\teff$ and integrating the
equations of stellar structure from the surface to the base of the
convective envelope. This procedure is iterated until the mass
position of the base of the convective envelope, $\mce$, is the same
as the position of the hydrogen-exhausted core, $\mc$. The output of
each envelope integration is the $\teff$ of the model and the
structure of the envelope, which are then used to calculate some
important parameters such as mass loss and also the envelope
nucleosynthesis.


The code follows the abundances of $\he$, $\ctw$, $\cth$, $\nit$,
and $\oxy$ in the surface layers of the star. The following nuclear
reactions are followed: 
\begin{itemize}
\item{$\ctw ({\rm p},\gamma )^{13}{\rm N}({\rm e}^{+}\nu _{\rm e})\cth$}
\item{$\cth ({\rm p},\gamma )\nit$}
\item{$\nit ({\rm p},\gamma )^{15}{\rm O}({\rm e}^{+}\nu _{\rm e})
^{15}N({\rm p},\alpha )\ctw$}
\item{$^{15}{\rm N}({\rm p},\gamma )\oxy ({\rm p},\gamma )^{17}{\rm
F}({\rm e}^{+}\nu _{\rm e})^{17}{\rm O}({\rm p},\alpha )\nit$.}
\end {itemize}
The rate for each reaction was taken from 
Caughlan \etal\ (1988) and references therein. The envelope is assumed
to be mixed instantaneously. 
 
The 1995 updated OPAL opacities ($\kappa _{OPAL}$) which are described
in 
Rogers and Iglesias (1992) are used when T$>$10000~K 
and the molecular opacities ($\kappa _{\rm mol}$) of 
Alexander and Ferguson (1994) are used for T$<$6000~K. At intermediate
temperatures a weighted average of the both types of opacities is
used. Points not on the grids are computed using the quadratic
interpolation scheme of Rogers and Iglesias. Some of the molecular
opacities were checked using the PHOENIX code (Hauschildt 1992ab,
1993) and found that they agree to within 1-4\% . 

\subsection{Mass Loss on the AGB}
The {\em relationship} between the zero-age main sequence (ZAMS) mass and
the white-dwarf (WD) mass M$_{\rm i}-$M$_{\rm f}$ of 
Weidemann (1987) indicates that low and intermediate mass stars must
lose $\sim0.3-7.0\msun$ of mass between the main sequence and the white
dwarf stages. AGB stars are observed to be losing mass with rates
ranging between $10^{-8}-10^{-4}\msolyr$ (e.g. Whitelock \etal\
1994). Therefore the AGB is probably the evolutionary phase where most
of the mass loss takes place. 

Observations of AGB stars indicate that the mass-loss rate probably
goes through two phases: a period of moderate mass-loss where an
ordinary wind operates ($\dot{\rm M}<10^{-5}\msolyr$), then a rapid
transition occurs and the star loses mass via a ``superwind'' where
$\mdot\sim10^{-4.3}\msolyr$ 
(Renzini 1981). The
mass-loss during the ordinary wind phase is often represented by the
Reimers (1975) mass-loss relation given by
\begin{equation}
\mdot_{R}=-(4\times 10^{-13})\eta{\rm L}{\rm R}/{\rm M}
\end{equation}
where $\mdot$ is in $\msolyr$, the luminosity L, the radius R, and the
mass M are in solar units. $\eta$ is a parameter of order unity. 

Bazan (1991) and 
Vassiliadis and Wood
(1993) have derived relations between $\mdot$ and pulsation period
P. Both find that the observational data can be fit with two
components; Below P$\approx$600 days there is a linear relation
between $\mdot$ and P and for periods longer the mass loss rate is
constant ($\sim 5\times 10^{-5}\msolyr$). 

There is considerable scatter in the data with any given period having
stars with mass-loss rates that vary by an order of magnitude. In fact,
comparison of the relations of Bazan and of Vassiliadis and Wood
indicate a qualitative similarity but quantitatively, $\mdot$ varies by
up to a factor of 5. Iben (1995) 
also notes that the
brightest galactic OH/IR stars are optically invisible but long-period
variables in the Magellanic Clouds with similar pulsation periods are
optically visible, indicating a smaller mass loss rate. Clearly, there
is no unique $\mdot -$P relationship.

Bowen (1988), 
Bowen and Willson (1991), and 
Willson \etal\ (1995) have developed a promising theory of AGB
mass-loss rates. The pulsations of the long period variable stars
causes shocks in the atmosphere. These shocks levitate the atmosphere,
increasing the scale height considerably. The outer parts of the
atmosphere have higher densities than they would otherwise have in a
static atmosphere, this allows larger grain nucleation further from
the star. The grains are accelerated outward by radiation
pressure. The grains transfer momentum to the gas and this drives the
mass loss. In this formulation, the mass-loss goes through two stages;
mass is lost by an ordinary wind until a certain critical luminosity
is exceeded at which point the mass-loss rate increases rapidly to
superwind like rates. The critical luminosity increases as the
metallicity of the star is decreased.

In this model three different mass-loss rates are used:
\begin{enumerate}
\item{The Reimers mass-loss rate, $\mdot_{\rm R}$, given above.}
\item{A Pulsation period mass loss rate, $\mdot_{\rm PP}$, given by 
\begin{eqnarray}
\log{\dot{\rm M}}(\msolyr)=-11.4+0.0123{\rm P}\\
\log{P~}(days)=-2.07+1.94\log{R/R_{\odot}}-0.9\log{M/\msun}
\end{eqnarray}
where R is the radius of the star. Note that we do not include their
modification for M$>2.5\msun$.}
\item{A superwind mass-loss rate, $\mdot_{\rm SW}$, which we take as
$5\times 10^{-5}\msolyr$.}
\end{enumerate}
Relation (1) is followed until $\mdot_{\rm PP}>\mdot_{\rm R}$, after
which relation (2) is used. Relation (2) is used until $\mdot_{\rm
PP}>\mdot_{\rm SW}$, after which a constant mass-loss rate of $5\times
10^{-5}\msolyr$ is used.

\subsection{Surface Luminosity}
In synthetic TPAGB modeling, one of the most important parameters to
model is the surface luminosity during the interpulse phase. The
surface luminosity is one of the outer boundary conditions of the
model star. This is not simple as the luminosity depends on mass,
core-mass, composition, and pulse number. The luminosity can also vary
significantly during each interpulse phase.

The ability in the past to calculate the interpulse luminosity as a
function of core-mass is a very useful property for modeling TP-AGB
stars. 
Paczy\'nski (1971) and 
Uus (1970)
discovered that maximum luminosity during the interpulse phase could
be represented by the following ``core-mass luminosity'' relation:
\begin{equation}
\ls = 60000({\rm M}_c-0.495)
\end{equation}
where $\mc$ is the mass of the hydrogen exhausted core or the
core-mass. 
Iben (1975) discovered the following
relationship for the luminosity: 
\begin{equation}
\ls = 63400(\mc -0.44)({\rm M}/7\msun)^{0.19}
\end{equation}
where M is the mass of the star. 
The above relations were derived from
stellar models of mass greater than 5$\msun$, 
Wood and Zarro (1981) derived a relation for low-mass stars
(M$<3\msun$). 
Boothroyd and Sackmann (1988) derived this core-mass, ${\rm M}_{\rm
c}$ luminosity relation for low-mass stars (M$<3\msun$)of solar
metallicity:
\begin{eqnarray}
{\rm L}_{\rm s}=52000(\mc -0.456)& 0.52\le{\rm M}_{\rm c}\le 0.7
\end{eqnarray}
and the following relation for all metallicities:
\begin{eqnarray}
{\rm L}_{\rm s}=238,000\mu^3({\rm Z}_{\rm
CNO})^{0.04}(\mc^2-0.0305\mc-0.1802)& 0.5<\mc <0.66
\end{eqnarray}
where ${\rm Z}_{\rm CNO}$ is the mass fraction of carbon, nitrogen,
and oxygen and $\mu$ is the mean molecular weight. 

Core-mass luminosity relations only give the luminosity at the local
``asymptotic'' limit. It is well known that the luminosity during the
first interpulse does not correspond to the core-mass luminosity
relation, and in general 5-10 pulses are needed to reach it. The first
thermal pulses occur when the helium burning shell still produces a
significant fraction of the luminosity ($\sim$50\%), but after a
few pulses the helium burning shell only produces a few percent of the
luminosity.  

In recent years, however, it has become apparent that a simple
core-mass luminosity relationship (with or without a metallicity
dependence) is not appropriate for intermediate mass stars
(M$>3.5\msun$) The luminosity now appears to depend on the stellar
mass as well. \nocite{tgb83} Tuchman \etal\ (1983) showed using
semi-analytic arguments that a core-mass luminosity relation holds
for AGB stars only when the hydrogen burning shell is separated from
the convective envelope. They found that a core-mass luminosity
relationship is not appropriate if the convective shell penetrates the
hydrogen burning layer. 
\nocite{bs91} Bl\"{o}cker and Sch\"{o}nberner (1991) modeled a
7$\msun$ star and found that it did not follow any kind of core-mass
luminosity behavior because the convective envelope penetrated the
hydrogen burning layer. This effect has been confirmed by the TP-AGB
models of \nocite{b95} Bl\"{o}cker (1995), \nocite{bs92} Boothroyd and
Sackmann (1992), \nocite{bsa93} Boothroyd \etal\ (1993), \nocite{l92}
Lattanzio (1992), and \nocite{vw93} Vassiliadis and Wood (1993). 


We have derived a new relation for the surface luminosity for
intermediate mass stars (M$>3.5\msun$). This relation is dependent on
mass, core-mass, metallicity, and pulse number. It consists of three
parts: 1) a description of the luminosity at the first pulse, 2) the
steep rise in luminosity from the first pulse until asymptotic
behavior is achieved, and 3) the asymptotic luminosity. The models of
Boothroyd and Sackmann (1992), and Boothroyd \etal\ (1993) with Sharp
(1992) molecular opacities, hereinafter Boothroyd and Sackmann models,
were used for this purpose. These models were chosen since they
compute a reasonably large grid of models in mass and metallicity and
the Sharp molecular opacities are probably better than the Los Alamos
opacities. Dr. Boothroyd (1995) kindly supplied me with machine
readable tables of these models.  

As mentioned earlier, the luminosity at the first pulse is less than
the asymptotic value. 
\begin{figure}
\centerline{\hbox{\psfig{figure=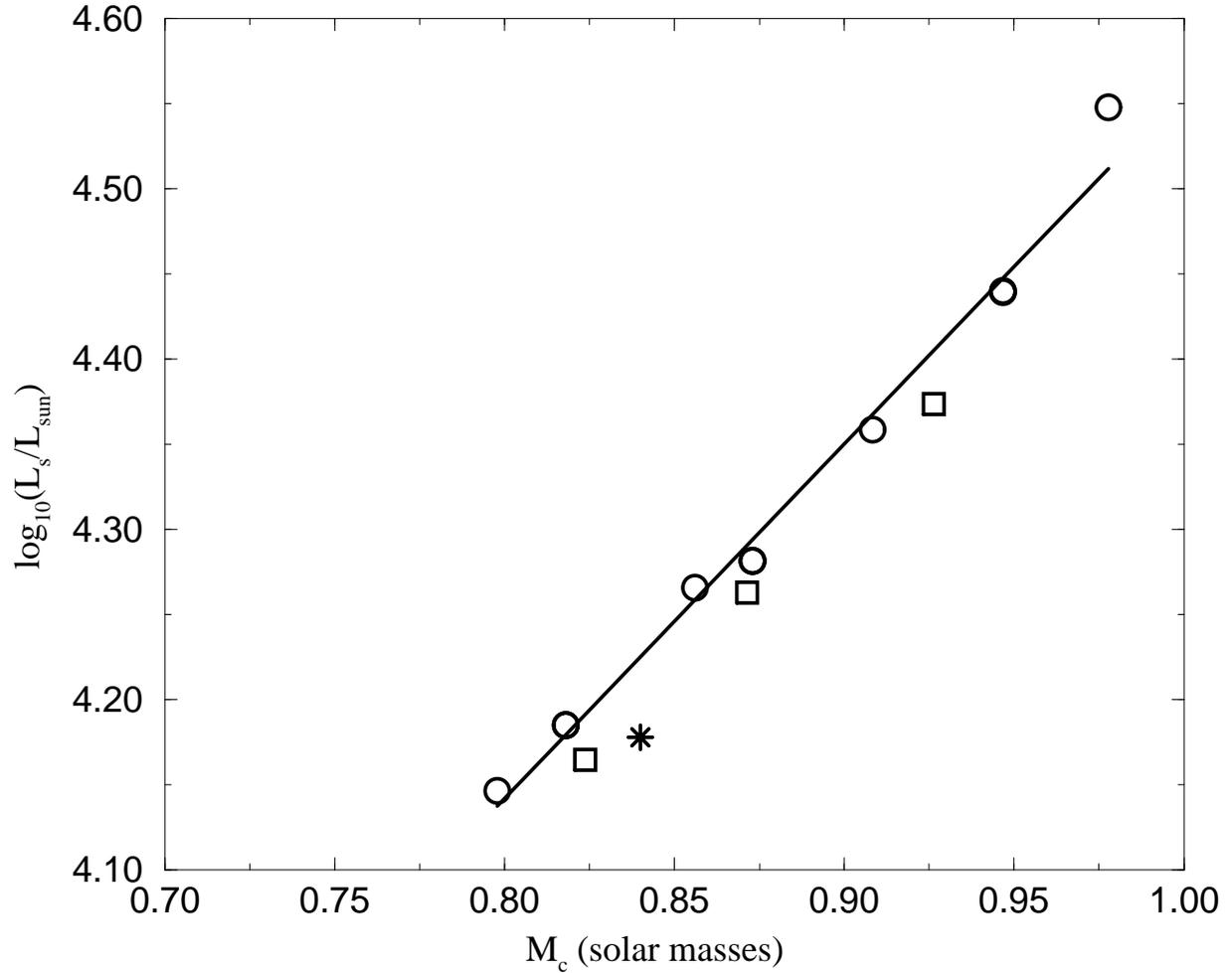,height=6truein}}}
\caption{Shown in the figure is the interpulse luminosity at the time
of the first thermal pulse as a function of core-mass. The Z=0.02,
0.01, and 0.0044 models of Boothroyd and Sackmann (1992) are
respectively shown by circles, squares, and stars. The
solid line is a least squares fit to the Z=0.02 models.}
\label{fig:lumzero.eps}
\end{figure}
In figure \ref{fig:lumzero.eps}, the luminosity at the first
interpulse of the Boothroyd and Sackmann models are shown. The line is
a fit to the Z=0.02 models. The Z=0.01 models also clearly follow a
linear relation with a slope nearly identical to the Z=0.02
models. There appears to be a slight trend of decreasing $\ls$ with
metallicity. For M$_{\rm c}>0.7\msun$ we adopt the following relation
for the luminosity at the first pulse; 
\begin{equation}
\log{\rm L_s(0)}=2.07\mc +2.48-3.(.02-Z)
\end{equation}
where Z is the metallicity of the model. For models with M$_{\rm
c}\le0.7\msun$ the expressions of 
Lattanzio (1986) are
used: 
\begin{eqnarray}
L(0)=29000(M_c-0.5)+1000&Z=0.001\\
L(0)=27200(M_c-0.5)+1300&Z=0.02,
\end{eqnarray}
where values at other metallicities are found by linearly 
extrapolating/interpolating in $\log{\rm Z}$.

A function describing the rise from the luminosity at the first pulse
to the local asymptotic limit would be very complicated if it could be
described at all. However, in this paper we choose to approximate it
with two linear fits, the first describing the rise from the first
pulse to a point approximately halfway to the asymptotic value, and
the second describing the rise from the halfway point to asymptotic
values. The first part, the rise from the first pulse to the halfway
point is described by this equation: 
\begin{equation}
\ls =A({\rm M}_{\rm c}-{\rm M}_{\rm c,0})+{\rm L_s(0)},
\end{equation}
where A is the slope of the relation. For $\mc >0.7$ the slope A was
found by fitting a line to the luminosity of the first few ($\sim
5-7$) interpulses for each Boothroyd and Sackmann model. 
\begin{figure}
\centerline{\hbox{\psfig{figure=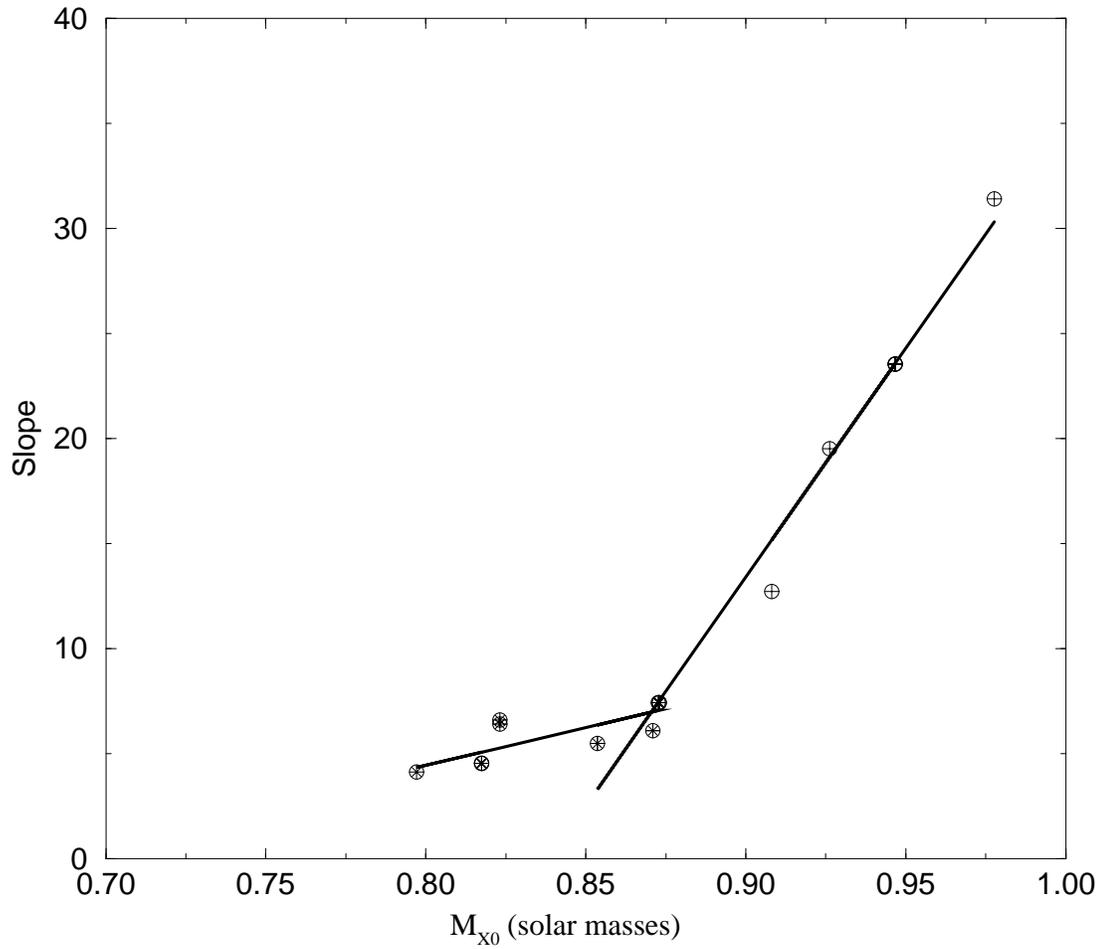,height=5truein}}}
\caption{The initial rate of rise is shown as a
function of core-mass. No metallicity dependence is shown. The two
component fit to the data is shown.}
\label{fig:lsrise.eps}
\end{figure}
In figure \ref{fig:lsrise.eps} the derived slopes are shown as a
function of core-mass and these are fit by two lines. A transition
takes place at $\mc \approx 0.87\msun$ where the line becomes
steeper. We use the following relationship to describe A:
\begin{eqnarray}
A=(218\mc -182)\times 10^5&{\rm M}_{\rm c}>0.87\\
A=(35.8\mc -24.2)\times 10^5&0.71<{\rm M}_{\rm c}\le 0.87\\
A=1.2\times 10^{5}&{\rm M}_{\rm c}\le 0.71\msun.
\end{eqnarray}
For the lowest core-masses we estimated the rate of rise in the low
mass models of Boothroyd and Sackmann (1988abcd). When the luminosity
is halfway to the asymptotic values, the slope is reduced to half its
previous value and is then allowed to continue until asymptotic values
are reached. This is done so the shape more closely approximates the
behavior of $\ls$ with $\mc$ exhibited by models.

The asymptotic luminosity for all models is found by finding a
core-mass luminosity, L$_{\rm cm}$, and then multiplying L$_{\rm cm}$
by a correction factor, $f$, which depends on mass. For the Boothroyd
and Sackmann models, L$_{\rm cm}$ is calculated from the following:
\begin{equation}
{\rm L}_{\rm cm}=52000({\rm M}_{\rm c}-0.456)
\end{equation}
For each interpulse model $f$ was found by dividing the model
luminosity, $\ls$, by the core-mass luminosity, L$_{\rm cm}$. 
\begin{figure}
\centerline{\hbox{\psfig{figure=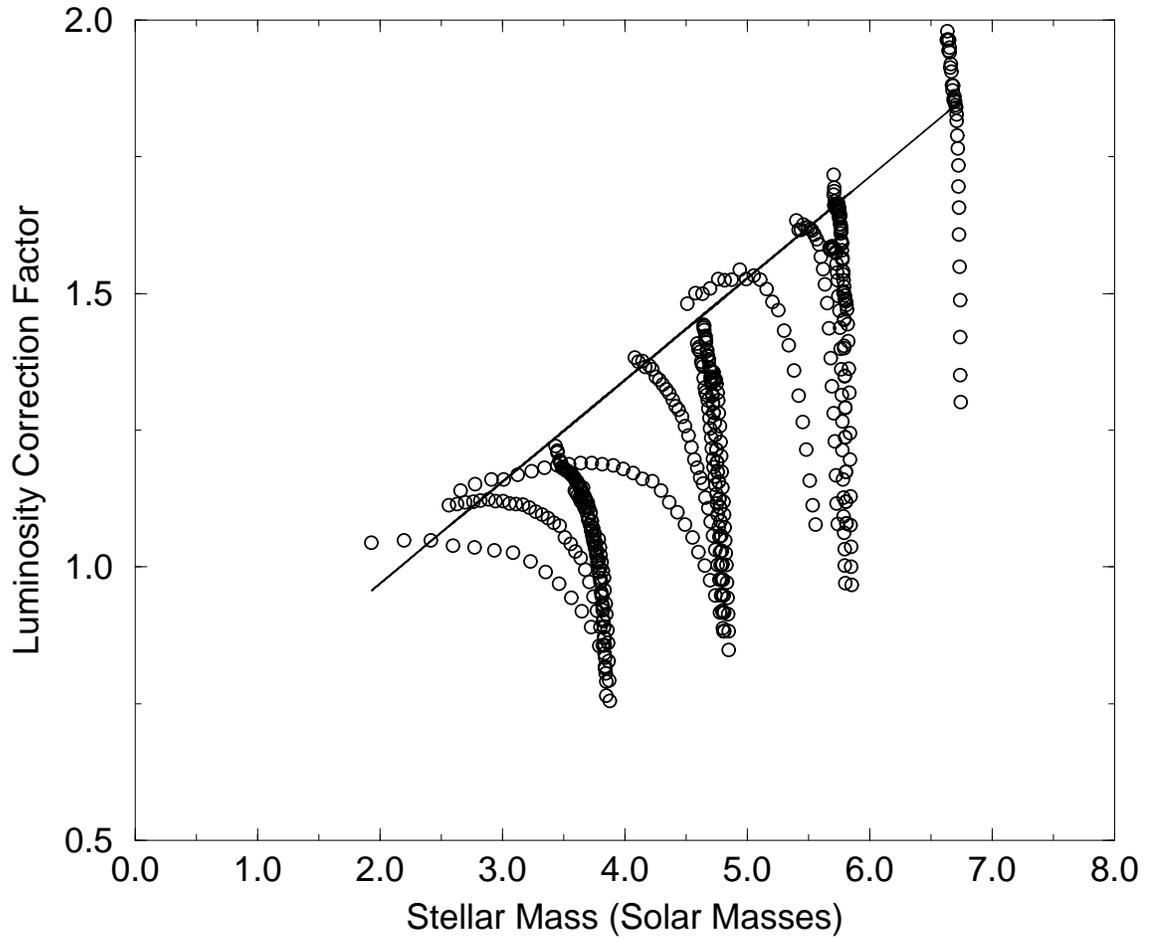,height=5truein}}}
\caption{The mass correction factor $f$ is plotted
versus mass for the Boothroyd and Sackmann models. The solid line is
our fit to the asymptotic value of $f$.}
\label{fig:lsasy.eps}
\end{figure}
The
results are plotted in figure \ref{fig:lsasy.eps} as a function of
mass. The line is a fit to $f$ of the last few interpulses of each
model. The fitted line roughly corresponds to the upper luminosity
limit of each Boothroyd and Sackmann model. Also, we note the line
corresponds to the last few interpulses of several of the Boothroyd
and Sackmann models, therefore, we propose to use the line as the
asymptotic correction factor.

The asymptotic surface luminosity is found from:
\begin{eqnarray}
L_s =f{\rm L}_{\rm cm} &\\
f=1+0.186({\rm M}-2.17) &{\rm M}>2.17\msun\\
f=1 &{\rm M}\le 2.17\msun 
\end{eqnarray}
where M is the mass of the model star and the line fitted in figure
\ref{fig:lsasy.eps} gives the equation for the correction factor,
$f$. For M$_{\rm c}<0.70\msun$ equation~7 is used to find L$_{\rm cm}$
and for $\mc$$>0.75\msun$, it is determined from equation~15. Between
0.70$\msun$ and 0.75$\msun$ we linearly interpolate in $\mc$ between
the values.

\begin{figure}
\centerline{\hbox{\psfig{figure=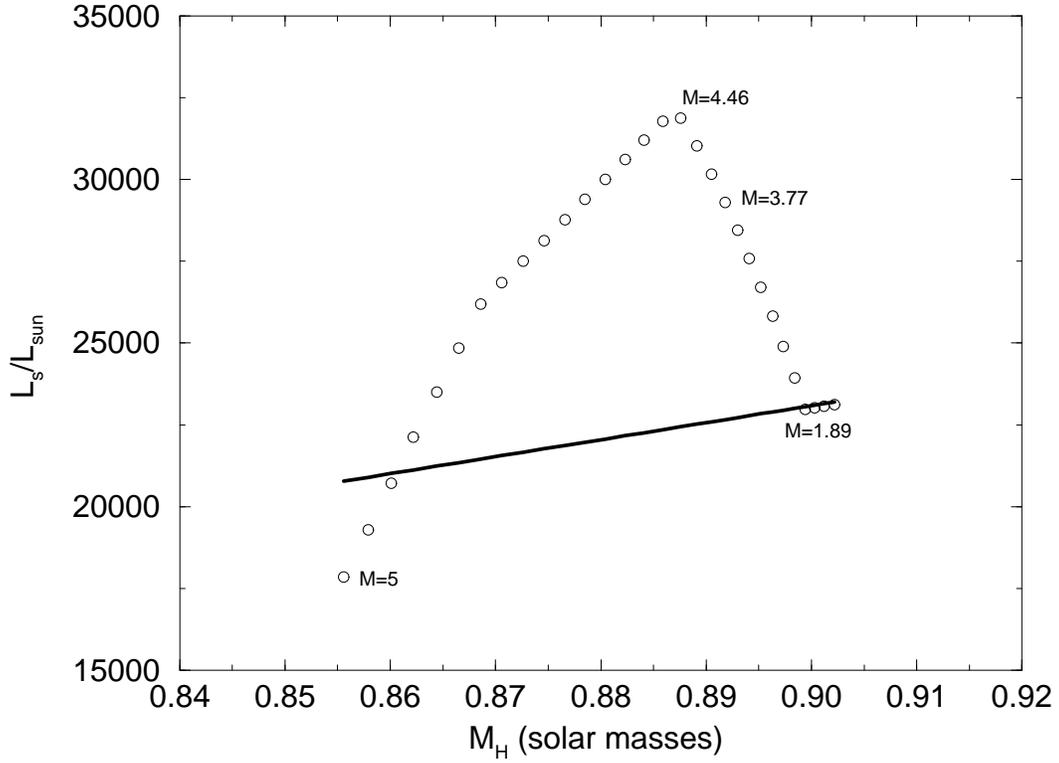,height=4.5truein}}}
\caption{The behavior of the interpulse surface luminosity of a
5~M$_{\odot}$, solar metallicity model, and mixing length parameter
$\alpha$=2.0 model. $M_{\rm H}$
is the mass of the hydrogen exhausted core. The open circles indicated
the luminosities of each interpulse. The solid line indicates the
low-mass core-mass luminosity relationship of Boothroyd and Sackmann
1988b. The text on the page indicates the model stars total mass at
various points.} 
\label{fig:m5a2a2oasd.eps}
\end{figure}

How does the surface luminosity, $\ls$, behave as a model evolves? In
figure \ref{fig:m5a2a2oasd.eps} the evolution of the luminosity of our
5$\msun$ model with a mixing length parameter, $\alpha =2.1$, solar
metallicity, and a pulsation-period mass-loss law is shown.  
\noindent
The model
starts with a luminosity slightly below the reference core-mass
luminosity relation but it rapidly evolves to nearly twice the
core-mass luminosity relation. When the mass reaches 4.77$\msun$ the
superwind begins and the star rapidly loses its envelope. During this
process the luminosity rapidly decreases and eventually during the
last few pulses the luminosity follows a core-mass luminosity
relation. This is qualitatively in agreement with the results of
Bl\"ocker (1995) and 
Vassiliadis and
Wood (1993) who computed TP-AGB models with mass loss up to envelope
ejection. In both of these studies, the models exhibited steep
luminosity increases until the superwind began, after which the
luminosity dropped until a core-mass luminosity relation was again
followed. This occurred because as the mass of the envelope is reduced
by the wind, hot-bottom burning contributes less to the total
luminosity and eventually the temperature at the base of the envelope
drops so that no significant luminosity is produced in the convective
envelope. This is the necessary condition for the star to obey a
core-mass luminosity relationship. 

\begin{figure}
\centerline{\hbox{\psfig{figure=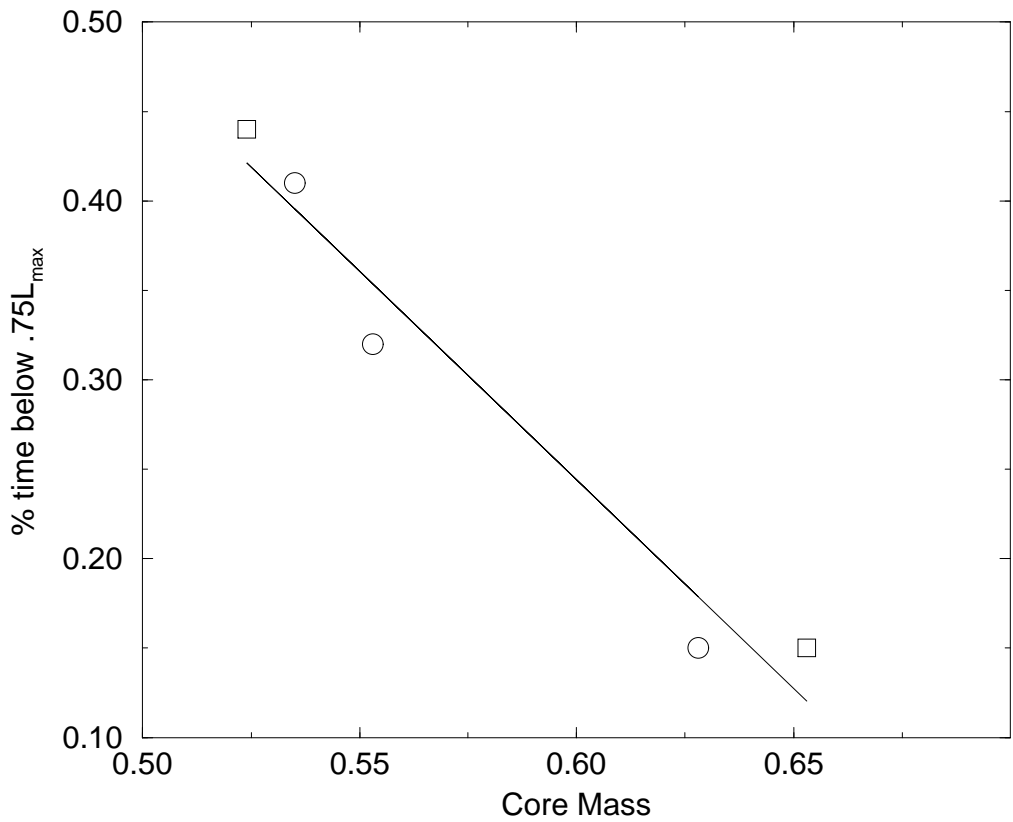,height=5truein}}}
\caption{The fraction of each interpulse spent below 75\% of the
maximum luminosity achieved during each interpulse. The line is a
least squares fit to the data, and the points are models obtained from
table 3 of BS88a} 
\label{fig:iplvmc.eps}
\end{figure}

Another important feature is the variation of the luminosity during
the interpulse phase. Examination of the models of Boothroyd and
Sackmann (1988a, hereinafter BS88a) and Vassiliadis and Wood (1993)
indicate low-mass AGB stars spend an appreciable fraction of the
interpulse phase at a luminosity less than the value indicated by our
luminosity relations. Using the models of BS88a, plotted in figure
\ref{fig:iplvmc.eps} is a measure of the time spent at lower luminosities
plotted against core mass.
The equation of the straight line is fitted to the data is given by:
\begin{equation}
g=-2.33\mc +1.64. 
\end{equation}
In the program the luminosity variation is
given by a step function with the star spending $g\tip$ at 50\% of the
interpulse luminosity, $\ls$ and $(1-g)\tip$ time at 100\% of $\ls$. 

\subsection{The Third Dredge Up}
At the end of each TP it is possible for $\he$ and $\ctw$ rich material
to be mixed from the core into the convective envelope. This event is
known as the third dredge up. The amount of material mixed upward is
determined by the dredge-up parameter
\begin{equation}
\lambda=\frac{\Delta \rm M_{dredge}}{\Delta M_c},
\end{equation}
where $\Delta{\rm M}_{\rm dredge}$ and $\Delta\mc$ are respectively
the mass of material dredged up and the advance of the core during the
preceding interpulse phase. Recent synthetic models of
Groenewegen and de Jong (1993, 1994a, 1994b), 
van den Hoek and
Groenewegen (1996), and 
Marigo \etal\ (1996) use a
constant $\lambda$ in their dredge up calculations. While, the use of
a constant $\lambda$ is certainly useful for constraining the amount
of dredge up, it is almost certainly too simplistic. 
Bazan (1991, B91) showed that for dredge-up to occur the peak
luminosity of the helium burning shell during the shell flash, ${\rm
L}_{\rm He,max}$, must exceed a certain minimum, ${\rm L}_{\rm
He,min}$, which is dependent on stellar mass. Bazan derived the
following equation for the dredge up parameter:
\begin{equation}
\lambda=0.90(\log{{\rm L}_{\rm He,max}}-\log{{\rm L}_{\rm He,min}})
\end{equation}
with the constraint $0\le\lambda <1$.

The peak He shell luminosity, $\lhemax$, during each pulse follows a
pattern similar to the surface luminosity. At the first pulse the
maximum He shell luminosity is well below the local asymptotic
value. In the following pulses it experiences a steep rise to the
local asymptotic limit. After achieving the asymptotic limit the star
continues to follow it. The result is an initially steep rise in
$\lhemax$ followed by a leveling off or even a decrease once the
asymptotic region is reached.

 For the peak helium luminosity at the first pulse, ${\rm
L}_{\rm He,max,0}$, we have adopted the expressions in Bazan (1991)
with no modification. Note that in this case ${\rm L}_{\rm He,max,0}$
depends on the initial helium abundance with lower He leading to
higher initial luminosities.  For the asymptotic values of ${\rm
L}_{\rm He,max}$, different expressions were used for high and low
core-masses. For $\mc\le 0.96\msun$, we use the following expression
taken from Bazan (1991):
\begin{equation}
\log{{\rm L}_{\rm He,max}}=48.1-125.7\mc
+84.5\mc^2+62.1\mc^3-63.1\mc^4.
\end{equation}
For $\mc >0.96\msun$, the expression from 
Iben (1977)
is used:
\begin{equation}
\log{{\rm L}_{\rm He,max}}=3.79+3\mc.
\end{equation}
For those pulses between the first and the attainment of the
asymptotic values, we approximate ${\rm L}_{\rm He,max}$ with the
following:
\begin{equation}
\log{{\rm L}_{\rm He,max}}=A(\mc -\mc_0)+\log{{\rm L}_{\rm
He,max,0}}
\end{equation}
where $\mc_0$ is the core-mass at the first pulse. Bazan (1991) adopts
A=34.44, however this is based on models with $\mc\lesssim
0.7\msun$. For $\mc\gtrsim 0.8\msun$, we estimated that A=65 from the
models of 
Wagenhuber and Weiss (1994). For
0.7$\msun\le\mc\le$0.8$\msun$, A was found by linearly interpolating
in $\mc$. 

A typographical error was found in Bazan's equation for ${\rm L}_{\rm
He,min}$, however we were able to use his model parameters and
rederive ${\rm L}_{\rm He,min}$ as a function of mass, and for M$>3\msun$
we use the following: 
\begin{equation}
{\rm L}_{\rm He,min}=6.20+0.600{\rm M}-0.116{\rm M}^2+0.00477{\rm M}^3
\end{equation}
while for smaller masses we have created a table of values and
interpolate quadratically in mass to get ${\rm L}_{\rm He,min}$. 
\begin{figure}
\centerline{\hbox{\psfig{figure=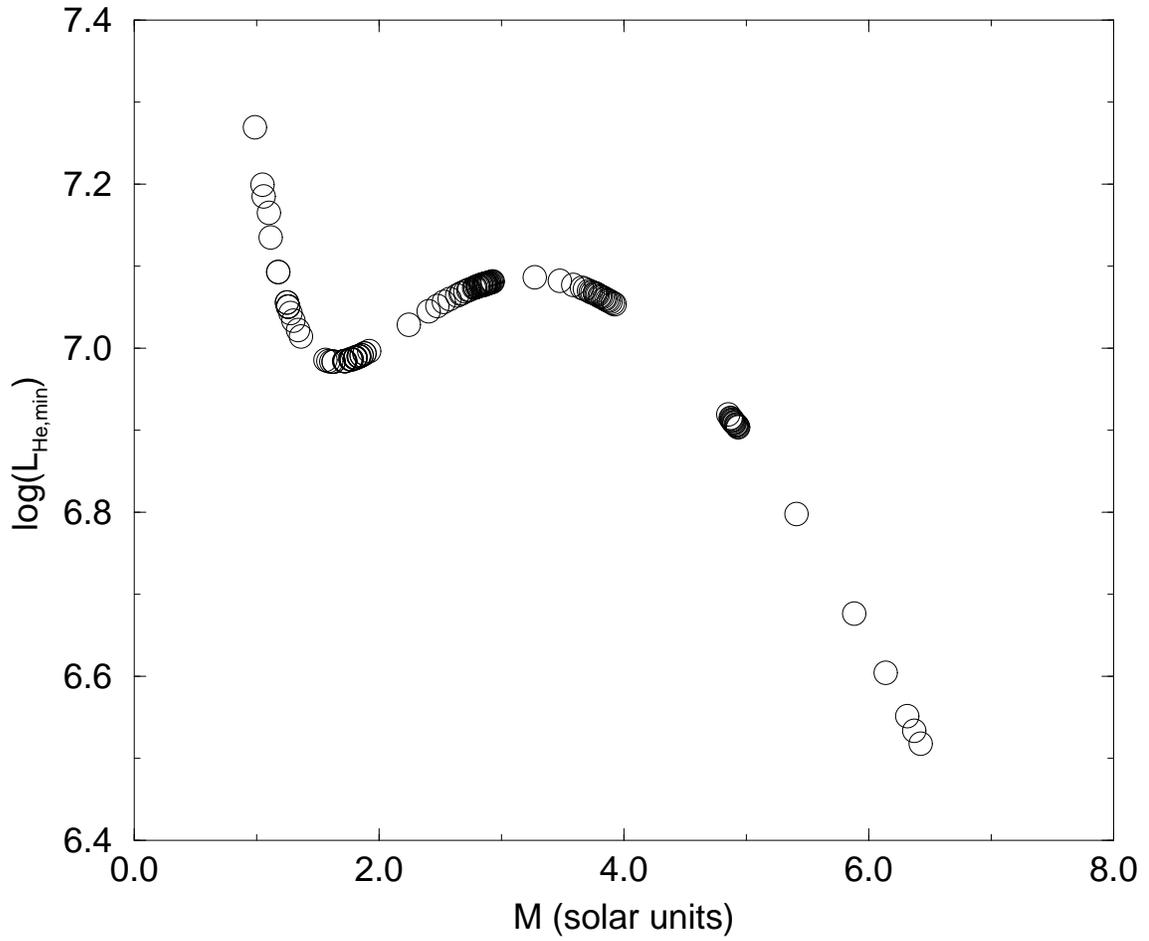,height=5truein}}}
\caption{The minimum luminosity of the helium burning shell during a
thermal pulse required for dredge-up to occur.} 
\label{fig:lheminmass.eps}
\end{figure}
Shown
in figure \ref{fig:lheminmass.eps} is $\log{{\rm L}_{\rm He,min}}$ as a
function of mass. The luminosity of the helium burning shell needed
for dredge-up to occur is a decreasing function of mass. This
result is confirmed by the latest results of Straniero \etal\
(1997). The results of Straniero \etal\ indicate $\lhemin$'s similar
to those of Bazan (1991). Straniero's relation between $\lambda$,
$\lhemax$, and $\lhemin$ is similar to Bazan's.

This model has several desirable features and one less than desirable
feature.
Some of the desirable features are:
\begin{enumerate}
\item{Dredge-up does not occur during the first few thermal pulses
which corresponds to real models.}
\item{Once dredge-up begins it starts with a low $\lambda$ and then
grows, which is also predicted by models.}
\item{The amount of dredge-up is mass dependent, which is also
predicted by models.}
\end{enumerate}
The undesirable feature is the problem that sometimes dredge-up occurs
when the envelope has a low mass. This is a very effective method to
pollute the envelope, however, it produces models which do not
correspond to observed PNe. In particular, it would produce models
of PNe with Galactic metallicity which have high N/O produced by
hot-bottom burning, but also high C/O where the enhancement of C/O
occurs on the final pulse. 

To eliminate the very effective enhancement that can occur at the last
thermal pulse, we surpress this pulse if 
\begin{equation}
\lambda_{\rm LP}<0.5\lambda_{\rm max}
\end{equation}
where $\lambda_{\rm LP}$ and $\lambda_{\rm max}$ are respectively the
dredge-up parameter of the last pulse and the maximum strength dredge.
While this is an ad hoc method of handling this problem, it does to
some degree simulate what happens in other models (Straniero \etal
1997, Vassiliadis and Wood 1993).

\subsection{Other Third Dredge-Up Parameters}
We determine the composition of the dredged up material from the
formulas in 
Renzini and Voli (1981), with $Y\approx
0.75$, $^{12}C\approx 0.23$, and $^{16}O\approx 0.01$ being the
approximate mass fractions. 

The mass advance during the preceding interpulse, $\Delta\mc$, is
related to the mass of the convective shell formed during the thermal
pulse, $\Delta{\rm M}_{\rm CSH}$ by this following expression:
\begin{equation}
\Delta\mc = (1-r)\Delta{\rm M}_{\rm CSH}
\end{equation}
where $r$ is the fraction of the current convective shell that will be
incorporated into the next pulse. To determine $r$ and $\Delta{\rm
M}_{\rm CSH}$ the following expressions from 
Bazan (1991) are used:
\begin{eqnarray}
r=11.921\mc^4-48.383\mc^3+72.288\mc^2-47.533\mc +12.136\\
\log{\Delta{\rm M}_{\rm CSH}}=-1.2172\mc^2-1.1953\mc -0.47444.
\end{eqnarray}
The above expression for the overlap gives $r>1$ which is unphysical,
therefore for $\mc\lesssim 0.52\msun$, so we also use the following
expression from 
Iben (1977):
\begin{equation}
r=1.5120736-1.45932\mc+0.346\mc^2
\end{equation}
which is used when the Bazan $r$ exceeds the Iben $r$.

The duration of each interpulse, $t_{ip}$, is important for
determining the mass-loss and the nucleosynthesis during hot-bottom
burning. The interpulse duration is simply the time needed for the
core to advance by the proper amount in mass and is therefore given by
the following expression:
\begin{equation}
\Delta t_{ip}=3.14\times 10^{18}\Delta\mc X_e/{{\rm L}_{\rm H}}
\end{equation}
where $X_e$ is the mass fraction of hydrogen in the envelope and ${\rm
L}_{\rm H}$ is the average luminosity of the hydrogen burning
shell. If hot-bottom burning is taking place then some of the
luminosity from the hydrogen burning shell is generated in the
convective envelope, any helium produced by hydrogen burning in this
region will be mixed up and does not contribute to the advance of the
core, therefore ${\rm L}_{\rm H}$ is the hydrogen burning luminosity 
generated beneath the convective shell. 

\subsection{Conditions at the First Pulse}
Any AGB code must determine the conditions at the onset of the first
thermal pulse as these determine much of the subsequent evolution of
the model. The conditions at the first pulse are determined by the
preceding evolutionary phases. 

The surface abundances at the first thermal pulse are determined by
combining published main sequence (MS) levels and changes due to the
first and second dredge-ups. The first and second dredge-ups can
modify the surface abundances of $\he$, $\ctw$, $\cth$, $\nit$, and
$\oxy$. 

The MS levels of the elements are established by scaling the solar
abundances of everything except the alpha elements: oxygen, neon,
magnesium, sulfur, and silicon to the appropriate
metallicity. We chose to set the abundances in the Sun to the levels
of Anders and Grevesse (1989). 
Oxygen, Neon, and Magnesium alpha elements are set as follows: 
\begin{eqnarray}
[A/Fe]=0.4&[Fe/H]<-1.0
\end{eqnarray}
\begin{eqnarray}
[A/Fe]= -0.5 [Fe/H]-0.1&-1.0\le [Fe/H]<-0.2
\end{eqnarray}
\begin{eqnarray}
[A/Fe]=0.0&[Fe/H]>-0.2
\end{eqnarray}
where $A$ is the abundance by number of O, Ne, and Mg.
The levels of $^{28}Si$ and $^{32}S$ were set to 
\begin{eqnarray}
[Si/Fe]=[S/Fe]=0.2&[Fe/H]<-1.0
\end{eqnarray}
\begin{eqnarray}
[Si/Fe]=[S/Fe]=-0.25[Fe/H]-0.05&-1.0\le [Fe/H]<-0.2
\end{eqnarray}
\begin{eqnarray}
[Si/Fe]=[S/Fe]=0.0&[Fe/H]>-0.2\
\end{eqnarray}
The values for oxygen, magnesium, and silicon were chosen from an
examination of the trends in the data of 
Edvarsson \etal\ (1993). 
Henry (1989) (also see Kaler(1975))
showed that neon and oxygen vary in lockstep in PNs and in HII regions,
therefore we assume that neon has a similar pattern to oxygen. Other
alpha elements such as Ca and Ti have a pattern similar to Si so we
assume that S follows the same pattern as Si. 

The helium abundance was calculated from the following:
\begin{equation}
Y=Y_0+\frac{\Delta Y}{\Delta Z}Z
\end{equation}
where $Y_0$ is the primordial mass fraction of helium. We chose
$Y_0=0.237$ as this value is close to the recent determinations of
Olive and Steigman (1995) and 
Pagel \etal\ (1992). We chose the slope so that $Y=Y_{\sun}$ at
$Z=Z_{\sun}$. The ZAMS hydrogen abundance was chosen so that 
\begin{equation}
X=1-Y-Z.
\end{equation}

For the Small Magellanic Cloud (SMC) and the Large Magellanic Cloud
(LMC), we scaled the metal abundances to solar. The mass fractions of
hydrogen, helium, and metals (X, Y, Z) were set to
(0.7352, 0.2554, 0.0094) and (0.7491, 0.2462, 0.0047) respectivelly
for the LMC and SMC. 

The first dredge-up occurs when the star reaches the first giant
branch and the convective envelope penetrates into regions where
hydrogen burning has occured. This affects the levels of $^1{\rm H}$,
$\ctw$, $\cth$, $\nit$, and $\oxy$. The abundance changes due to the
first dredge-up are calculated from the following formulae taken from  
Groenewegen and deJong (1993).
For helium, the change in the mass fraction, $\Delta{\rm Y}$, is
dependent on the mass and intial helium content. The change is
calculated for two different helium contents shown below:
\[\Delta{\rm Y}=\left\{\begin{array}{lcc}
-0.0170{\rm M}+0.0425&{\rm M}<2&{\rm Y}=0.3\\
-0.0068{\rm M}+0.0221&2\le{\rm M}<3.25&{\rm Y}=0.3\\
0&{\rm M}\ge 3.25&{\rm Y}=0.3\\
-0.0220{\rm M}+0.0605&{\rm M}<2.2&{\rm Y}=0.2\\
-0.0078{\rm M}+0.0293&2.2\le{\rm M}<3.75&{\rm Y}=0.2\\
0&{\rm M}\ge 3.75&{\rm Y}=0.2.
\end{array}\right. \]
These results are interpolated linearly to get $\Delta{\rm Y}$. The
destruction of hydrogen is computed from this formula:
\begin{equation}
\Delta{\rm X}=-\Delta{\rm Y}.
\end{equation}
\begin{equation}
\Delta\ctw =\ctw (g-1)\\
\end{equation}
\[
g=\left\{ \begin{array}{ll}
0.64-0.05({\rm M}-3)&{\rm M}<3\\
0.64&{\rm M}>3 ,
\end{array} \right. 
\]
\begin{equation}
\Delta \nit =-1.167\Delta\ctw ,
\end{equation}
\begin{equation}
\Delta \oxy=-0.01\oxy ,
\end{equation} 
with the equations indicating the changes in the mass fractions of the
different elements.

For stars in which the second dredge up occurs we use the formulation
of 
Becker and Iben (1980) where the mass of the core
before, ${\rm M}_{\rm cb}$, and after, ${\rm M}_{\rm ca}$, the second dredge up
are given by:
\begin{eqnarray}
{\rm M}_{\rm cb}={\rm AM}+{\rm B}
{\rm M}_{\rm ca}={\rm CM}+{\rm D}
\end{eqnarray}
where
\[
A=0.2954+0.0195{\rm L}_{\rm Z}+0.377{\rm L}_{\rm Y}-1.35{\rm L}^2_{\rm
Y}+0.289{\rm L}_{\rm Z}{\rm L}_{\rm Y}
\]
\[
B=-0.5-30.6{\rm D}_{\rm Z}-412{\rm D}_{\rm Z}^2-1.43{\rm D}_{\rm
Y}+29.3{\rm D}_{\rm Y}^2-204{\rm D}_{\rm Z}{\rm D}_{\rm Y}
\]
\[
C=0.0526+0.754{\rm D}_{\rm Z}+54.4{\rm D}_{\rm Z}^2+0.222{\rm D}_{\rm
Y}-1.07{\rm D}_{\rm Y}^2+5.53{\rm D}_{\rm Y}{\rm D}_{\rm Z}
\]
\[
D=0.59-10.7{\rm D}_{\rm Z}-425{\rm D}_{\rm Z}^2-0.825{\rm D}_{\rm
Y}^2-44.9{\rm D}_{\rm Y}{\rm D}_{\rm Z}
\]
and 
\[
{\rm L}_{\rm Z}=\log{{\rm Z}_{\rm i}}
\]
\[
{\rm L}_{\rm Y}=\log{{\rm Y}_{\rm i}}
\]
\[
{\rm D}_{\rm Z}={\rm Z}_{\rm i}-0.02
\]
\[
{\rm D}_{\rm Y}={\rm Y}_{\rm i}-0.28
\]
with ${\rm Y}_{\rm i}$ and ${\rm Z}_{\rm i}$ being the inital He and metal
abundances. 

The material mixed out by the second dredge-up has experienced
complete hydrogen burning. Therefore, all the hydrogen will have been
burned to helium and most of the CNO nuclei will have been converted
into $\nit$. So, the composition of the material in the second dredge
is determined by
\[
Y_{\rm 2dr}={\rm X}+{\rm Y}
\]
and
\[
X_{14,{\rm 2dr}}=14(\frac{{\rm X}_{12}}{12}+\frac{{\rm
X}_{13}}{13}+\frac{{\rm X}_{16}}{16})+{\rm X}_{14} 
\]
where the mass fractions on the right-hand side of the equations are
those of the star before the second dredge-up. All the other mass
fractions of the dredged-up material are set to zero.

\subsection{Core Mass}
When convective overshoot is ignored. For low mass stars the mass of
the hydrogen exhausted core ($M_H$) at the first thermal pulse is
given by the expression found in 
Lattanzio (1986).
\[
M_c(0)=\left\{ \begin{array}{lc} 0.53-(1.3+\log{Z})(Y-0.20)&Z\ge 0.01\\
0.524+0.58(Y-0.20)+(0.025-20Z(Y-0.20))M&0.01>Z\ge 0.003\\
(0.394+0.3Y)\exp{(0.10+0.3Y)M/M_{\odot}}&Z<0.003.\end{array} \right.
\]
If the second dredge occurs the core mass at the first
pulse is given by ${\rm M}_{\rm ca}$. The realtion above is used until
it reaches ${\rm M}_{\rm cb}$. 

\newpage
\chapter{Model Predictions of Element Production in Intermediate-Mass Stars}
\label{modreschap}

Using the methods described in chapter 3, a large grid of TP-AGB
models in mass, metallicity, and mixing length parameter has been
computed. The results of these models are presented in
Appendix~\ref{modresapp}. Tables of these models are available by
anonymous ftp (some computer address to be determined) or via the
World Wide Web at (some site to be determined). The model parameters
were chosen to form a dense grid in mass. Table \ref{params} lists the
parameters used in the model grids.

\begin{center}
\begin{table}
\begin{tabular}{|c|c|l|}\hline
$\alpha$&[Fe/H]&Masses\\\hline
2.3& 0.0&1-8$\msun$\\\hline
2.3& 0.1&1-8$\msun$\\\hline
2.3& 0.2&1-8$\msun$\\\hline
2.3&-0.1&1-8$\msun$\\\hline
2.3&-0.5&1-4$\msun$\\\hline
\end{tabular}
\caption{Table of Input Model Parameters}
\label{params}
\end{table}
\end{center}

A large number of interesting results can be derived from these
models, but the ones we are interested in here are the resulting model
planetary nebulae (PNe). The output of each model PN is a set of
abundances (He, C, N, O, Ne) and a stellar core-mass which can be
compared to the PN data. 

\section{Effect of Mass and Metallicity on Element Production}
\label{mass_met}

The model PN abundance ratios result from the combined effect of
mass-loss, the three dredge-ups, and hot-bottom burning. The different
nucleosynthetic events will make different relative contributions of
the elements listed below: 
\begin{enumerate}
\item{The first dredge-up occurs when a star enters the first giant
branch and the convective envelope dips into regions of the star where
CNO reactions have taken place. The material mixed to the surface
during the first dredge-up has experienced only partial hydrogen
burning, primarily transforming $\ctw$ to $\cth$ and $\nit$. 
Significant amounts of hydrogen is not converted into helium in these
zones. Therefore, the expected result is a significant increase in the
$\nit$ abundance (as well as $\cth$) at the expense of $\ctw$, but
only a minimal increase in the helium abundance. The abundance of
$\oxy$ experiences only minor changes ($\sim$1\%).}
\item{The second dredge-up occurs at the entrance onto the early
asymptotic giant branch. The convective envelope reaches into the
hydrogen exhausted core and mixes He and N into the surface layers of
the star, increasing both the helium and nitrogen abundances.}
\item{At the end of each thermal pulse, the convective envelope can
reach into the He-burning shell. Therefore, the third dredge-up mixes
the products of partial He burning to the surface. Nitrogen is
destroyed during He burning and will not be dredged-up. The dredged-up
material will contain $\ctw$ and $\he$ resulting in increases in the
carbon and helium abundances.}
\item{During the interpulse phase of the TP-AGB, the base of the
envelope can reach temperatures high enough for CNO reactions to
occur. This is known as hot-bottom burning. This will result in the
conversion of $\ctw$ to $\cth$ and $\nit$. Some $\oxy$ will also be
converted to $\nit$. Given a long enough time to act some $\he$ will
be produced.}
\end{enumerate} 
Each of these processes has a strong mass dependence and each may also
have a strong dependence on metallicity, therefore, PN models should
reflect this. Mass-loss is important because it determines the length
of time during which the third dredge-up and hot-bottom burning can
operate.

\begin{figure}
\centerline{\hbox{\psfig{figure=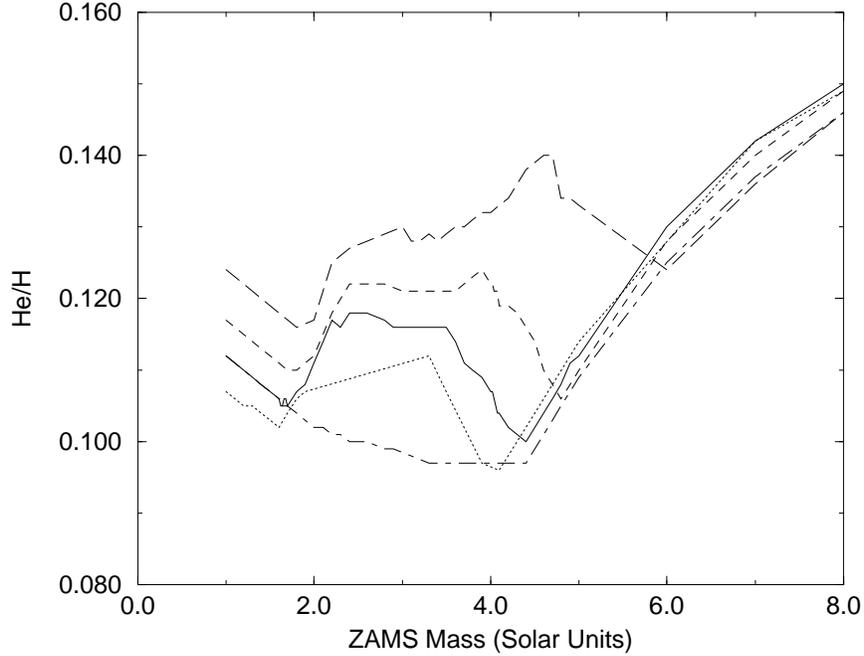,height=4truein}}}
\caption{Expected value of He/H in PN for model mass between 1M$_{\odot}$
and 8M$_{\odot}$. The solid line, dashed line, long dashed line and
the dotted line respectively indicate models calculated with
[Fe/H]=0.0, 0.1, 0.2, and -0.1. The dashed-dotted line indicates the
abundance of He/H at the first pulse for the [Fe/H]=0.0 model.}
\label{fig:hehmass.eps}
\end{figure}

As expected, our model results show the resulting abundances have a
strong dependence on mass and metallicity. Figures \ref{fig:hehmass.eps},
\ref{fig:nomass.eps}, \ref{fig:ohmass.eps}, and \ref{fig:comass.eps}
respectively plot the predicted PN abundance ratios for the
models of He/H, N/O, log(O/H)+12, and C/O as a function
of the ZAMS mass for several different values of [Fe/H]. The mixing
length parameter, $\alpha$, was held fixed at 2.3, since it also has
important effects on the model results which are described in section
\ref{mixlensec}. To facilitate comparisons of the effects of pre-AGB
nucleosynthesis and TP-AGB nucleosynthesis, the model abundances at
the first pulse for the [Fe/H]=0.0 models have been included. 

\begin{figure}
\centerline{\hbox{\psfig{figure=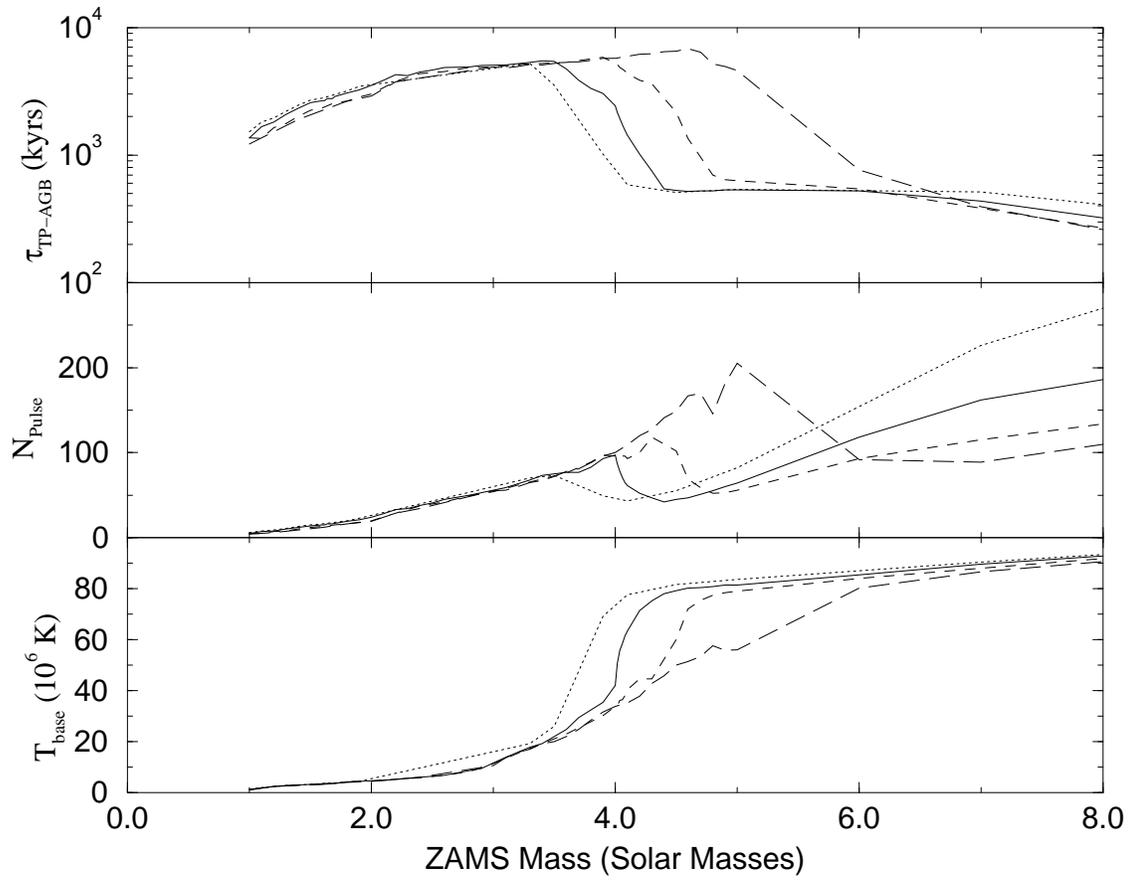,height=5truein}}}
\caption{The panels from top to bottom indicate the time the model
spends on the TP-AGB, the number of pulses, and the maximum base
temperature achieved at the bottom of the convective envelope. The lines have the same meaning as figure~\ref{fig:hehmass.eps}}
\label{fig:mass_age_all.eps}
\end{figure}

\begin{figure}
\centerline{\hbox{\psfig{figure=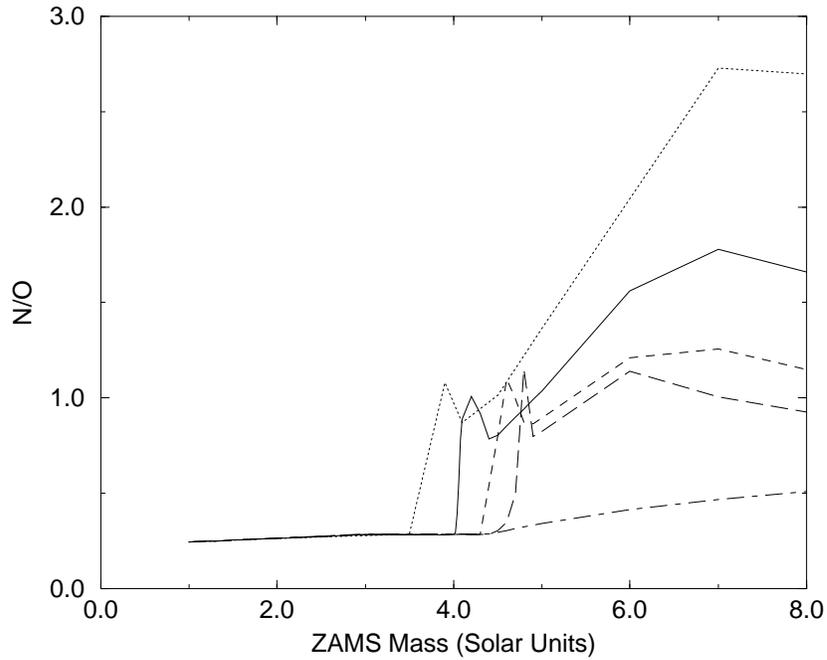,height=4truein}}}
\caption{Same as figure \ref{fig:hehmass.eps} except N/O is compared to
ZAMS mass.}
\label{fig:nomass.eps}
\end{figure}

\begin{figure}
\centerline{\hbox{\psfig{figure=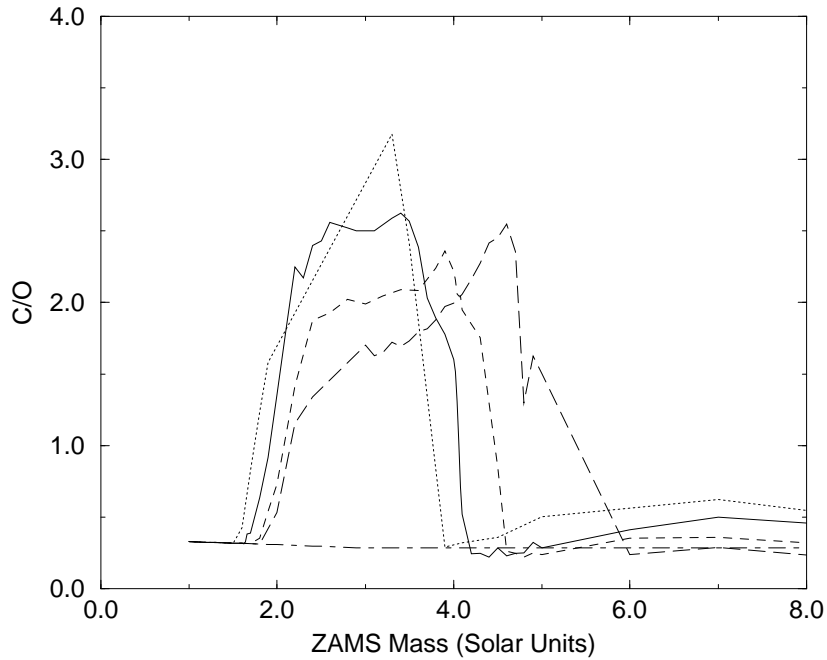,height=4truein}}}
\caption{Same as figure \ref{fig:hehmass.eps} except C/O is compared to
ZAMS mass.}
\label{fig:comass.eps}
\end{figure}

\begin{figure}
\centerline{\hbox{\psfig{figure=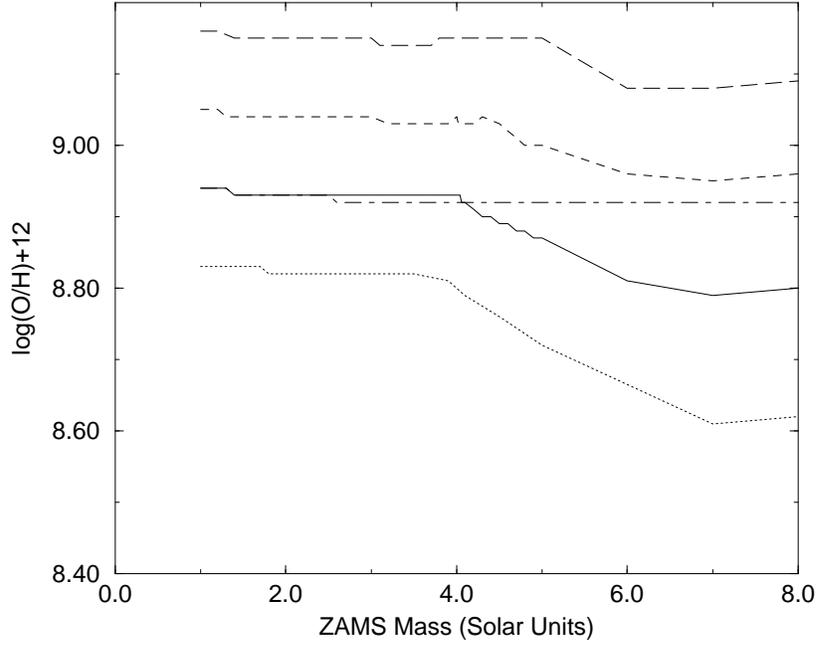,height=4truein}}}
\caption{Same as figure \ref{fig:hehmass.eps} except $\log{\rm O/H}$+12 is
compared to ZAMS mass.}
\label{fig:ohmass.eps}
\end{figure}

The two physical parameters which directly affect the resulting model
PN abundances are the mass-loss rate and the temperature at the base
of the convective envelope. Mass-loss determines the rate at which the
envelope is removed and limits the number of thermal pulses. Mass-loss
also determines the envelope mass as a function of time and the mass
of the envelope determines the dilution of the material
dredged-up. The thermally pulsing lifetime, $\ttpagb$, and the number
of pulses, $\np$, are related to the mass-loss rate. Hot-bottom
burning, which occurs when the base of the envelope is hot enough for
CNO reactions to occur, determines the amount of both CN and ON
cycling. An indicator of which of the CNO reactions occurred in
the hot-bottom burning of each model is the maximum attained
temperature at the base of the convective envelope, $\tbase$. In some
of our models, temperatures of nearly 100$\times 10^6$K were sometimes
attained. Different types of nuclear reactions will be important in
different temperature regimes.
\begin{enumerate}
\item{If $\tbase < 30$K then no hot-bottom
burning occurs.}
\item{If 30K$<{\rm T}_{\rm 6,base}<$50K then $\ctw$ is converted
to $\cth$.}
\item{If $\tbase\gtrsim 50$K then $\ctw$ is converted to $\nit$. Also
ON cycling begins converting $\oxy$ to $\nit$.}
\end{enumerate}
The panels of figure \ref{fig:mass_age_all.eps} respectively show the age
of each model in years, the number of thermal pulses a model
experiences, and the maximum temperature achieved at the base of the
convective envelope. 

Figure~\ref{fig:hehmass.eps} shows the He/H ratio to be a complicated
function of mass and metallicity. Since helium is produced by each of
the three dredge-ups and by hot-bottom burning, it is not surprising
that helium shows complex behavior. The resulting PN He/H also
appears to depend on the initial helium mass fraction. For each
metallicity, He/H has three local maxima and two local minima. The
behavior of He/H in between adjoining maxima and minima reflect the
dominance of one or more of the acting processes. The first maximum is
at $\sim$1.0$\msun$ for each metallicity. The second maximum should be
regarded as a plateau which stretches from $\sim$2-4$\msun$. The third
maximum is located at 7-8$\msun$ and corresponds to the highest mass
model calculated for each [Fe/H] sequence. The first minimum in He/H is
located at $\sim$1.7-1.8$\msun$ and is weakly metallicity
dependent. The second minimum is located at $\sim$4.0-5.5$\msun$ and 
is strongly dependent on metallicity. For the purposes of discussion
we divide this mass range into four parts:  
\begin{enumerate}
\item {The region between the first maximum in He/H and the first
minimum which we define as the low low mass range,}
\item{The region between the first minimum and the second maximum which
we define as the high low mass range,}
\item{The region between the second maximum and the second minimum which
we define as the low intermediate mass range,}
\item{The region between the second minimum and the second maximum which
we define as the high intermediate mass range.}
\end{enumerate}

In the low low mass range (M$\lesssim$1.7$\msun$), there is no
difference between the abundance ratios at PN ejection and the first
pulse because no nucleosynthesis occurred during the TP-AGB phase of
these models. The base of the envelope never gets hot
enough for nuclear reactions to occur. No third dredge-up events
occurred in any model in this range, primarily because mass-loss
removed the envelope before the strength of the pulses was sufficient
to cause it to occur.

In the low low mass range, the third dredge-up does not occur because
mass-loss removes the envelope before $\lhemax$ reaches values high
enough for it to occur. As noted in chapter~\ref{modelchap}, the peak
luminosity of the helium burning shell during the thermal pulse,
$\lhemax$, controls the occurrence of the third dredge-up. For all
TP-AGB stars, $\lhemax$ is too small at the first pulse for dredge-up
to occur. In subsequent pulses, $\lhemax$ rises. and then rises to
values high enough for dredge-up to occur. Models in the low-low mass
range do not last long enough for $\lhemax$ to grow sufficiently to
allow dredge-up. 

The decrease of He/H with stellar mass in this range is a consequence
of the behavior of the He/H ratio with mass in the first
dredge-up. Models of the first dredge-up (e.g. Sweigert \etal\ 1992,
Boothroyd and Sackmann 1997) predict that the He/H ratio decreases as
a function of mass. Such models also predict in this region an
increase in the ratio of N/O. This reflects the transition from the pp
cycle to the CNO cycle in stars. Although the scale of figures
\ref{fig:nomass.eps} and \ref{fig:comass.eps} does not clearly show
it, in the low low mass range, there is a slight positive correlation
between N/O and mass and a corresponding negative correlation between
C/O and mass.


The metallicity dependence of the first minimum is tied to the
mass-loss scheme. Mass-loss is metallicity dependent. For example, if
two AGB stars have the same mass and luminosity but different
metallicities, the lower metallicity star will have the smaller
radius. In our mass-loss scheme $\dot{\rm M}\propto{\rm R}^{2.7}$, and
therefore the lower metallicity star will take longer to reach the
superwind phase and lose its envelope. A longer lifetime on the TP-AGB
allows more thermal pulses in which to have a dredge-up. No
nucleosynthesis occurs on the TP-AGB for low low mass range
objects. The enhancements of abundances over the ZAMS level are due to
pre-AGB nucleosynthesis.  

Between 1.7$\msun$ and 4.0$\msun$, both the levels of He/H and C/O are
modified between the first pulse and PN ejection, but the N/O and O/H
ratios remain approximately constant throughout the TP-AGB phase. The
elevation of C/O and He/H results from the action of the third
dredge-up. The main components of third dredged-up material are $\he$
(Y$\approx$0.75) and $\ctw$ (X$_{12}\approx$0.23), which leads to
significant increases of both He/H and C/O. There is a small decrease
in N/O ratio and slight increase in the O/H ratio because of the
dredge-up of oxygen on the TP-AGB, which are not visible on figures
\ref{fig:nomass.eps} and \ref{fig:ohmass.eps}, occurs because of the
small amounts of $\oxy$ dredged-up. The level of $\nit$ is not
affected by hot-bottom burning, since the maximum base temperature in
this mass range is not sufficient for CN cycling.

According to figure~\ref{fig:hehmass.eps}, the level of He/H found at
the second maximum increases with metallicity. The change in the
helium mass fraction from the first to final pulse is similar for
different metallicities. Therefore the difference in He/H is due to
He/H at the first pulse. Pre-AGB mixing episodes in the high-low mass
range leave the He/H ratio essentially unchanged. Therefore, the most
important factor in determining the level of the second maximum is the
ZAMS abundance of He/H.  

Between 2 and 4$\msun$, figure~\ref{fig:mass_age_all.eps} shows
metallicity differences have only a minor effect on the number of
thermal pulses and $\ttpagb$. The number of thermal pulses is an
approximate measure of the number of third dredge-up events,
therefore, the mass of helium dredged-up to first approximation does
not vary with metallicity. 

At $\sim$4$\msun$, the beginning of the low intermediate mass range,
figures~\ref{fig:mass_age_all.eps} indicate a steep drop in both
$\ttpagb$ and N$_{\rm P}$, and a steep increase in the maximum base
temperature. Due to the onset of hot-bottom burning, models with mass
above $\sim$4$\msun$ spend an appreciable fraction of the TP-AGB stage
with luminosities in excess of the luminosity indicated by the
core-mass luminosity stage. This extra luminosity causes the star to
swell to radii larger than if the star followed a core-mass luminosity
relationship. Since the mass-loss prescription used here is strongly
radius dependent, the average mass-loss is higher and the superwind is
reached earlier. Therefore, there is a significant drop in $\ttpagb$.

The mass of the steep drop in $\ttpagb$ and N$_{\rm P}$ increases
with metallicity. Figure~\ref{fig:hehmass.eps} shows the high mass end
of the He/H plateau increases with metallicity. Figure
\ref{fig:mass_age_all.eps} shows that the mass of the maximum age
model increases with [Fe/H]. The high mass end of the second He/H
plateau corresponds to the maximum in age and also to the model with
the maximum number of thermal pulses shown. This is not surprising
since the number of pulses and the number of third dredge-up events
should be correlated.

The precipitous drop in $\ttpagb$ and N$_{\rm P}$ corresponds to the
drop in He/H between the plateau and the second He/H minimum. The
number of thermal pulses roughly correlates with the number of third
dredge-up events and the amount of He dredged-up. Therefore, the cliff
in He/H is due to the sharp drop-off in the number of dredge-up
events.

The second He/H maximum approximately indicates the lower mass limit
to models which undergo hot-bottom burning, $\mhbb$. In the lowest
panel in figure \ref{fig:mass_age_all.eps}, the rise in the maximum
base temperature corresponds to the drop in $\ttpagb$ and N$_{\rm
P}$. When the base temperature exceeds 3$\times10^7$K, significant
amounts of CNO burning take place at the base of the envelope, a
process known as hot-bottom burning. Clearly, hot-bottom burning is
very important in determining the structure of the star. Hot-bottom
burning shortens $\ttpagb$, thereby decreasing the number of third
dredge-up events and the mass of helium and carbon mixed up by the
third dredge-up. 

The position in mass of the second He/H minimum and the lower end of
the high intermediate mass range is dependent on the lowest mass at
which the second dredge-up occurs. On figure \ref{fig:hehmass.eps},
the second He/H minimum note that there is not much difference between
He/H at the first pulse and He/H at the last pulse. Also note, that
near the second He/H minimum, there is a steep turn up in He/H at the
first pulse (at approximately 4.3$\msun$). This turn-up indicates the
onset of the second dredge-up. We define this turn-up mass as the mass
of the onset of second dredge-up, M$_{2dr}$. So if the initial mass of
the star, M, is greater than M$_{2dr}$, than the second dredge-up
occurs. The amount of helium dredged up depends on M-M$_{2dr}$ in a
roughly linear manner.

The lowest mass star of a given mass which experiences
second dredge-up, M$_{2dr}$, is metallicity dependent. Models of
second dredge-up ( \nocite{bi79} Becker and Iben (1979), \nocite{bs97}
Boothroyd and Sackmann 1997) indicate that the minimum mass star which
experiences second dredge-up depends on the metallicity. Our second
dredge-up prescription is based on the models of Becker and Iben, so
the onset of second dredge in our models depends on metallicity. 

 
In the high intermediate mass range, the second dredge-up dominates
the other processes in determining the PN He/H. In figure
\ref{fig:hehmass.eps}, for models with [Fe/H]=0.0, there is only a
small difference in the models between He/H at the first pulse and at
PN ejection. Above 5$\msun$, the level of He/H at the first pulse is
due to the second dredge-up. The mixing of helium-rich material due to
third dredge-ups into the surface layers causes He/H to increase by a
few percent. However, for most of the TP-AGB lifetimes the envelope
mass is large compared to the mass dredged-up which strongly dilutes
the dredged-up helium.

The behavior of N/O is seen in figure \ref{fig:nomass.eps}. N/O is
essentially level for M$\lesssim$4$\msun$ at a low value and for
M$\gtrsim$4$\msun$ it jumps up to high values. The sudden shift from
low N/O to high N/O occurs just before the second He/H
minimum. This indicates that the sudden increase in N/O is due to
hot-bottom burning. Reference to figure \ref{fig:mass_age_all.eps}
indicates N/O begins its steep increase when the maximum base
temperature exceeds 50$\times$10$^6$K. Using figure
\ref{fig:nomass.eps} and comparing N/O at the first pulse to N/O at
the last pulse for the solar metallicity models, N/O has clearly been
modified during the TP-AGB for models which experience hot-bottom
burning. Therefore, the important factor in determining the N/O
abundance is the presence or absence of hot-bottom burning.
	 
It should be noted that N/O can be modified before the TP-AGB phase by
both first and second dredge-up. The first dredge-up operates in all
models and N/O after the first dredge-up is approximately double its
ZAMS value. In figure \ref{fig:nomass.eps}, there is a noticeable
upturn in the first pulse N/O at $\approx$4.5$\msun$. This is due to
the fact that in these models second dredge-up occurs. Those models
which experience second dredge-up also experience hot-bottom
burning. Hot-bottom burning dominates nitrogen production when it
occurs, therefore, nitrogen can not be used as an indicator of the
second dredge-up.

C/O reaches its maximum value in a rather broad plateau in the
high-low mass range. In figure \ref{fig:comass.eps} for all
metallicities, C/O is significantly larger than 1 between 2 and
$\sim$4$\msun$. The C/O plateau occurs in the same mass range as the
second He/H maximum. The C/O plateau has the same origin as the He/H
maximum, third dredge-up. In the region from 2-4$\msun$, TP-AGB stars
have more thermal pulses than in any other mass range, leading to the
largest amount of third dredge-up.

In the low-intermediate and high-intermediate, C/O is much less than
in the broad plateau. The drop in C/O corresponds to the increase in
N/O in the same range. This is not surprising since carbon will be
converted to nitrogen via hot-bottom burning. 

In the low-low mass range C/O is low relative to the plateau in the
high-low range. Recall, He/H is low in this region because the third
dredge-up does not occur. Therefore C/O will be low for the same
reason, no carbon is mixed up to the surface layers. This means that
level of C/O in the low-low mass range will be determined by the first
dredge-up.

Our models indicate that in addition to the CN cycle being active at
the base of the envelope during hot-bottom burning, the ON cycle also
contributes to the nucleosynthesis. In figure \ref{fig:ohmass.eps},
for all metallicities, there in the low-low and high-low mass ranges
O/H is constant. In contrast, O/H drops significantly in the
low-intermediate and high-intermediate mass ranges. The drop occurs
in models which experience hot-bottom burning, and it is due to oxygen
being converted to nitrogen via the ON cycle. In some of our models
the base temperature reaches 60-80$\times$10$^6$K, which is hot enough
for the ON cycle to operate. Some small amount of oxygen is dredged-up
at each thermal pulse, however, in our model this effect does not
significantly modify the surface abundance of oxygen.

Note, the decrease in O/H is of the order of 0.1dex (30\%). Such a
small amount would be difficult to detect unambiguously
observationally. 

\section{Summary of Results}

The PNe levels of the important abundance ratios He/H, N/O, C/O, and
O/H show important changes as a function of the mass of the progenitor
star. High C/O and He/H show up in models in the high-low mass range,
which indicates they have undergone multiple third dredge-up
events. In the low-intermediate mass range, the models exhibit high
N/O but low C/O and low He/H indicating hot-bottom burning but not
much dredge-up. In the high-intermediate mass range, models give high
PN N/O and He/H but low C/O, indicating hot-bottom burning and second
dredge-up. The abundance ratio O/H reflects the presence or absence of
hot-bottom burning, in the low-mass range, models exhibit no ON
cycling and in the intermediate mass range, ON cycling is indicated.

Some of these results are qualitatively similar to those of other
investigators. The models of Renzini and Voli (1981) with hot-bottom
burning indicate that C/O is enhanced in the low-mass region and N/O
and O/H are respectively enhanced and depleted when hot-bottom
burning occurs. Similar results have also been reported by Boothroyd
\etal\ (1993), Forrestini and Charbonnel (1997), Groenewagen and
deJong (1993, 1994), and Marigo \etal\ (1996). 

There is general agreement between our models and others on the He/H
abundance, however, we believe we are the first to report the
possibility of the second He/H minimum.

\subsection{The Importance of Mixing Length}
\label{mixlensec}

Important uncertainties result from the use of the mixing length
theory. To account for convection, we use the mixing length
theory. The distance a convective element travels relative to 
the pressure scale height before being it is destroyed is a free
parameter. Different choices of mixing lengths have significant
effects of stellar structure.

The mixing length parameter, $\alpha$, has important effects on TP-AGB
evolution. The stellar radius, $\rstar$, depends inversely on
$\alpha$. In our mass-loss scheme, the mass-loss rate, $\mdot$, varies
directly with $\rstar$ to some power. The time on the TP-AGB,
$\ttpagb$, depends inversely on the mass-loss rate. Therefore,
$\ttpagb$ is directly proportional to $\alpha$, i.e. as $\alpha$
increases so does the mean lifetime.

\begin{figure}
\centerline{\hbox{\psfig{figure=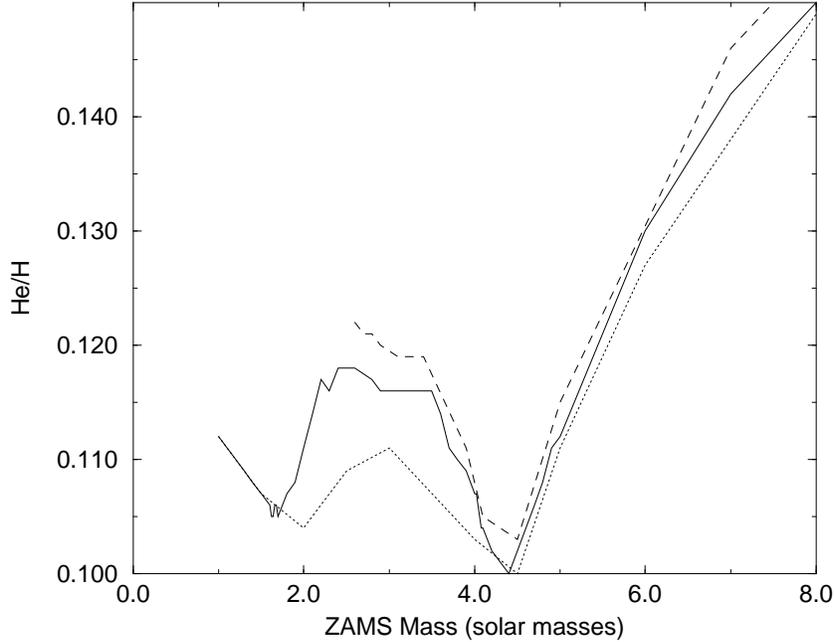,height=4truein}}}
\caption{Behavior of He/H as a function of $\alpha$ and ZAMS mass. all
models have [Fe/H]=0.0. The dotted, solid, and dashed lines indicate
respectively model grids with the mixing length parameter, $\alpha$,
respectively set to 1.9, 2.3, and 2.5.}
\label{fig:hehalfamass.eps}
\end{figure}

\begin{figure}
\centerline{\hbox{\psfig{figure=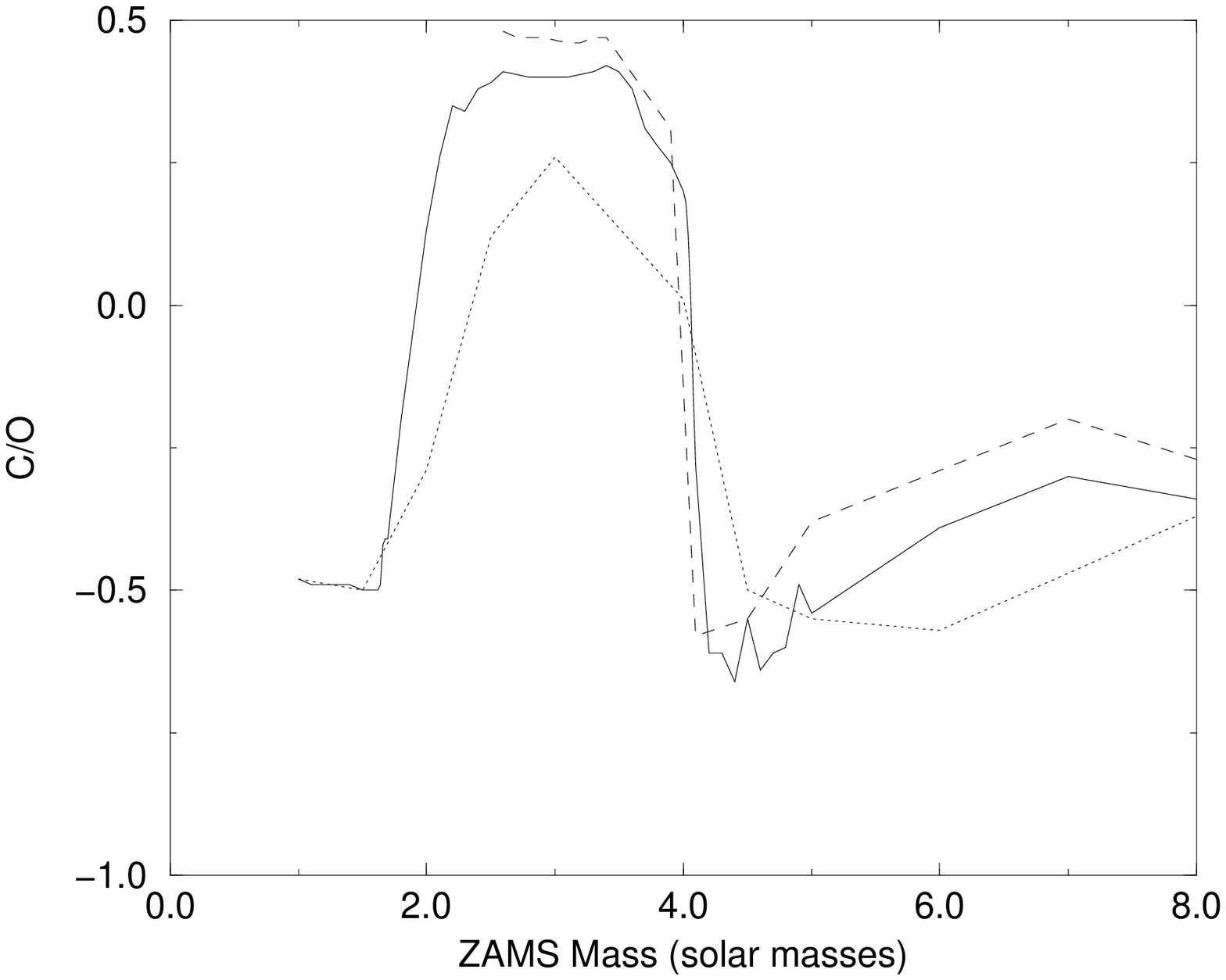,height=4truein}}}
\caption{Same as figure \ref{fig:hehalfamass.eps} except C/O is compared to
$\alpha$ and ZAMS mass.}
\label{fig:coalfamass.eps}
\end{figure}

\begin{figure}
\centerline{\hbox{\psfig{figure=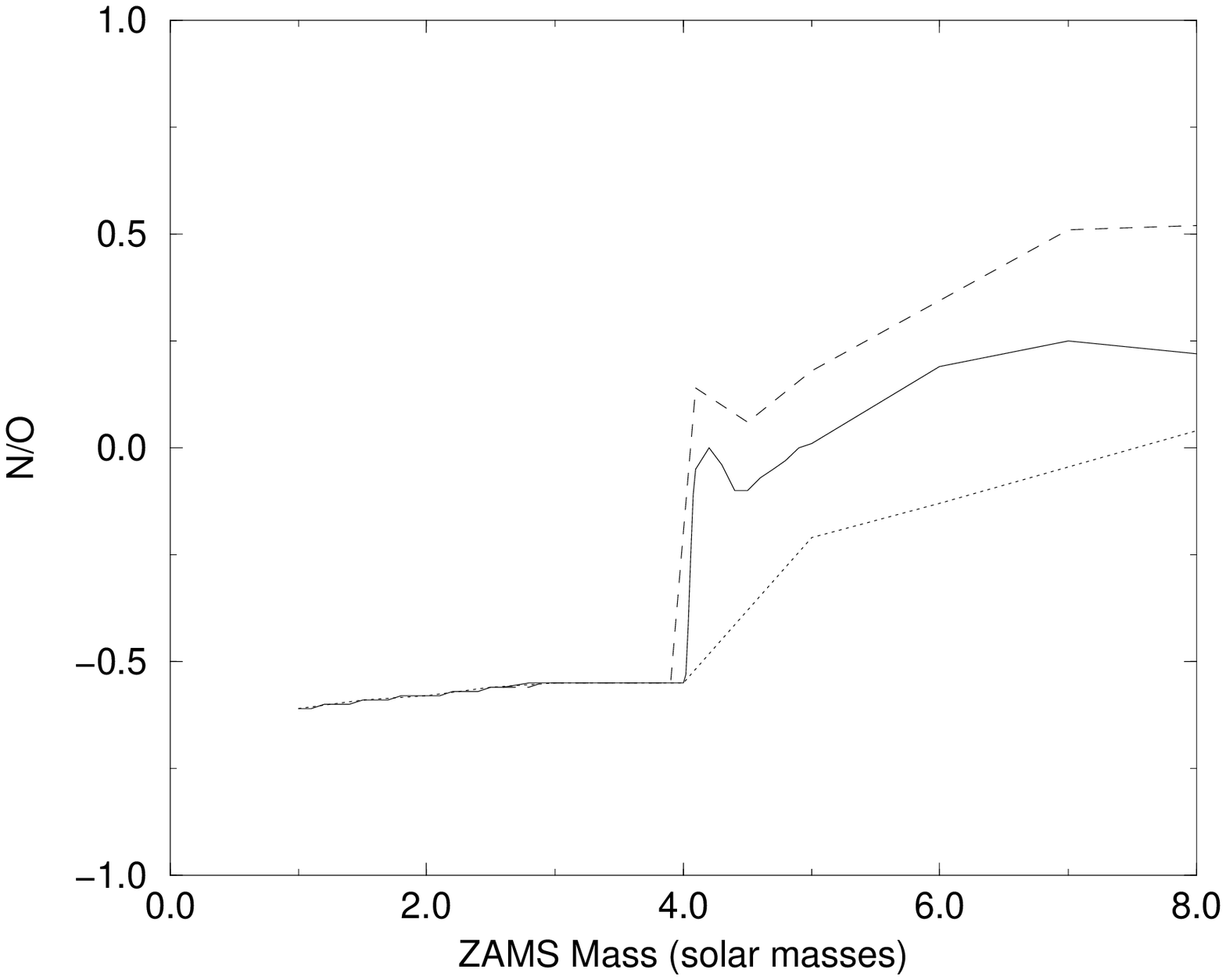,height=4truein}}}
\caption{Same as figure \ref{fig:hehalfamass.eps} except N/O is compared to
$\alpha$ and ZAMS mass.}
\label{fig:noalfamass.eps}
\end{figure}

\begin{figure}
\centerline{\hbox{\psfig{figure=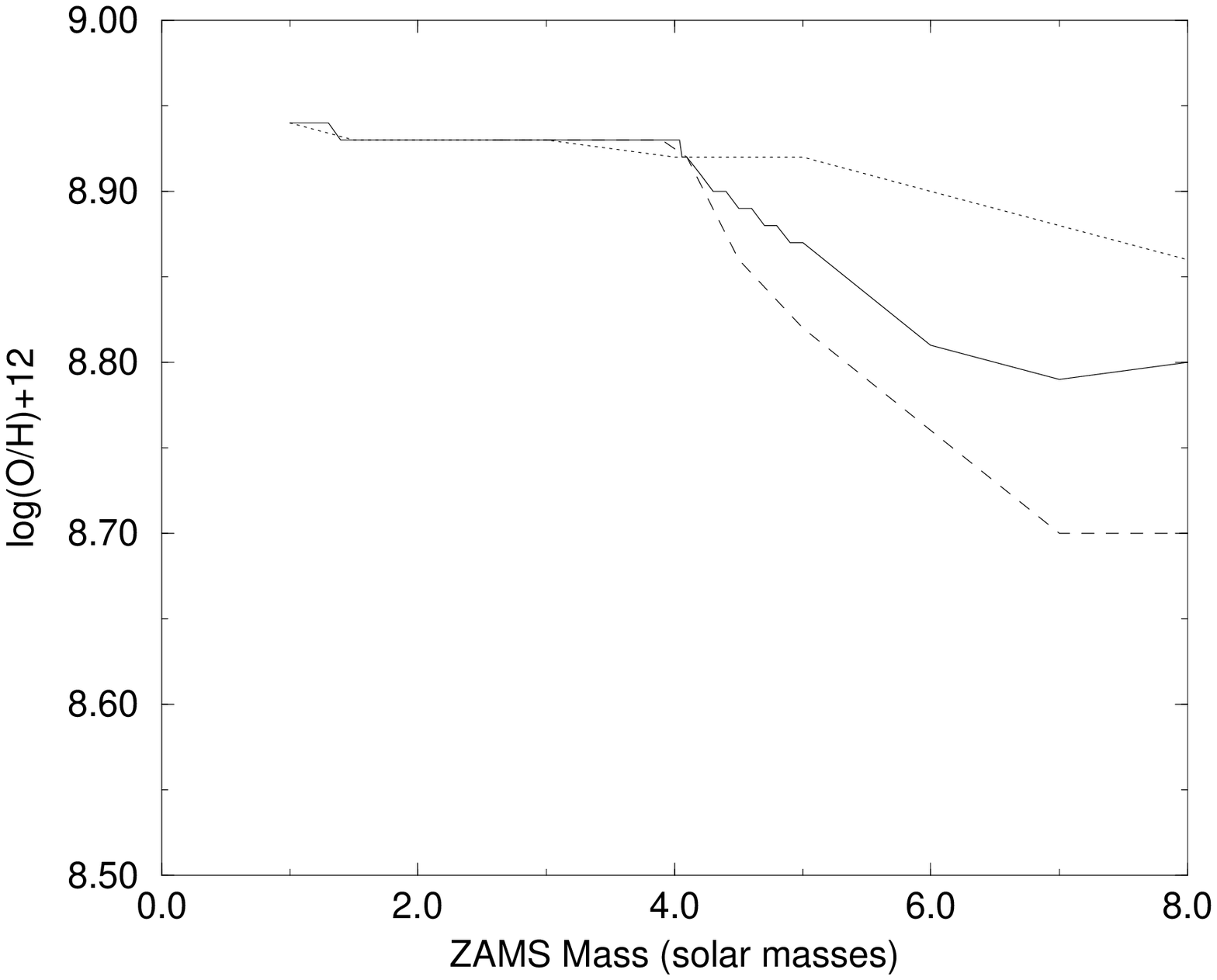,height=4truein}}}
\caption{Same as figure \ref{fig:hehalfamass.eps} except O/H is compared to
$\alpha$ and ZAMS mass.}
\label{fig:ohalfamass.eps}
\end{figure}

The increase in lifetime with mixing length parameter, allows more
time on the TP-AGB for nucleosythesis. A longer lifetime allows for
more thermal pulses and thus more third dredge-up events. This should
lead to higher final C/O and He/H in low mass models for higher
$\alpha$. As expected, in figures \ref{fig:hehalfamass.eps} and
\ref{fig:coalfamass.eps} both He/H and C/O increase with $\alpha$. In
the models with M$\gtrsim$4$\msun$ this leads to higher N/O, because
more carbon is dredged up and then converted to nitrogen. As expected,
N/O increases with $\alpha$ in figure \ref{fig:noalfamass.eps}. By
increasing $\alpha$, the time a model experiences hot-bottom burning
increases. If the time of hot-bottom burning increases than so should
the amount of ON cycling, this effect of this can be seen in figure
\ref{fig:ohalfamass.eps} where O/H decreases with $\alpha$.

Since there is no simple physical limit on $\alpha$, the differences
in $\alpha$ should be regarded as the model errors. Clearly changing
$\alpha$ makes important differences in the predicted abundances. We
determined the ``best'' value of $\alpha$ by deciding which set of
solar metallicity models provided the closest match to the PNe
data. We came up with $\alpha=2.3$. This is similar to the
$\alpha=2.1$ used by Boothroyd and Sackmann on which our luminosity
model is based. 

\newpage
\chapter{The Galaxy}

This chapter provides a comparison of the model predictions in chapter
\ref{modreschap} with the chemical abundances of Galactic
PNe. The comparisons are made on abundance ratio versus abundance ratio
plots.

\section{Data Sets Used}

For our comparison between the PNe data and our models, we have chosen
two data sets because both have carbon abundances determined from IUE
data:
\begin{enumerate}
\item{The set of Henry and Kwitter described in 
\nocite{hkh96} 
Henry \etal\ (1996), 
\nocite{kh96} 
Kwitter and Henry (1996), and Kwitter
and Henry (1997). This set contains 17 objects for which the
abundances of helium, nitrogen, oxygen, neon, and especially carbon
have been carefully determined. This data set will be referenced
hereinafter as HK.}
\item{The sample of Kingsburgh \& Barlow (1994) which contains 80
southern Galactic PNe, for which the abundances of helium, nitrogen,
oxygen, neon, sulfur, and argon were determined. For some PNe the
abundance of carbon was also been determined. This data set will be
referenced hereinafter as KB.}
\end{enumerate}
The abundances in both samples have been determined by using
photoionization models to derive ionization correction factors. Both
groups determined the abundances of NGC 2440 and NGC 7009. Good
agreement is found in the abundance ratios He/H and O/H for NGC2440
and good agreement is found for He/H, N/O, Ne/O, and O/H for
NGC7009. When the two samples are placed on abundance ratio-abundance
ratio plots, both samples seem to have similar patterns. The halo
objects (K648, BB1, DDDM1, and H4-1) are not considered in this
chapter since we examine them in chapter 6.


\section{Metallicity of Galactic PNe}
\label{sec-metalpn}
Samples of Galactic PNe usually exhibit a large range in the
abundances of O, Ne, and Ar, each of which approximately measures the
progenitor's ZAMS composition. In chapter \ref{modreschap}, we noted
two processes which can affect the surface levels of $\oxy$:
hot-bottom burning and third dredge-up, which respectively lower and
raise this mass fraction by $\sim$0.15 dex. The abundance of neon can
be modified by the production of $\nett$ via the reaction chain
$\nit$($\alpha$,$\gamma$) $^{18}{\rm F}(\beta^+,\nu )$ $^{18}{\rm
O}(\alpha,\gamma )$ $\nett$ which occurs during thermal pulses. The
$\nit$ is supplied from the ashes of the hydrogen burning shell. The
mass fraction of the dredged-up $\nett$ is limited by the mass
fractions of the CNO elements and we estimate an upper limit of 0.02
in the case of solar metallicity. Therefore, the surface abundance of
$\nett$ could have increased by as much as 0.15dex. Such a small
change would be difficult to detect although some investigators have
claimed it (Perinotto 1991). Sulfur and argon are probably not
significantly affected, although $s$-process nucleosynthesis might
make small modifications to each element. The possible changes in O/H
and Ne/H make them only approximate indicators of progenitor
metallicity. However, S/H and Ar/H are better indicators because their
surface abundances are less likely to have been modified.

\begin{figure}
\centerline{\hbox{\psfig{figure=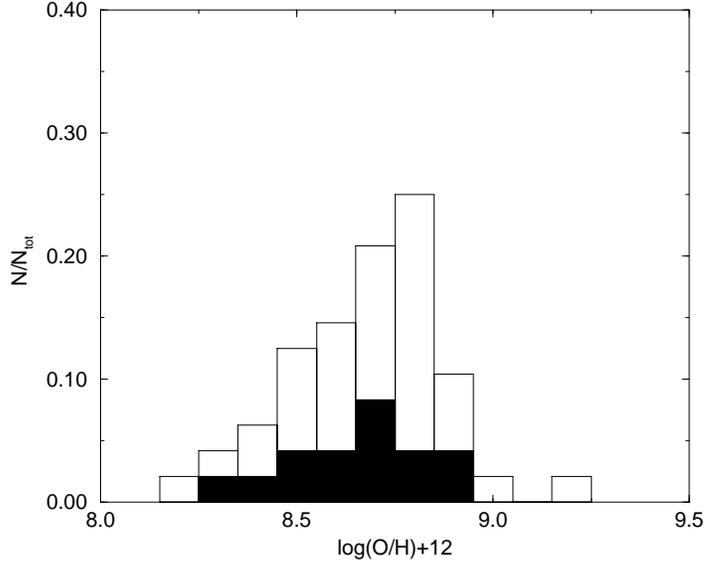,height=3truein}}}
\caption{Distribution of the O/H abundance ratio of the Kingsburgh and
Barlow (1994) and the HK PNe. The open bars indicate the distribution
of all the PNe. The solid bars indicate the distribution of PNe with
N/O$>$0.5.}
\label{fig:ohhist.eps}
\end{figure}

The sample was divided into two subsamples: those with N/O$\le$0.5 and
those with N/O$>$0.5. From the results of chapter \ref{modreschap},
note that models with PN N/O$>$0.5, the excess nitrogen was produced
by hot-bottom burning and second dredge-up. This also appears to be a
division in mass with high N/O objects having M$\ge$4.0$\msun$. Later, we
will briefly consider a third subset with intermediate N/O objects,
which includes objects with 0.5$<$N/O$<$0.7. 

\begin{figure}
\centerline{\hbox{\psfig{figure=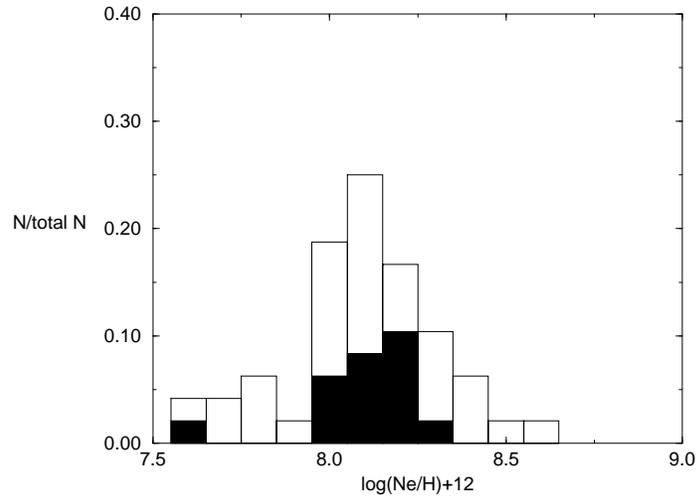,height=3truein}}}
\caption{Same as \ref{fig:ohhist.eps} except the distribution of Ne/H is
shown.} 
\label{fig:nehhist.eps}
\end{figure}

\begin{figure}
\centerline{\hbox{\psfig{figure=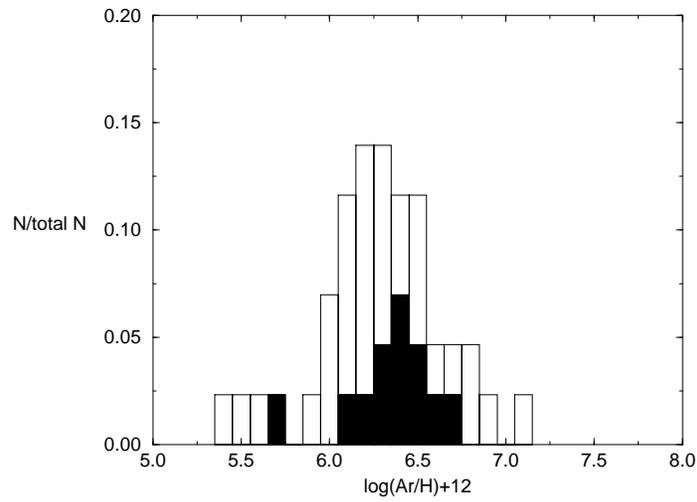,height=3truein}}}
\caption{Same as \ref{fig:ohhist.eps} except the distribution of Ar/H
is shown.}
\label{fig:arhhist.eps}
\end{figure}

Shown respectively in figures \ref{fig:ohhist.eps},
\ref{fig:nehhist.eps}, and \ref{fig:arhhist.eps} are the O/H and the
Ne/H distributions of the KB and HK samples, and Ar/H distributions of
the KB sample. Histograms of the full sample and the subsample with
N/O$>$0.5 are indicated on each figure. The majority of high N/O
objects lie in a narrow metallicity range: O/H is generally between
8.6 and 8.9, Ne/H is generally between 7.95 and 8.25, and Ar/H is
between 6.25 and 6.55. The distribution of sulfur is not shown because
the errors in the sulfur abundances are much larger than those of
other abundances. If we assume that the random error in the O/H, Ne/H
and Ar/H abundances is 0.15dex, then the Ar/H, Ne/H, and O/H data are
consistent with a narrow abundance range ($\approx 0.1$dex). The high
N/O objects have a narrower range of X/H than the low N/O
objects. High N/O objects from the sample need to be fit by a narrow
metallicity range (width=0.1dex) and the low N/O objects need to be
fit by a wide metallicity range.

\section{Comparison of Models to Data}

Comparing the results of TP-AGB star models to PNe data on abundance
ratio versus abundance ratio plots gives us insight into what
processes are operating. As noted earlier, each abundance ratio
results from a unique combination of the nucleosynthetic processes,
the initial abundances, and the mass-loss rate. Therefore, by plotting
two ratios against each other, one would expect to see features which
resulted from the different processes producing the elements. 

\subsection{He/H vs. N/O}

Helium and nitrogen are products of hydrogen burning, but models show
that each mixing process mixes different relative amounts of each
element to the surface. This is different from the simple assumption
that one would expect these to vary in lockstep. In fact only the
second dredge-up simultaneously produces both substantial amounts of
helium and nitrogen. As shown in chapter \ref{modreschap}, mass is the
most important factor determining which mixing processes operate and
to what degree each contributes. Therefore, we expect the signatures
of all the different processes to be visible on the He/H-N/O plane.

The abundance ratio N/O is expected to trace the nucleosynthesis of
nitrogen. Our models suggest that between ZAMS and PN ejection, the
level of oxygen does not change by more than 30\% whereas the nitrogen
abundance can increase by up to a factor of 4. Therefore, any change
in N/O will primarily reflect a change in N and not a change in
O. Also, using N/O avoids the difficulty of not knowing the initial N
abundance since the ZAMS N/H cannot be easily determined for any
individual PN.

Also, using the N/O ratio avoids the potential problems of temperature
fluctuations or dust, both of which may reduce the abundances inferred
from gaseous nebulae. Up to 20\% of the carbon, nitrogen, and oxygen
present can be incorporated onto grains (Meyer 1985) \nocite{mey85},
reducing the inferred abundance by 0.08dex. Temperature fluctuations
(Peimbert 1967) can also reduce the inferred nebular abundances. Using
N/O avoids these problems since the magnitude of the effects on N and
O should be comparable.

The ratio He/H is a measure of the helium production, since He
represents the ashes of hydrogen burning.

\begin{figure}
\centerline{\hbox{\psfig{figure=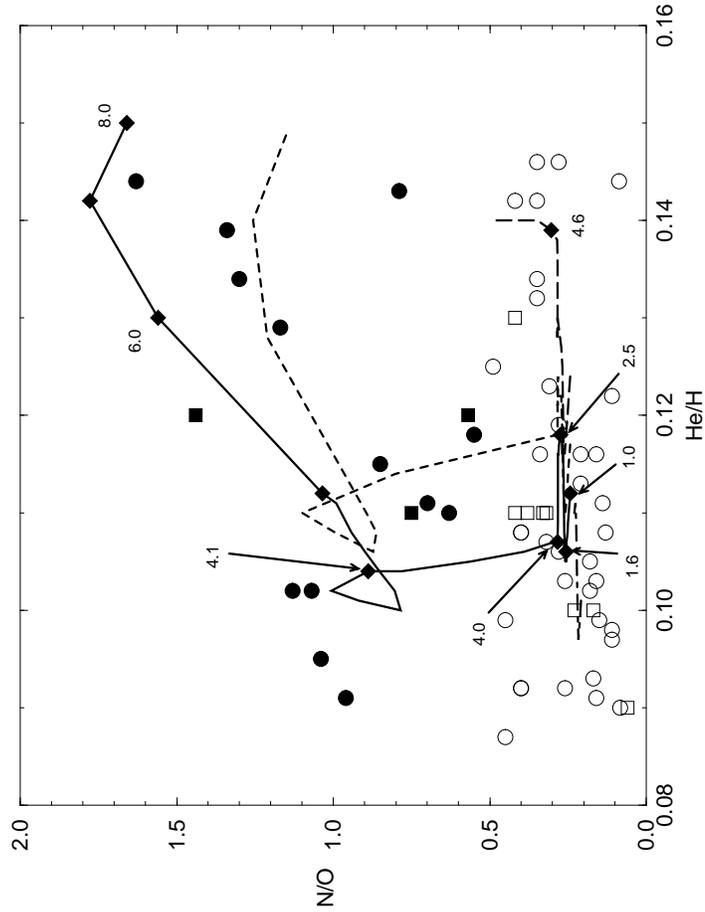,height=5.5truein}}}
\caption{The circles and squares represent the KB and HK data sets,
respectively. Open and closed symbols respectively indicate PNe
with N/O$\le$0.5 and N/O$>$0.5, respectively. The solid, dashed,
long-dashed, and dash-dotted lines refer to models calculated
respectively with [Fe/H]=0.0, 0.1, 0.2, and -0.5. The mixing length
parameter, $\alpha$, of each model was set to 2.3. Only models with
M$<$4.6M$_{\odot}$ are shown for the [Fe/H]=0.2 and only models with
M$<$2M$_{\odot}$ are shown for the [Fe/H]= -0.5. The solid diamonds
indicate the results of models with [Fe/H]=0.0 and masses of 1.0, 1.6,
2.5, 4.0, 4.1, 6.0, and 8.0 M$_{\odot}$. Also the position of the
model with [Fe/H]=0.2 and mass of 4.6 M$_{\odot}$ is indicated.}
\label{fig:heno_comp_all.eps}
\end{figure}

In figure~\ref{fig:heno_comp_all.eps}, we compare our models to the
PNe data on the He/H-N/O plane. Each line consists of several models
with different masses but the same [Fe/H]. For the models with
[Fe/H]=0.0 to 0.1 we have plotted all of the available models for ZAMS
masses between 1 and 8$\msun$. We expect objects with low
metallicities to be in the thick disk or to be very old thin disk
objects, therefore we have included a series of models with
[Fe/H]=-0.5. Since we expect objects with low metallicities to be old
we have only included models with a M$_{\rm init}\lesssim2\msun$. The
metallicity distribution shows tails on both sides of the peak. To
account for the high metallicity tail, we included a grid of objects
with [Fe/H]=0.2. For the [Fe/H]=0.2 line we have only included objects
with M$\le$4.7$\msun$. Since each line is really a grid of different
mass models, we have indicated the positions of certain mass models on
the diagram. The range of metallicities is reasonably comparable to
the range of metallicities found in dwarf F and G stars by
\nocite{edv93} Edvarsson \etal\ (1993).

Inspection of figure~\ref{fig:heno_comp_all.eps} suggests that these
models fit most of the data reasonably well. We expect the [Fe/H]=0.0
and 0.1 grids to overlap the majority of the PNe.  In section
\ref{sec-metalpn} we showed there are clear peaks in the distributions of
metallicity indicators (O, Ne, and Ar) and if we assume the PNe near the
peaks have solar metallicity, then most PNe should be explained by the
models with [Fe/H]=0.0 and 0.1, particularly those with high N/O. As
expected the [Fe/H]=0.0 and 0.1 curves reach most of the high N/O
PNe. The near solar metallicity models also get to most of the low N/O
objects. Only two areas where PNe exist cannot be reached by the two
near solar metallicity curves:
\begin{enumerate}
\item{PNe with N/O$<$0.5 and He/H$<$0.105, and}
\item{PNe with N/O$<$0.5 and He/H$\ge$0.125.}
\end{enumerate}
The objects in the first category can be fit with models with less
than solar metallicity. Edvarsson \etal\ (1993) \nocite{edv93}
demonstrated that significant numbers of low-mass low-metallicity
stars exist in the solar neighborhood, which are probably the
progenitors of the PNe in the first category. In figure
\ref{fig:heno_comp_all.eps} the low-mass [Fe/H]=-0.5 models reach most
of the low N/O and low He/H PNe. The figure also shows that objects in
the second category can be reached by models with [Fe/H]=0.2 and
masses between 2.5 and 4.5$\msun$.

To simplify the comparison between models and data, we have divided
the sample and data into 3 subsamples: 1) N/O$<$0.5, 2)
0.5$\le$N/O$<$0.7, and 3) N/O$\ge$0.7. Subsamples 1 and 3 are
equivalent to the low and high N/O PNe discussed earlier. We call the
lowest N/O PNe type IIb, the intermediate group type IIa, and the high
N/O group type I. This classification system resembles the
well known Peimbert system (Peimbert 1978 \nocite{pei78}), but it is
not the same since helium is not included.

\begin{figure}
\centerline{\hbox{\psfig{figure=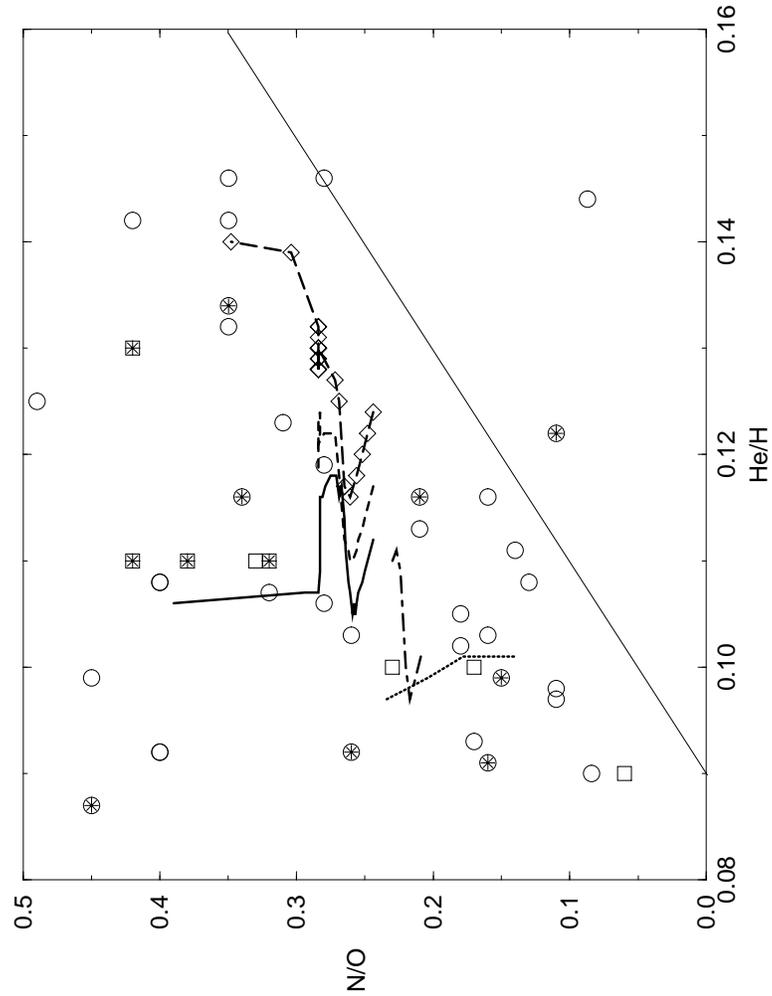,height=6truein}}}
\caption{The symbols and lines have the same meaning as they do in
figure \ref{fig:heno_comp_all.eps}. Also included is a dotted line,
which shows the predicted abundances for stars with [Fe/H]=-0.5 and
masses of 0.9, 1.0, 1.1, and 1.2 M$_{\odot}$ using a first dredge-up
model based on the recent results of Boothroyd and Sackmann (1997) as
explained in the text. The light solid line is a rough approximation
of the divide between the allowed and forbidden regions of the
plane. The symbols with stars in them indicate PNe with
C/O$>$0.8. Note that many objects do not have an observed C/O ratio.} 
\label{fig:heno_lowno.eps}
\end{figure}

To make more detailed comparisons in the models to type IIb PNe, we
have ``zoomed-in'' on the region of figure~\ref{fig:heno_comp_all.eps}
with N/O$<$0.5 in figure~\ref{fig:heno_lowno.eps}. For the type IIb
PNe, a weak correlation exists between He/H and N/O. Similar
correlations have been reported by Henry (1990) \nocite{hen90} and
Perinotto (1991) \nocite{per91}. In general, He/H increases as N/O
increases for the type IIb PNe. There is an apparent ``forbidden''
region on the He/H-N/O plane. The region below and to the right of the
light solid line on figure \ref{fig:heno_lowno.eps} is only sparsely
populated with PNe.

Our models qualitatively explain both the correlation between He/H
and N/O and the ``forbidden'' region seen in the type IIb PNe. Note
that the N/O level of [Fe/H]=-0.5 models is significantly less than
that of the solar metallicity models. Also in general, He/H for the
thick disk metallicity models are generally lower than those of the
solar metallicity models. Examining figure~\ref{fig:heno_lowno.eps}
and noting the position of the [Fe/H]=0.0 and -0.5 models in
particular, the models appear to duplicate the correlation. Each model
also predicts a maximum He/H with each given [Fe/H] which explains the
existence of the ``forbidden'' region. In chapter \ref{modreschap}, it
was noted that the level of the second He/H maxima is a function of
[Fe/H]. Since the level of N/O increases with [Fe/H] and there is a
maximum He/H for each N/O we expect that the maximum He/H will
increase with N/O giving the diagonal boundary of the forbidden
region.

The ZAMS level of C/O and N/O have significant effects on the N/O
abundance at PN ejection. The most important factor in determining the
model type IIb PN N/O is the first dredge-up. The first dredge brings
material to the surface from layers where the CN cycle has
operated. However, the first dredge can at most double the surface
nitrogen abundance. Therefore, if the ZAMS level of C/O and N/O are
less than the solar levels, the resulting PN N/O will also be lower. A
consequence of our initial abundance is the level of N/O at the first
dredge-up will correlate with model [Fe/H] below [Fe/H]=-0.2.

Another minor effect is the increase in the oxygen abundance during
the third dredge-up. Increasing the oxygen abundance will cause a
corresponding decrease in N/O. The amount of this N/O depletion will
decrease with increasing metallicity. This occurs because it is easier
to dilute the N/O ratio if the abundance of N is lower. 

The first dredge-up increase of N/O with metallicity and the increased
impact of the dredge-up of $\oxy$ qualitatively explain the He/H-N/O
correlation for type IIb PNe, but do they explain it quantitatively?
Our diagram indicates reasonable agreement between the models and the
data. However, if we consider the models at the extreme metallicities
([Fe/H]=-0.5 and 0.2) they do not have enough range in N/O to reach
the lowest or highest N/O PNe. In the case of the high range in N/O,
the disagreement is slight being no more than a 30\% difference. Two
possibilities exist which would allow the extension of the models to
reach the highs and lows of N/O:
\begin{enumerate}
\item{A slightly different choice of a model for ZAMS C/O and N/O
would increase the range of N/O.}
\item{Our model of first dredge-up needs modification at the lowest
masses.} 
\end{enumerate}

In our model the ZAMS C/O and N/O are constant for [Fe/H]$>$-0.2. As
noted earlier, this reflects the trends in Edvarsson \etal\ (1993)
\nocite{edv93}. This is responsible for the fact that the N/O level in
the low mass models (M$\lesssim$4$\msun$) with [Fe/H]$\ge$0.0 is
nearly constant on the diagram. However, other models exist
(e.g. Boothroyd and Sackmann 1997) in which C/O and N/O never level
off but continue to rise as a function of [Fe/H]. If such a model were
used it would lower N/O for the [Fe/H]=-0.5 models and raise N/O for
the [Fe/H]=0.2 models giving better agreement.

The other possibility was impossible to study until after this work
was essentially completed when the tables for the first dredge-up
calculations of Boothroyd and Sackmann (1997) \nocite{bs97} became
available. They have recalculated the first dredge-up for masses from
0.85-9$\msun$ and in general their results agree with those of
previous investigators. However, they made first dredge calculations
at metallicities never before used and also at lower masses than other
investigators. Using their results, we estimated that the change in
He/H at low mass ranges is unchanged. However, we found that the
average change in the nitrogen abundance showed significant
differences. In table \ref{nit_enh}, we compare the enhancements
predicted by our models and those of Boothroyd and Sackmann. Clearly
at the lowest masses significant differences exist. On figure
\ref{fig:heno_lowno.eps}, we have plotted an estimate of the change in
[Fe/H]=-0.5 models with M=0.9, 1.0, 1.1, and 1.2 $\msun$ using the
results of table \ref{nit_enh}. These very low mass models clearly
reach low enough in N/O to explain the lowest N/O models. A
recalculation of first dredge-up using the models of Boothroyd and
Sackmann (1997) \nocite{bs97} may result in slight but important
changes in the N/O level. This may bring the high range of N/O into
better agreement with the models.

\begin{table}
\begin{center}
\begin{tabular}{|c|c|c|}\hline
Mass&BS97&Our Calc.\\\hline
0.9&+0.10&+0.26\\\hline
1.0&+0.20&+0.27\\\hline
1.1&+0.23&+0.28\\\hline
1.3&+0.29&+0.29\\\hline
\end{tabular}
\end{center}
\caption{Comparison of our model enhancements of $\log{\rm N/O}$ against
those of Boothroyd and Sackmann (1997). Their calculations indicate
enhancements in N/O depend mainly on mass and appear to depend only
very weakly on metallicity at least in the mass range considered.}
\label{nit_enh}
\end{table}

The PNe with He/H$\ge$0.125 and N/O$<$0.5 deserve more attention. We
noted earlier that these objects can be fit by models with
[Fe/H]=0.2. However, there is some question whether or not these
abundances are real. For one, these objects may be the result of
uncertainties in the abundance determination processes, in particular
He/H. All but one of the objects in this mass range comes from the KB
subsample. Some other samples do not have such objects or have only a
few such objects, e.g. Costa \etal\ (1996) does not have any
corresponding objects, Henry (1990) has only one Galactic object with
N/O$<$0.5 and He/H$\approx$0.14. In the HK subsample, the highest
He/H is 0.130, which can probably be fit by a model with [Fe/H]=0.1.

Another possibility is those PNe with He/H$>$0.125 and low N/O have
experienced a helium shell flash on the post-AGB, the so-called ``born
again'' phenomena of Iben \etal\ (1983). If this occurs additional
helium-rich material would be ejected into the nebula at the beginning
of the PNe stage, substantially increasing the helium abundance in the
PN. Can the models of solar metallicity PNe be made to reach
He/H=0.140 before the onset of hot-bottom burning? For this to occur
He/H must go from 0.100 to 0.140, an increase of 40\%. In our models,
even with average lambdas of 0.7, this does not seem
possible. However, helium is most effectively enhanced during the last
few pulses when the mass of the envelope has been reduced. During the
last pulse, if the envelope has been reduced to a few tenths of a
solar mass or less and a dredge-up occurs it might be possible helium
could be enhanced enough to allow a solar metallicity model to reach
very high He/H.

\subsection{He/H vs. C/O}

\begin{figure}
\centerline{\hbox{\psfig{figure=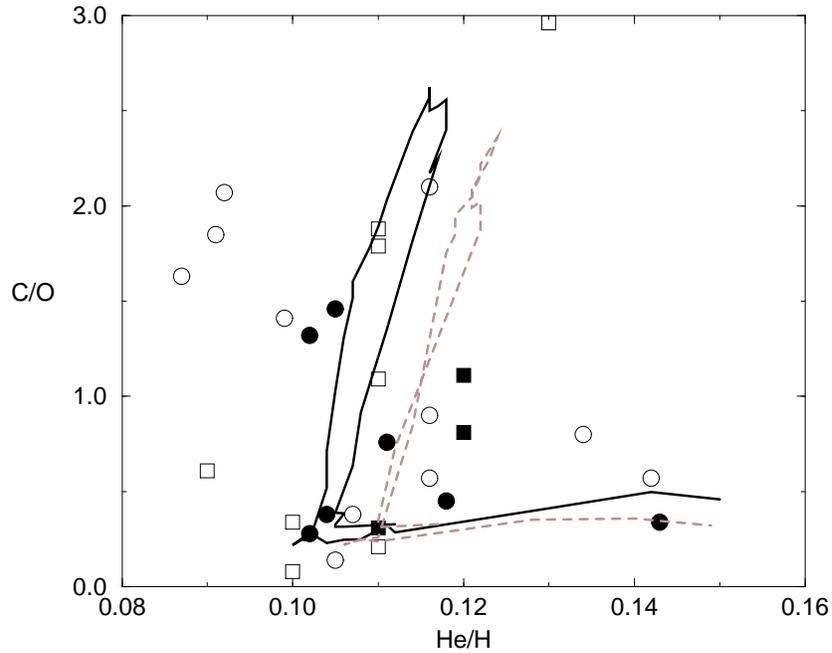,height=4truein}}}
\caption{The symbols have the same meaning as figure
\ref{fig:heno_comp_all.eps}. The dotted line are the [Fe/H]=-1.0
models.}
\label{fig:hecocomp.eps}
\end{figure}


In figure \ref{fig:hecocomp.eps}, a comparison is made between the
models and the Galactic PNe on the C/O versus He/H plane. The
agreement between models and data is good. Some points
of agreement:
\begin{enumerate}
\item{The near solar metallicity models explain most of the data.}
\item{The models with the highest C/O seem to correspond to the
highest C/O found in the sample (Except for the obvious outlier
J900).}
\item{The models reproduce the overall trends in the data.}
\end{enumerate}

The objects with low helium and high C/O (He/H$\lesssim$0.09 and
C/O$>$0.9) present some problem for our model:
\begin{enumerate}
\item{These could be low metallicity PNe, in which case a model with
lower [Fe/H] could be used to fit them. The diagram indicates that
these PNe can be reached by models with [Fe/H]=-1.0, suggesting a low
metallicity.}
\item{The measured helium abundances in these objects could be systematically
low. For instance, a low helium abundance could be generated if
neutral He corrections were required. For two of these PNe, N5873 and
I2448, the abundance of He$^+$ was determined from only a weak
$\lambda$4471 line. Also, a 10\% discrepancy between determinations of
He$^+$/H$^+$ for the various lines is common in the KB sample. A
systematic shift upward by 10\% would bring these objects into
agreement with the rest.}
\end{enumerate}
Either explanation is possible, although we favor the second one,
since the objects have extremely low He/H implying a ZAMS He/H of
0.08, which is nearly primordial and therefore seems unlikely.

\begin{figure}
\centerline{\hbox{\psfig{figure=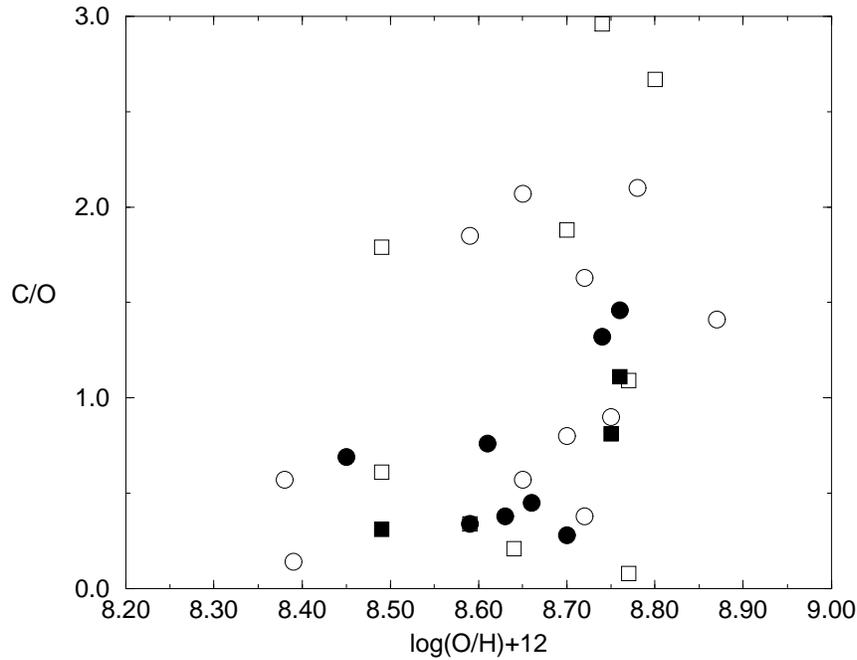,height=4truein}}}
\caption{Superimposed on the grid of [Fe/H]=0.0 $\alpha$=2.3 models
are the masses of those models. }
\label{fig:cooh.eps}
\end{figure}

The masses of selected models are shown in figure
\ref{fig:cooh.eps}. The objects with C/O$>$1 are low mass objects. The
objects with the highest C/O lie between 2 and 3.5 $\msun$. The PNe
with intermediate mass progenitors have low C/O due to hot-bottom
burning. One discrepancy is the low N/O objects with
He/H$>$0.13. These objects are predicted by these models to have high
N/O. These objects could potentially be due to errors in abundance
determinations, in particular that of N/O or He/H. For instance Kaler
(1983) determined the abundance of N2438 as 0.103 which is
significantly different from the value used here. Also if the inferred
N/O is too low these objects become high mass objects. Therefore,
although these objects are a problem, it is possible they are simply
due to errors in abundance determination.

\subsubsection{He/H vs. X/H}
\begin{figure}
\centerline{\hbox{\psfig{figure=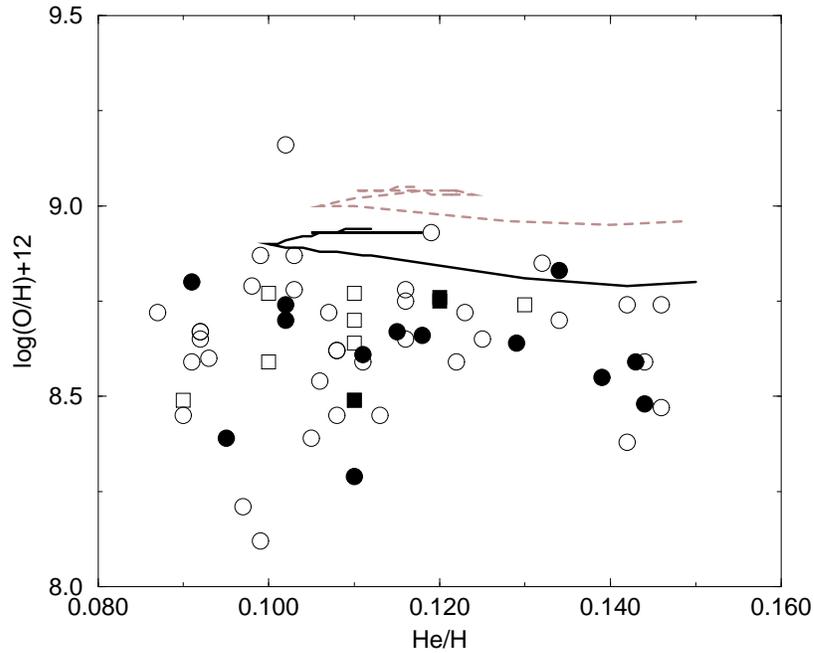,height=4truein}}}
\caption{Galactic PNe plotted on the He/H-O/H plane. The symbols have
the same meaning as in figure \ref{fig:heno_comp_all.eps}.}
\label{fig:heocomp1.eps}
\end{figure}

Our models predict, for low N/O PNe, He/H should reach a different
maximum at each metallicity. In figure \ref{fig:heocomp1.eps}, a
comparison is made between models and PNe data on the
He/H-log(O/H)+12 plane. Clearly, the models do not agree with the
data. However, note that the majority of the PNe have observed O/H
less than the solar value (8.93, Anders and Grevesse 1989). This is
the well known problem that the abundances inferred from lines in
gaseous nebulae do not agree with the values determined in the
Sun. For instance, several investigators have determined the Orion
Nebula to be $\sim$2 less in metallicity than the Sun. Two
explanations have been advanced to explain this potential discrepancy.
\begin{enumerate}
\item{Strong temperature fluctuations exist (Peimbert 1967) which
could reduce the apparent abundances by a factor of 2. Walter \etal\
(1992) accounted for temperature fluctuations and found the abundances
of CNO in the Orion Nebulae matched those of the Sun; but when they
did not invoke such fluctuations, the oxygen abundance was 0.45$dex$
less.}
\item{The solar abundances are not representative of the typical
interstellar medium. Snow and Witt (1996) make a good case for this
possibility. Snow and Witt show the average abundances of B-stars and
F and G stars are less than those found in the Sun. In particular, the
average O abundance is 0.3dex less than in the Sun. Therefore it may not
be necessary to invoke temperature fluctuations to explain the
discrepancy.}
\end{enumerate} 

Snow and Witt did not look into the average He abundance in the
interstellar medium, although it is very important for determining PNe
abundances. Kilian (1992) and Gies \& Lambert (1992) investigated the
He abundance in B-stars. Excluding a few exceptionally large helium
abundances, the average He abundance of these samples is approximately
solar while the average O abundance is reduced by a factor of
$\sim$2. There may also be a real scatter in the initial He/H for any
given O/H.

\begin{figure}
\centerline{\hbox{\psfig{figure=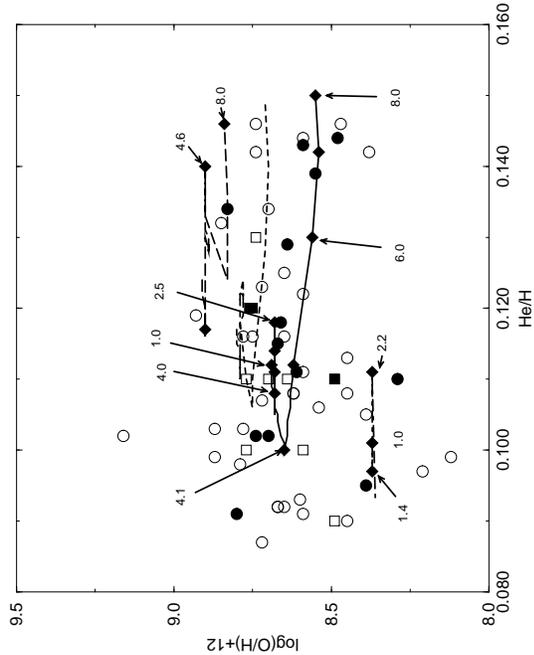,height=4truein}}}
\caption{Same as figure \ref{fig:heocomp1.eps} except the model curves
have been shifted down by 0.25{\it dex} as indicated in the text.}
\label{fig:heocomp2.eps}
\end{figure}

No resolution to the discrepancy shall be proposed here. However,
corrections need to be made for it. Therefore, in figure
\ref{fig:heocomp2.eps} the model tracks have been shifted downward in
log(O/H)+12 by 0.25{\it dex} in order to force agreement between data
and models. In figure \ref{fig:heocomp2.eps}, we show the effect of
the above shift. This is admittedly an ad hoc method, however, we feel
it is reasonable given the possible discrepancies mentioned above.

Referring to figure \ref{fig:heocomp2.eps}, it can be seen that the
high and low N/O objects both fit the expected pattern. We show in
figure \ref{fig:heocomp2.eps} the position in mass of the various
models. Our models predict that the progenitors of
M$\gtrsim$4.0$\msun$ should produce high N/O objects. Since the models
from $\sim$4.5$\msun$-$\sim$8.0$\msun$ exhibit a slight downward slope
on the He/H-log(O/H) plane, we expect the high N/O objects to exhibit
a similar pattern and it appears to match the observations. However,
the drop in O/H between 4.1 and 8.0 $\msun$ is less than 0.1$dex$,
therefore it would be almost impossible to detect. Also, our models
predict that for the low N/O, that the upper limit of He/H should be a
function of O/H (also Ne/H and Ar/H). Examination of the PNe in figure
\ref{fig:heocomp2.eps} shows this to be true.

\begin{figure}
\centerline{\hbox{\psfig{figure=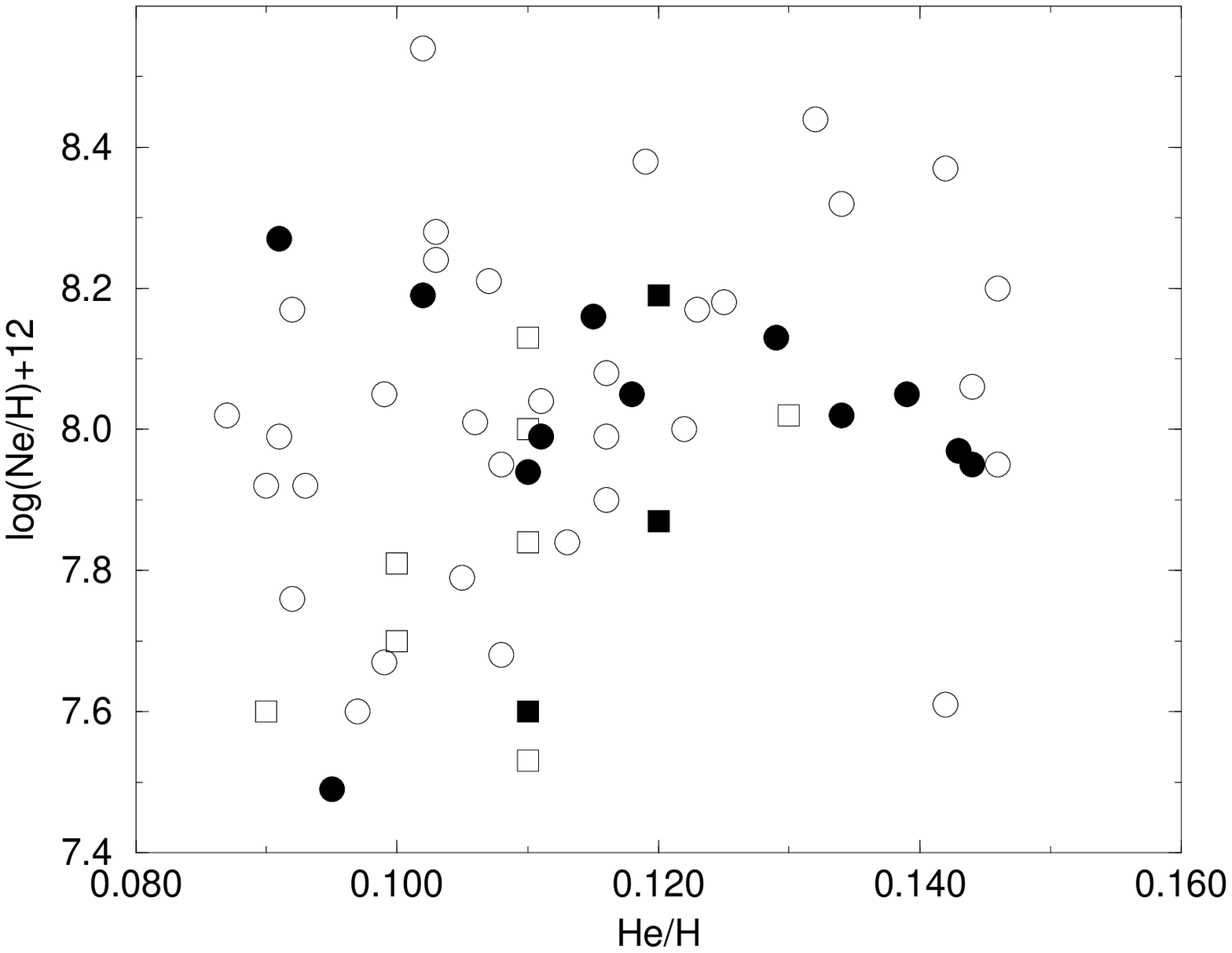,height=4truein}}}
\caption{Comparison of He/H vs Ne/H. The low and high N/O PNe are
indicated by open and closed circles, respectively. Note that the
maximum He/H for low N/O objects increases with Ne/H.}
\label{fig:hene.eps}
\end{figure}

\begin{figure}
\centerline{\hbox{\psfig{figure=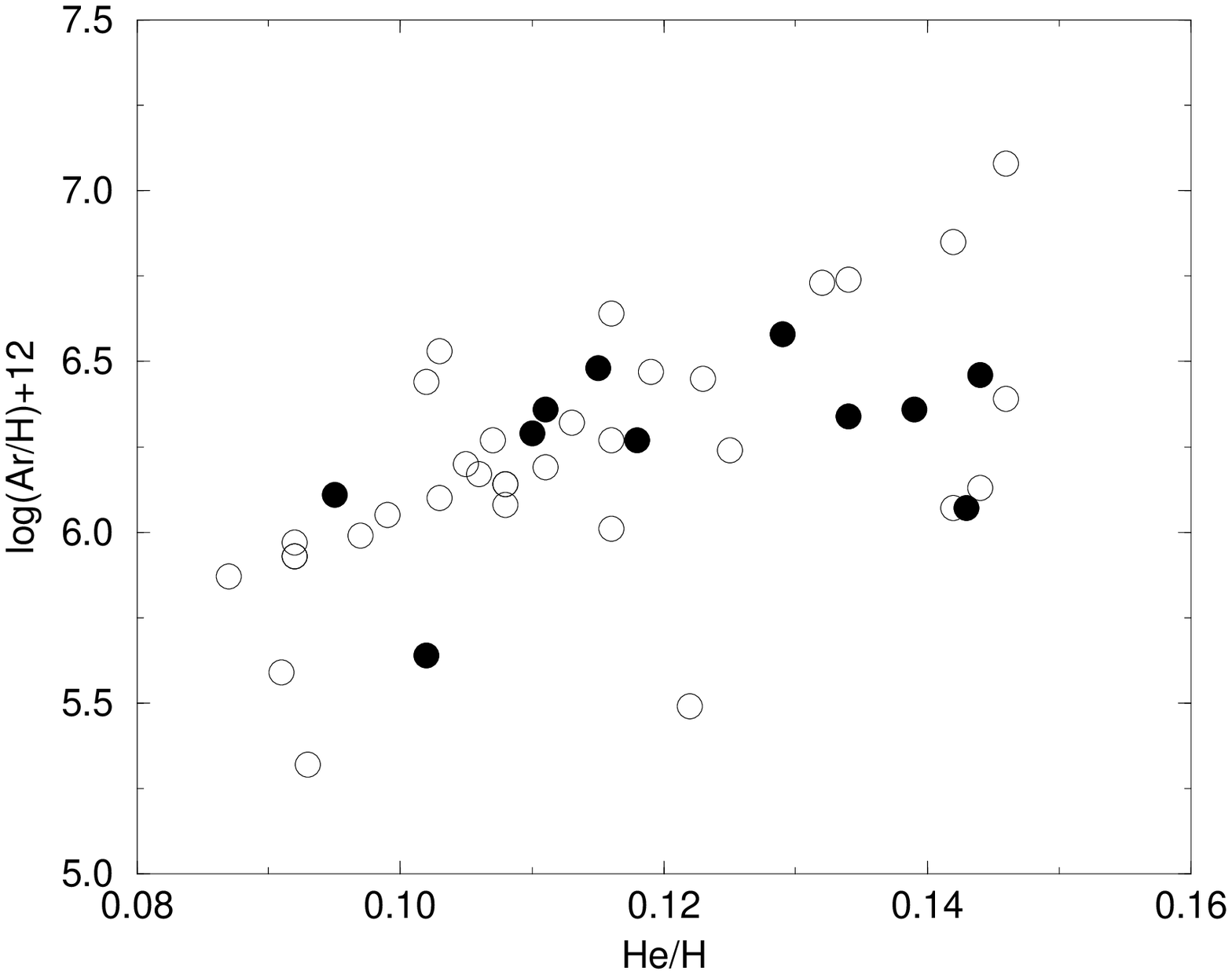,height=4truein}}}
\caption{The symbols have the same meaning as in figure
\ref{fig:hene.eps}. Note the maximum He/H for low N/O objects
increases with Ar/H.}
\label{fig:hear.eps}
\end{figure}

The same effect should be evident with other metallicity indicators. In
figures \ref{fig:hene.eps} and \ref{fig:hear.eps}, the low N/O PNe
both seem to have a clear upper limit in He/H as a function of Ne/H
and Ar/H, respectively. 

\begin{figure}
\centerline{\hbox{\psfig{figure=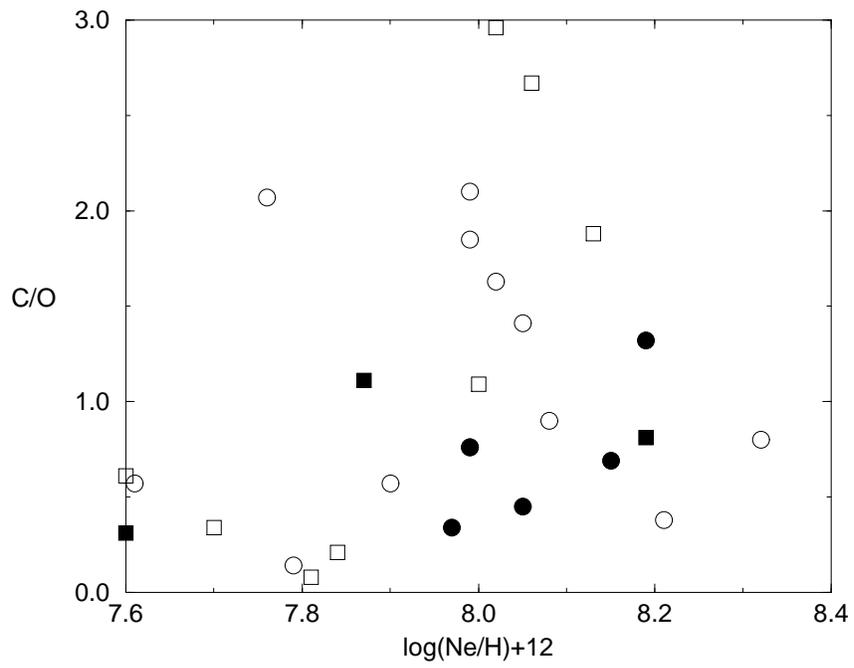,height=4truein}}}
\caption{Galactic PNe plotted on the C/O-Ne/H plane. The symbols have
the same meaning they do in figure \ref{fig:heno_comp_all.eps}. Note
the narrow range in Ne/H of PNe with C/O$>$1.0.}
\label{fig:cone.eps}
\end{figure}

Before we further examine figure \ref{fig:heocomp2.eps}, an
examination of the metallicity of the high C/O PNe is in order. In
figure \ref{fig:cone.eps}, we have plotted C/O vs log(Ne/H) and a simple
trend clearly jumps out, the high C/O (C/O$>$1) PNe tend lie between
$12+\log{\rm Ne/H}=$8.0-8.2. This very narrow range in Ne/H suggests
these objects all came from essentially the same metallicity
progenitors. 
Also in figure \ref{fig:cone.eps} many of the high N/O objects lie in this
same region, suggesting that both PNe with C/O$>$1 and PNe with
N/O$>$0.5 come from the {\em same} narrow range of metallicity ($\sim
0.1-0.2dex$).

The narrow metallicity range for high N/O and high C/O objects
suggests that nearby objects with M$\gtrsim$2.0$\msun$ come from
progenitors with a range in metallicity of $\sim$0.1-0.2$dex$. Our
models and those of several other investigators (Bazan 1991, Boothroyd
and Sackmann 1988abcd, Boothroyd and Sackmann 1992, Boothroyd \etal
1993, Busso \etal 1992, Busso \etal 1995, Forestini and Charbonnel
1996, Groenewegen and deJong 1993, Groenewegen and deJong 1995, Marigo
\etal 1996, Lattanzio 1986 \nocite{latt86}, Vassiliadis and Wood 1993)
suggest that the minimum mass for carbon star formation,
$\mcs\gtrsim1.5\msun$ and for the formation of nitrogen rich stars via
hot-bottom burning, M$_{\rm hbb}\gtrsim4.0\msun$. Models by Schaller
\etal\ (1992) suggest that a 1.5$\msun$ star requires 1.8-2.6Gyrs to
reach the He core flash depending upon metallicity. This can be
regarded as the lower limit to the age of the high N/O and high C/O
objects.

Since all of these objects appear to come from such a narrow
range in metallicity, this range may indicate an upper limit to the
enhancement of the local interstellar medium. The limited range of
lifetimes available to progenitors with M$_{ZAMS}\gtrsim1.5-2.0\msun$
and the narrow metallicity range combined suggest the placement of
limits on the rate of enhancement of the interstellar medium with 
\begin{equation}
\frac{\Delta\log{\rm O/H}}{d{\rm t}}<0.11dex\cdot Gyr^{-1}.
\end{equation}

This appears to contradict the results of Edvardsson \etal\ (1993), who
claim that there is no unique age-metallicity relationship. They
showed in their paper that at any given age a large spread of
metallicities exists. However, the sample of high C/O and high N/O
objects probably is not from the same population as their G and F
stars. The probable mass of the PNe progenitors indicates that on the
main sequence, they were probably A and B stars. Therefore, since the
samples are different, there is no contradiction.

The difference in the metallicity spreads between our sample and the
sample of Edvardsson \etal (1993) may point to why their sample has
an age-metallicity degeneracy. If stars with M$\gtrsim$2.0$\msun$ come
from a narrow range of metallicities and those with
M$\lesssim$2.0$\msun$ come from a wide range, some process must cause
stars that are degenerate in age and [Fe/H] to appear in the solar
neighborhood. It is reasonable to suppose that the initial
metallicity is a function of the time and place of birth, which can be
expressed as:
\begin{equation}
[Fe/H]=f(R,t).
\end{equation}
However, as stars get older they are less likely to be found near
their point of origin. A possible explanation of the trend in F and G
stars is that they originated at many different positions in the
Galaxy with different [Fe/H]'s and migrated to the stellar
neighborhood. The reason the high N/O and high C/O PNe do not show
this trend is because there is not sufficient time for the migration
to occur. Note that the low C/O-low N/O PNe do show a very wide range
in O/H, Ne/H, and Ar/H which is consistent with the stars of the
Edvardsson sample.

\subsection{Other Diagrams}

\begin{figure}
\centerline{\hbox{\psfig{figure=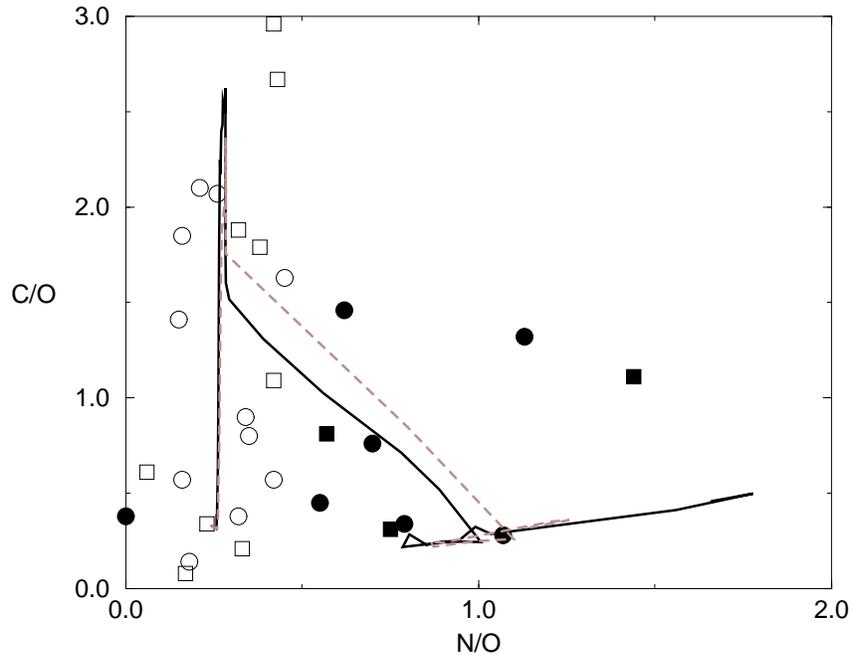,height=4truein}}}
\caption{Comparison of models to data on the C/O to N/O plane.}
\label{fig:noco.eps}
\end{figure}

\begin{figure}
\centerline{\hbox{\psfig{figure=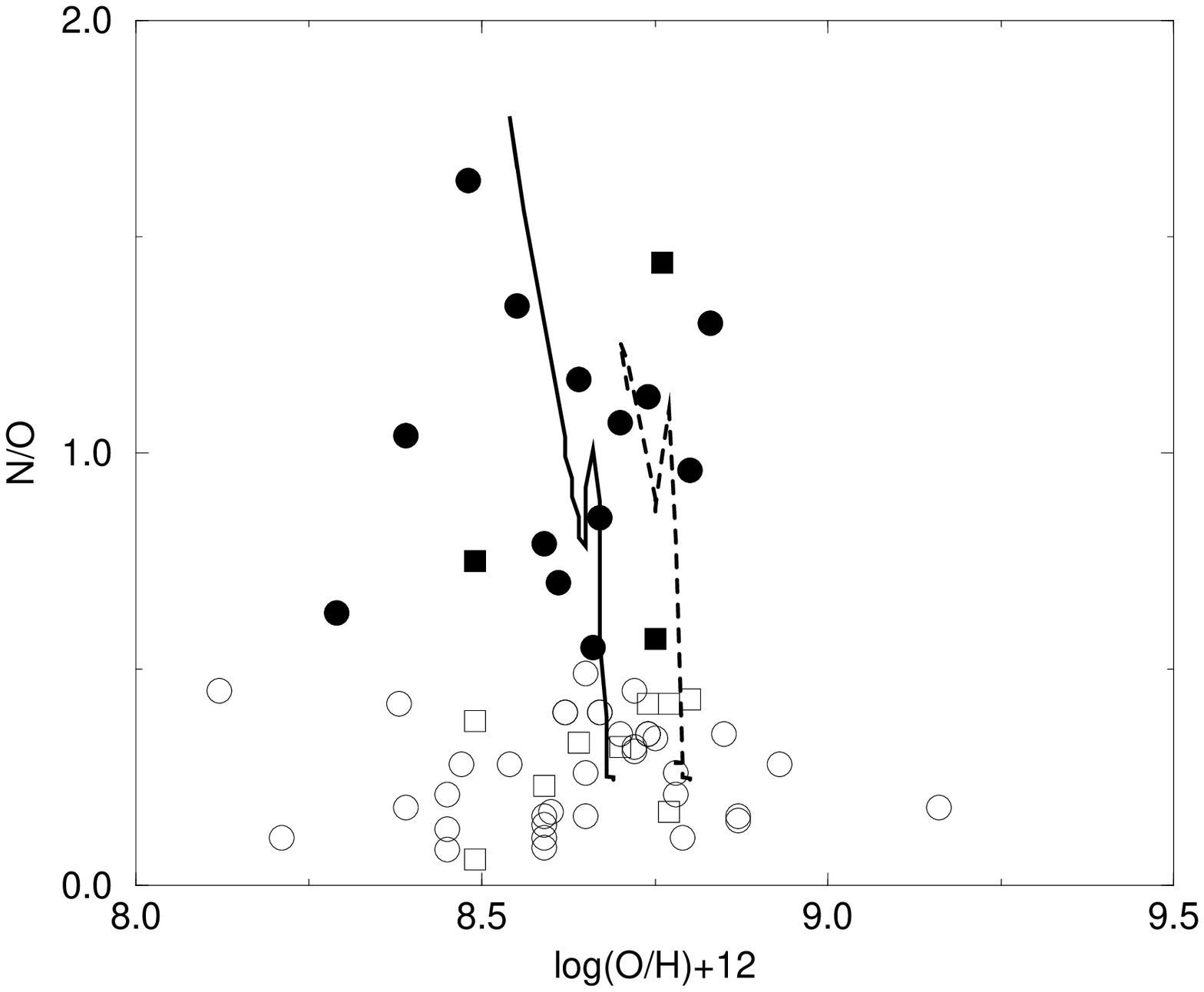,height=4truein}}}
\caption{}
\label{fig:nooh.eps}
\end{figure}

Hot-bottom burning converts carbon to nitrogen, so when N/O is high
C/O should be low and vice-versa. Figure
\ref{fig:noco.eps} compares the models to the data on the N/O-C/O
plane. The agreement between models and data is reasonably
quantitative. C/O and N/O are related as suggested above.

An often used method to look for ON cycling is to plot PNe on the
O/H-N/O plane and to look for an anticorrelation between the two
quantities. Figure \ref{fig:nooh.eps} shows that our models can fit
the data. At high N/O, our models show a slight drop in O/H. However,
the magnitude is very small at near solar metallicities, therefore,
any anticorrelation that might exist is weak.

\section{Summary}

We have found that our models match the observed features of PNe
abundance ratio data. In particular:
\begin{enumerate}
\item{We have found examples of helium being produced by both the
second and third dredge-up. In high N/O PNe, the helium is produced by
the second dredge-up. In low N/O-high He/H PNe, helium is produced via
the third dredge-up.}
\item{Nitrogen is produced via the first dredge-up and via a
combination of second dredge-up and hot-bottom burning. In high N/O
PNe, some of the N/O comes from the second dredge and then hot-bottom
burning converts the remaining C to N. There is evidence of first
dredge-up in low N/O PNe.}
\item{Carbon is produced in low-mass stars and is destroyed in
intermediate mass stars. In low-mass stars carbon that is dredged-up
is not destroyed via hot-bottom burning while in intermediate mass
stars it is.}
\item{For low N/O PNe, the maximum possible He/H is controlled by the
metallicity.}
\item{We can match very closely the maximum N/O and C/O seen in PNe.}
\end{enumerate}
We find that to match all the PNe data, a large range of metallicity
progenitors are necessary. This is consistent with stellar
abundances. However, to match PNe with high C/O ($>$1) or high N/O
($>$0.5) a very narrow range of metallicity progenitors are necessary.

\section{Discussion}

Overall, our models with hot-bottom burning and dredge-up do a good
job of matching the PNe abundances. We can duplicate the high C/O and
N/O ratios seen in many PNe. The models also predict how He/H should
behave. 

\subsection{Correlations with Mass and Core-Mass}

An interesting possibility is using the abundances of PNe to get a
rough idea of both their core-mass and the progenitor masses. The two
masses are not unrelated. Weidemann and Koester (1987) have shown a
relation exists between the ZAMS mass and the mass of the resulting
white dwarf, a so called initial-final relation (${\rm M}_i-{\rm
M}_f$). Some observational efforts have been made along these lines
(Ratag 1991, Kaler and Jacoby 1990, Kaler \etal 1989) who attempted
successfully to correlate N/O with core-mass.

Can He/H be used to determine the mass of PNe? Clearly, there is no
unique relation between He/H and mass. Our models do a reasonable job
of fitting the data and predict that He/H rises with mass between
$\sim$1.5$\msun$ and $\sim$2.5$\msun$, but between $\sim$2.5$\msun$
and $\sim$4.5$\msun$ He/H is anticorrelated with mass. Above
4.5$\msun$, He/H again rises. However, it maybe possible using N/O and
He/H in conjunction to determine the mass. For high N/O PNe
(N/O$\gtrsim$0.8), figure \ref{fig:heno_comp_all.eps} shows that He/H
is a function of progenitor mass. For high N/O PNe, N/O and He/H are
correlated. Therefore, it should be in principle possible to get a
rough idea of progenitor mass from He/H for high N/O PNe. More work
needs to be done on the second dredge-up before predictions are made
as to the actual mass of the high N/O progenitors. Also, we predict
that the core-masses of PNe with N/O$>$0.8 will be correlated with
He/H.

For the low N/O PNe, the use of He/H to determine mass is
probably impossible. There is a ``degeneracy'' in mass
for He/H, in the sense that a different mass model for each [Fe/H]
will give the same He/H. In principle, a plot of He/H vs Z/H where Z
is O, Ne, S, or Ar overlayed with a grid showing the masses could be
used to determine masses. However, the errors in the abundances are
such that it is probably impossible to accomplish this with any
confidence. On the other hand, C/O could probably be used as a rough
mass indicator for the low N/O PNe. The ratio of C/O seems to vary
with mass and also core-mass. 

\subsection{Hot-bottom Burning}

There is some debate as to whether the enhancements
of nitrogen seen in PNe are due to the CN cycle, the ON cycle, or a
combination of both. 

The CN cycle is clearly in operation. Peimbert (1973) was the first to
show that N/O is much greater than the typical HII region values. Many
abundance studies (Henry 1990\nocite{hen90}, Kingsburgh and Barlow
1994\nocite{kb94}, Leisy and Dennefeld 1996\nocite{ld96})have
demonstrated convincingly that $\ctw$ has been converted into $\nit$
by the PN progenitors.

Must hot-bottom burning occur to produce the high N/O ratios? For
instance, the second dredge-up can produce nitrogen. However, figure
\ref{fig:nomass.eps} suggests that N/O at the end of the second dredge
does not match the highest N/O PNe. Kingsburgh and Barlow (1994)
suggested the maximum N/O from first and second dredge-up is 0.8. They
reasoned that the first two dredge-ups can only convert carbon
initially present in the star and by converting all the carbon to
nitrogen they get 0.8. We suggest that a more realistic limit based on
our calculation of first and second dredge-ups is $\approx$0.6. Clearly,
several PNe have N/O larger than this, and the only other source is hot-bottom burning converting carbon to nitrogen. 

More questions revolve around the operation of the ON cycle. Some
investigators (Dufour 1991\nocite{d91}, Henry 1990\nocite{hen90},
Perinotto 1991\nocite{per91}) have argued that the ON cycle must occur
because they detected an anti-correlation between O/H and N/O. Other
investigators (Kingsburgh and Barlow 1994\nocite{kb94}) have argued
against it claiming the O/H-N/O anticorrelation is not
significant. Kingsburgh and Barlow also argue that there should be an
anticorrelation between O/H and He/H which they do not observe.

Our results suggest however that for Galactic PNe, that ON cycling can
occur and still bring into agreement the results of the many
investigators. 
\begin{enumerate}
\item{The predicted amount of ON cycling will only lead to a small
reduction of O/H by 0.1-0.2$dex$. Since this is comparable to the
errors, any existing anticorrelation between O/H and N/O should be
weak. This explains both the weak detections and the no detections.}
\item{Our comparison of models and data predict that the behavior of
He/H as a function of mass is more complicated than previously
assumed. In other words He/H and O/H are probably not correlated. A
more subtle test is to look for a correlation between He/H and O/H for
PNe with N/O$>$0.8 which our models do predict. This correlation maybe
detectable but it will be very weak.}
\end{enumerate}
Our models suggest that things that have been taken as evidence both
for and against ON cycling are not conclusive due to the weakness of
the correlations. On the other hand, the weakness of the correlations
means that the case for ON cycling is still open. 

Since our models fit on the plots of O/H versus N/O and O/H versus
He/H and the difference between model ZAMS O/H and PNe O/H is
0.1$dex$, we feel that this represents an upper limit to the O/H
depletion.

In contrast, the models of Renzini and Voli (1981) suggest that O/H
can be reduced by as much as 0.3$dex$ or a factor of 2 which should be
detectable with large samples. The difference is due to the time the
models undergo hot-bottom burning. In our models, which acheive higher
base temperatures than those of Renzini and Voli, the hot-bottom
burning epoch approximately lasts one-tenth as long as comparable
Renzini and Voli models. Therefore, their models burn more oxygen to
nitrogen. Since O/H reductions of this magnitude are not seen, we feel
our models better fit the observed data.

\section{Comparison to Other Models}
Our models mostly agree with the recent ones of Groenewegen and
deJong (1993, 1994) and Marigo \etal\ (1996). These two other models
indicate that the minimum mass model for hot-bottom burning is
approximately 4$\msun$, which agrees with our result. Our models
indicate PNe with low and intermediate mass progenitors will be C-rich
and N-rich, which agrees with their result. 

Our models give somewhat different results in some areas than those of
Groenewegen and deJong (1993, 1994) and Marigo (1996).
\begin{enumerate}
\item{Our intermediate mass star models reach do not reach as high N/O
ratios as either of these other models. This is due to our choice of
third dredge-up model. They use a constant dredge-up parameter,
$\lambda$, while ours is related to mass. For intermediate mass stars
they use a $\lambda$ which fits the LMC carbon star luminosity
function. However, the progenitors of carbon stars are probably
low-mass stars. A typical $\lambda$ for our intermediate mass stars is
0.4 while they use 0.6.}
\item{We do not reach as high C/O ratios as they do. Both Marigo and
Groenewegen and deJong achieve maximum C/O ratios greater than 5,
while ours are only 3. C/O calculated from the CIII]$\lambda$1909 line
is superior to that from the optical line (Rola and Stasinska
1994). Both groups however use C/O determinations using the optical
carbon line which gives misleading results. Therefore, we feel that we
have acheived a better fit to this important element.}
\end{enumerate}

\newpage
\chapter{On the Origin of Planetary Nebula K648 in Globular Cluster M15} 

\section{Introduction}
The globular cluster M15 contains the well studied planetary nebula
(PN) K648. This is one of the few galactic PNe with a reasonably
well-determined distance. Therefore, fundamental properties such as
the stellar luminosity can be determined with some confidence. Because
of its globular cluster membership, many of the progenitor properties,
such as the zero age main sequence (ZAMS) mass, can be inferred
reliably.

Due to the importance of K648 as a halo PN, it has been the focus
of several abundance studies, and all of these show the abundances of
most metals to be depleted relative to the sun, consistent with a
progenitor of low metallicity.  Carbon is an exception; studies which
determine the ratio (by number) of C/O in K648 infer values that range
from $4-11$ (Adams \etal\ 1984; Henry, Kwitter, and Howard 1996;
Howard, Henry, and McCartney 1997), which is far above C/O in the Sun
of 0.43 (\nocite{ag89}Anders and Grevesse 1989, hereafter
AG89). This is in fact much higher than the average C/O ratio of
$\approx 0.8$ for solar neighborhood PNe (\nocite{rs94}Rola and
Stasi{\'n}ska 1994).

Low and intermediate mass stars that have left the main sequence,
ascended the giant branch, and passed through the horizontal branch, 
then enter a thermally unstable phase where energy is generated by
shell He and H-burning called the thermally pulsing asymtoptic giant
branch (TP-AGB) stage, which is a very important yet not well understood
phase [Detailed reviews of this stage can be found in \nocite{i95} 
Iben (1995), \nocite{l93} Lattanzio (1993), and \nocite{ir83} Iben
and Renzini (1983)]. During the TP-AGB stage the star alternates
between a long stage where the luminosity is generated mostly by
quiescent hydrogen shell burning, with a helium burning layer
producing a minority of the energy, and a thermal runaway stage in the
unstable helium burning layer (\nocite{sh65} Schwarzschild and
H\"{a}rm 1965, 1967; and \nocite{w66} Weigert 1966).  The second
stage results in expansion of the outer layers and an extinguishing of
the H burning shell. This short stage, characterized by rapid changes,
with helium burning dominating the energy generation, is known as a
thermal pulse or a He shell flash.

TP-AGB stars exhibit large mass-loss rates ranging from
10$^{-7}-$10$^{-4}$ \msolyr\ . Indeed such high mass-loss rates
are predicted to result in the ejection of the envelope, at which
point the star leaves the AGB and becomes a planetary nebula central 
star (CSPN). The first models of CSPN tracks were made by
\nocite{p71} Paczynski (1971) who showed that the CSPNs evolve
horizontally on the HR diagram when nuclear burning is still taking
place and then as they cool the luminosity and temperature
decrease. \nocite{hs75} H\"arm and Schwarzschild (1975) showed that 
a CSPN could leave the AGB as either a helium burning or hydrogen
burning star. The observational consequences of hydrogen and helium
burning have been
studied in the more refined models
including mass loss showed that the subsequent evolution of the
central star depends on whether or not the star leaves the AGB as a
helium or hydrogen burner [Sch\"{o}nberner (1981, 1983) and
\nocite{i84} Iben (1984)]. 

Low mass stars (M$\lesssim$ 3 M$_{\odot}$) can experience two mixing
episodes or ``dredge-ups''. During dredge-up, material that has been
processed by nuclear burning is mixed into the surface layers. At the
entrance to the giant branch, the convective region can extend into
the core, leading to mixing of CNO products into the outer
layers. Similarly as shown by \nocite{i75} Iben (1975), after a
thermal pulse on the AGB, the convective region can extend into the
core, mixing He-burning products into the outer layers. These two
mixing events are known as first and third dredge up, respectively
(second dredge up will not concern us here). Therefore, a third
dredge-up is a natural explanation of the high carbon abundance found
in K648.  On the other hand, no carbon stars have been observed either
in M15 or in any other globular cluster, although such stars should be
the immediate progenitors of objects such as K648 if a third dredge-up
occurs. Thus, the lack of carbon stars in M15 weakens the argument for
a third dredge-up event.

One possible explanation for the absence of carbon stars is a delayed
scenario in which the third dredge-up of carbon rich material changes
the structure of the envelope during the following interpulse phase,
ultimately increasing the mass-loss rate significantly and driving off
the stellar envelope (Iben 1995). Thus envelope ejection is delayed
until the interpulse phase following this dredge-up of carbon rich
material. 

Another explanation supposes that the
envelope is removed during the quiescent He-burning stage that follows
a thermal pulse (\nocite{ren89}Renzini 1989 and
\nocite{rfp88}Renzini and Fuci-Pecci 1988).  The carbon then
originates in a fast wind from the central star (CSPN).  In addition,
the wind produces shock-heating in the nebula, which, if not properly
accounted for during an abundance analysis, may lead to the inference
of a spuriously high C/O ratio. In this case the envelope would be
ejected immediately after a thermal pulse while helium shell burning
still dominates the luminosity.  We refer to this mechanism as the
prompt scenario. 

In this paper we calculate detailed envelope models of thermally pulsing
asymptotic giant branch star envelopes to test the predictions of the
delayed mechanism, perform other calculations relevant to the prompt
mechanism, and compare output of each with observations of K648.
Section~2 describes the envelope code, section~3 presents the
observational data and the results for the delayed and prompt models,
and a brief discussion of our findings is given in section~4.

\section{Models}
The computer code used to calculate the delayed models is a
significantly updated and modified version of a program kindly
provided to us by \nocite{ren92} A.~Renzini for modeling
the envelope of  TP-AGB
stars during the interpulse phase.  Many of the basic details of the
method are enumerated in \nocite{it78} Iben and Truran (1978) and
\nocite{rv81} Renzini and Voli (1981) and references therein; in
this section we concentrate on those features which are different. In
a future paper (\nocite{buell97}Buell \etal\ 1997) we will provide
a more detailed description of the code. 

The mass of the hydrogen exhausted core ($M_H$) at the first thermal
pulse is given by the expression found in \nocite{latt86}Lattanzio
(1986).  During each interpulse phase the code follows the mass of the
hydrogen exhausted core and envelope, the evolution of envelope
abundances of $^4He$, $^{12}C$, $^{13}C$, $^{14}N$, and $^{16}O$, and
determines $T_{\rm eff}$ by integrating the equations of stellar
structure from the surface to the core.
Envelope abundances at the first pulse are determined by combining
published main sequence levels with changes due to the first
dredge-up. The former are established by scaling the AG89 solar
abundances of all metals except the alpha elements, i.e. oxygen, neon,
and magnesium, to the appropriate metallicity, and then setting
$[N_{\alpha}/Fe]=0.4$, where $N_{\alpha}$ is the number abundance of
O, Ne, and Mg. This last value is chosen from an examination of the
trends in the data of \nocite{edv93} Edvardsson \etal\ (1993) for
[Fe/H]$<$-1.0 and by assuming that neon and oxygen vary in lockstep in
PNe as shown by \nocite{hen89}Henry (1989). The abundance changes
due to the first dredge-up are calculated from the formulae of
\nocite{gwdj93}Groenewegen and deJong (1993). 

The mass-loss both before and during the TP-AGB phase is very
important, although the parameters are poorly understood. The
pre-TP-AGB mass-loss is a free parameter, while during the TP-AGB
phase, mass-loss is determined by using the expression of
\nocite{vw93}Vassiliadis and Wood (1993), which can be written as
\begin{eqnarray}
\log{\dot{M}=-11.43+1.0467\times
10^{-4}\left(\frac{R}{R_{\odot}}\right)^{1.94}
\left(\frac{M}{M_{\odot}}\right)^{-0.9}}& {\msolyr}.
\end{eqnarray}
The above rate is used until $\log{\dot{M}} = -4.5$, and then it is
held fixed. Equation 1 is a $\dot{M}\rm -Period$ relation based on
mass-loss from population I stars.  However, recent calculations by
\nocite{wbs96} Wilson, Bowen, and Struck (1996) suggest that the
mass-loss rates of low metallicity AGB stars are also strongly
dependent on radius. There is considerable uncertainty in this
equation.  For example, predicted mass-loss rates from other equations
with a similar form (e.g. Bazan 1991) differ from predictions of
eq.~(1) by up to a factor of five.

The luminosity of TP-AGB stars after the first few pulses can be
described by a linear relation between core-mass and luminosity as
first discovered by Paczynski (1970). Models of TP-AGB
stars have shown that for  $M\lesssim3.0\msun$ this
relation depends on metallicity (Lattanzio 1986, Hollowell and Iben
1988, Boothroyd and Sackmann 1988b). At the first pulse the luminosity
of TP-AGB stars is less than the asymptotic core-mass-luminosity
relation. The luminosity at the first pulse in our models is found by
linearly extrapolating in metallicity from the expressions found in
\nocite{bs88b} Boothroyd and Sackmann (1988b).  After the first
pulse, the luminosity of the AGB star rises steeply until it reaches a
value predicted by the core-mass luminosity relation (CML) of
\nocite{bs88b}Boothroyd and Sackmann (1988b).  This relation
predicts luminosity primarily as a function of core mass, although it
has a weak dependence on helium and metal mass fractions.
 

Carbon rich material can be dredged from the core into the envelope
following a thermal pulse. We assume that when the mass of the
hydrogen-exhausted core exceeds a minimum mass ($\rm M^{DU}_{min}$)
that a dredge-up occurs. The amount of material dredged up, $\Delta
\rm M_{dredge}$, is determined by the free parameter $\lambda$, where
\begin{equation}
\lambda=\frac{\Delta \rm M_{dredge}}{\Delta M_c}.
\end{equation}
In eq.~(2) $\Delta M_c$ is the amount of core advance during the
preceding interpulse phase. We determine the composition of the
dredged up material from the formulas in \nocite{rv81} Renzini and
Voli (1981), with $^4He\approx 0.75$, $^{12}C\approx 0.23$, and
$^{16}O\approx 0.01$ as the approximate mass fractions.

Finally, the code uses the opacities of \nocite{opal} Rogers and Iglesias
(1992) supplemented by the low temperature opacities of
\nocite{af94} Alexander and Ferguson (1994). 

\section{Results and Discussion}
\subsection{Observational Parameters}

\begin{table}
\begin{small}
\begin{tabular}{ccc}
Parameter &Value &ref.\\ 
T$_{\rm eff}$& $36000\pm 4000{\rm\ K}$ & 1,2\\
d& $10.0\pm 0.8{\rm\ kpc}$ & 3\\
$\theta$& $1.0-2.5 {\rm arcsec}$ &1,2,5\\
$v_{exp}$& $15-25{\rm\ km~s}^{-1}$& \\
$\rm n_e$& $1700-8000{\rm\ cm}^{-3}$ & 1,2\\
$\rm T_{e}$& 12000{\rm\ K}& 1,2\\
$\log{F_{H\beta}}$& $-12.10\pm 0.03$ & 4\\
\end{tabular}
\caption{Observational Data for K648:
This table is a summary of the observed and
inferred parameters for PN K648. The effective
temperature, T$_{\rm eff}$, refers to the central star, while the distance,
d, is the adopted distance to K648. The following nebular parameters
are also listed: the angular size of the nebula, $\theta$; the
expansion velocity, $v_{\rm exp}$; the electron density, $\rm n_e$;
the ionized gas temperature, $\rm T_{e}$; and the
log of the measured H$\beta$ flux in erg~cm$^{-2}$~s$^{-1}$. The large
range in $\theta$ and $n_e$ arise from differences between newer HST
data and ground based data, The HST data give higher a
value of $\theta$ and a lower  value for $n_e$. The references are as
follows: 
(1) Adams \etal\ 1984; (2) Bianchi \etal\ 1995; (3) Durell
and Harris 1993; (4) Acker \etal\ 1992; (5) Gathier \etal\ 1983
}
\label{obsparam}
\end{small}
\end{table}

Numerous observed and inferred parameters for K648 are listed in tables
\ref{obsparam} and \ref{paramcomp}, where the symbols in column~(1) are explained in the table
notes.  We comment here on the method of determination for several of
them. 

Radio images of K648 have been made by Gathier \etal\
(1983) \nocite{gpg83} and optical images were made by
\nocite{adams84} Adams \etal\ (1984), and recently by
\nocite{bianchi95} Bianchi \etal\ (1995) using the HST. The HST data
called into question the small size for the nebula inferred in the
radio studies of Gathier \etal\ (1983) \nocite{gpg83} and the
optical studies of Adams \etal\ (1984)\nocite{adams84}, since HST
was able to resolve the structure of the nebula. This leads to, e.g.,
a larger planetary mass and smaller electron density. In Tables
\ref{obsparam} and \ref{paramcomp}, we quote all results.

M$_{PN}$ was computed using equation V-7 in \nocite{pott84} Pottasch
(1984), while the dynamical age was estimated by dividing the nebular
radius by the expansion velocity ($v_{exp}$). Since no $v_{exp}$ is
available for K648 we use a range which represents typical values for
PNe. The central star mass for K648 was estimated by linearly
interpolating/extrapolating using both hydrogen burning and helium 
burning post-AGB tracks of \nocite{vw94} Vassiliadis and Wood (1994)
in the log~L-log~T plane.  The metallicity of M15 suggests using a low
Z track, although the carbon abundance of K648, if correct, would
increase the metallicity of the star, suggesting that a higher Z track
is more appropriate. Since the metallicity dependence is unclear, we
estimated the range of possible central star masses by performing the
interpolation for each metallicity considered by Vassiliadis and Wood.
Thus, the mass range of the central star is 0.55--0.58$M_{\odot}$ for
the H burning tracks and 0.56--0.61$M_{\odot}$ for the He-burning
tracks.  We adopt a final core mass of $0.58\pm0.03M_{\odot}$.

\begin{small}
\begin{table}
\begin{tabular}{ccccc}
Parameter &Observed Value &ref. &Delayed Scenario&Prompt Scenario\\ 
L&$3200-4700\ L_{\odot}$ & 1,2 & 4600 & 4000\\ $\rm M_{PN}$ &
$0.015-0.090{\rm M}_{\odot}$& 1,2&$0.048\pm 0.012{\rm
M}_{\odot}$&$0.064{\rm M}_{\odot}$ \\ 
$\rm M_c$ & $0.58 \pm 0.03$ & &$0.57 \pm 0.01$ & $0.58$ \\ 
$\tau_{dyn}$& $2000-8000{\rm\ yr}$& &12000{\rm\ yr}& 1800\ yr\\ 
He/H&$0.083-0.10$& 1,3,4 & $0.087-0.091$ & $0.9$ \\ 
C/O& $4-11$& 1,3,4& $4-25$& $4$ \\ N/O& $0.05-0.20$&1,3,4&
$0.17-0.19$& $0.17$\\
\end{tabular}
\caption{ Observational Data and Models Compared: This table compares
the observed and predicted parameters 
for PN K648. The observed luminosity, L, refers to the central star,
while the predicted luminosity is the luminosity on the AGB, but since
the tracks are nearly horizontal they should be comparable. The
following nebular parameters are also listed: the mass of ionized gas
in the nebula, M$_{PN}$; the mass of the central star, $M_c$; the
dynamic timescale, $\tau_{dyn}$; and the abundance ratios He/H, C/O,
and N/O by number. The abundances for the prompt scenario are
calculated assuming $0.00014\ {\rm M}_{\odot}$ of helium and carbon
rich material is removed by mass-loss from the CSPN. The observed
value for the dynamical timescale, $\tau_{dyn}$, corresponds to an
upper limit for the age of the nebula.  The theoretical values
correspond to evolutionary time scales required to reach a given
central star temperature. The large range in $L$ and $M_{\rm PN}$
arise from differences between HST data and ground based radio and
optical data.  The HST data give higher values for $L$ and $M_{\rm
PN}$.  References: (1) Adams \etal\ 1984; (2) Bianchi \etal\ 1995; (3)
Henry, Kwitter, and Howard 1996; (4) Howard, Henry, and McCartney
1997}
\label{paramcomp}
\end{table}
\end{small}

The theoretical age of the central star was estimated from the figures
of Vassiliadis and Wood and 
linearly interpolating in $\log{\rm L}$ between
tracks which closely match the core mass of K648, e.g., the hydrogen
burning M$_{c}$=0.56, Z=0.016 track and the helium burning
M$_{c}$=0.56 M$_{\odot}$, Z=0.004 give evolutionary ages of $\sim$
12000~yr and $\sim$1800~yr, respectively. Other tracks with M$_{\rm
c}\lesssim 0.6\msun$ and different metallicities give similar
results. When compared to the dynamical age a He burning track is
favored. 

The adopted abundances of K648 for He/H, C/O, and N/O ratios represent
a range of recent literature values.  Howard \etal\ (1997) find that
in six of the nine halo PNe they studied, the C/O ratio exceeds the
solar value.  Many of these nebulae have stellar temperatures much
higher than that of K648, implying that they are older and more
evolved. Since the high C/O ratios persist into the later stages of PN
evolution, this suggests that the inferred C/O is not influenced by
the presence of shock heating in the nebula.

The mass-loss rate at the tip of
the AGB was determined by dividing the nebular mass by the dynamical
age. This is in reality a lower limit since it assumes that the nebula
has a filling factor of 1, which is unrealistic. By this procedure we
calculate that the lower limit to the mass-loss is 9$\times$10$^{-6}$
\msolyr . The upper limit is assumed to be $10^{-4}$ \msolyr .

The composition of the central star is uncertain, as two recent papers
do not agree. McCarthey \etal (1996) find that the central star has a
normal helium abundance, whereas Heber \etal (1993) find that the central
star is helium and carbon rich.

\subsection{Delayed Scenario}

\begin{figure}
\centerline{\hbox{\psfig{figure=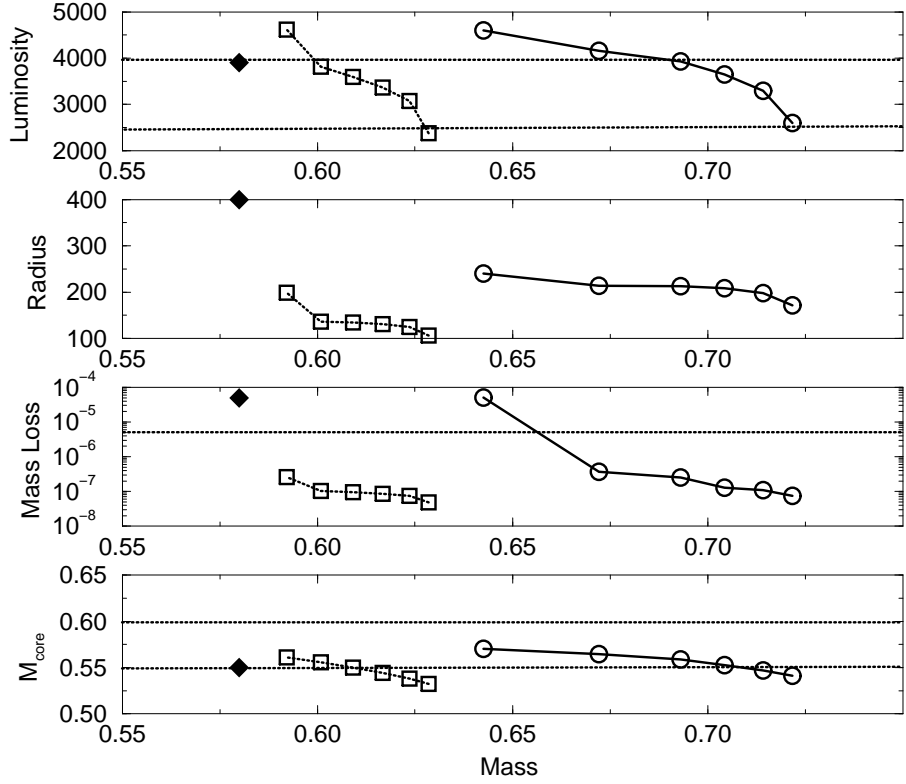,height=4truein}}}
\caption{Shown in the panels of this figure are the
evolution of our thermally pulsing AGB models and the parameters of
our prompt model. The dotted line and open squares track model 1,
the solid line and open circles track model 2, and the solid diamonds
are the parameters of the prompt model. The abscissa of the graphs 
tracks the mass of the models in solar masses. Due to  mass loss the
stars move from right to left on the graphs. The parameters in the
panels are for the interpulse phase. From top to bottom the parameters
are stellar luminosity (in L$_{\odot}$), radius (in $\rm R_{\odot}$),
the mass-loss rate (in \msolyr), and the mass of the core. The
observed upper and lower limits of the AGB tip luminosity are
indicated with dark long dashed lines, the lower limit on the AGB tip
mass-loss rate is indicated with a long dashed line, and the upper and
lower limits of the central star mass are indicated with long dashed
lines.}
\label{fig:k648modx.eps}
\end{figure}

\begin{figure}
\centerline{\hbox{\psfig{figure=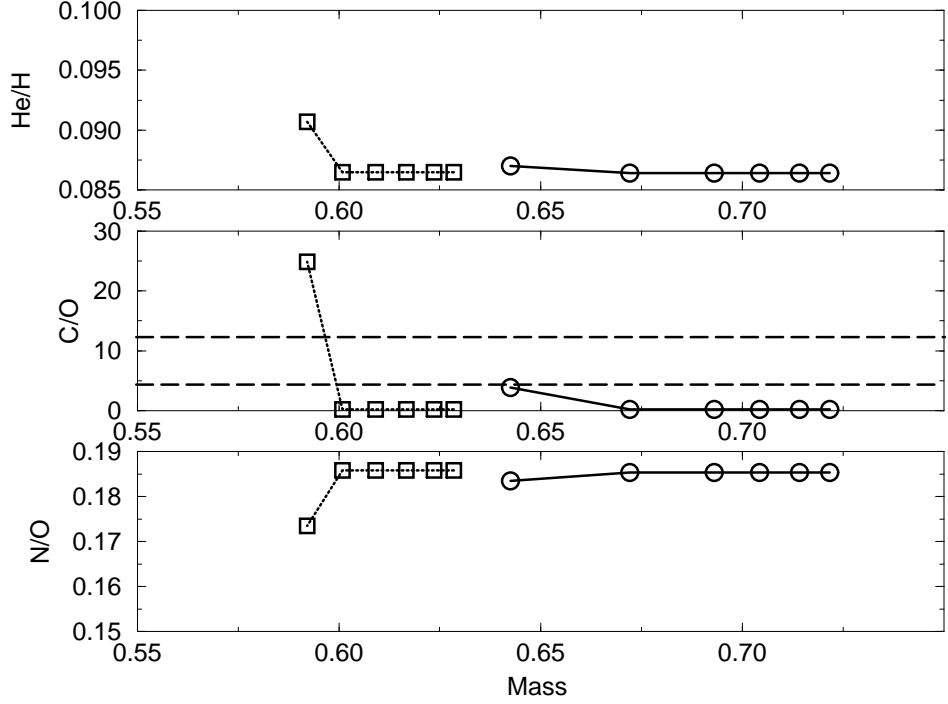,height=4truein}}}
\caption{Shown in the panels of this figure is the
evolution of the surface abundance ratios. The symbols have the same
meaning as the first figure. In the C/O panel the upper
and lower observational limits are shown on the figure with the dark
long dashed line. The range of possible He/H and N/O is encompassed by
the ordinates of these figures.}
\label{fig:k648abun.eps}
\end{figure}

We have calculated several low mass, low metallicity models, but here
we focus on the two models listed in table \ref{modpara}, where we present the
model input parameters: the ZAMS mass (M), the core mass at PN
ejection (M$_c$), the mass of the PN ($M_{PN}$), the ZAMS [Fe/H]
ratio, the adopted ratio of the mixing length to pressure scale height
($\alpha$), the mass of the model star at the first thermal pulse
(M$_{FTP}$), the adopted dredge-up parameter ($\lambda$), and the
minimum core mass for dredge-up (M$_{c,min}^{DU}$).  The panels of
Figure \ref{fig:k648modx.eps} show the evolution of the interpulse
luminosity, the stellar radius, the mass-loss rate, and the core mass
as a function of total mass. All quantities are expressed in solar
units. Figure \ref{fig:k648abun.eps} shows the evolution of the
chemical composition of the envelope as a function of total
mass. Table~\ref{paramcomp} compares the observed quantities to our
predicted ones. 

We note in figure~\ref{fig:k648modx.eps} that the interpulse radius of
each model star increases dramatically after the final pulse, as
compared to the preceding interpulse phase. The increase in radius
leads to a large increase in the mass-loss rate in each model during
the final interpulse phase; the mass-loss rate increases by almost a
factor of 100 in model 2 and by a factor of 5 in model 1. This is a
consequence of the steep dependence of our mass-loss law on the
stellar radius. The significant increase in the mass-loss rate causes
the star to lose its envelope in a few thousand years. The mass-loss
rate for model~1 is clearly too small relative to the observationally
derived value.  However, we have found that by reducing the mixing
length ($\alpha$) by a factor of two, as we have done in model 2, we
can make a model that essentially reproduces the observed AGB tip mass
loss rate.

The significant event that occurs during the final pulse is a
dredge-up of helium and carbon rich material. The mass of material
dredged up is a few times $10^{-5}M_{\odot}$. However, given the mass
of the envelope and the low initial abundances, the amount of carbon
dredged into the envelope is significant enough to increase the carbon
mass fraction by a large factor in each case.  Consequently, the
envelope opacity rises, causing a dramatic increase in the stellar
radius.

\begin{table}
\begin{tabular}{ccccccccc}
No. &M &${\rm M}_c$ &${\rm M}_{PN}$ &[Fe/H]&$\alpha$ &${\rm M}_{\rm
FTP}$&$\lambda$&${\rm M}^{\rm DU}_{\rm c,min}$\\ 
1& 0.88& 0.56& 0.037& -2.1& 1.6& 0.62& 0.10& 0.55 \\ 
2& 0.85& 0.58& 0.060& -2.1& 0.8& 0.72& 0.02& 0.56 \\
\end{tabular}
\caption{Input Parameters and Results for Delayed Models: Note, the
masses are in M$_{\odot}$}
\label{modpara}
\end{table}

The envelope of each model at the last thermal pulse is only a few
times $10^{-2} M_{\odot}$ and, after the final carbon dredging pulse,
is ejected on a timescale of a few hundred years. Each model star is a
carbon star for only a few hundred years, due to the rapid mass-loss
after a dredge-up of carbon. 
This short lifetime, coupled with the relatively low incidence of PNe
in globular clusters [two confirmed and three possible candidates
(Jacoby \etal\ 1995)], perhaps explains why carbon stars have not been
observed in globular clusters.


An important check on our models is to compare the predicted AGB tip
luminosity with its observed value. The predicted luminosity of our
models at the top of the AGB agrees fairly well with the tip of M15's
red giant branch (Adams \etal\ 1984\nocite{adams84}). Our models
suggest that the observed AGB tip will actually correspond to the
second-to-last pulse, since after the dredge-up event the star is
predicted to remain as an AGB star for only $\sim$1000~yr. The
luminosity of K648 in Adams \etal\ (1984) appears to be 0.1dex higher
than the tip of the giant branch, this may be due to the metallicity
enhancement due to the dredge up. As noted earlier, the core-mass
luminosity relationship depends on metallicity, with the luminosity at
a set core-mass increasing with increasing metallicity. If we lower
the luminosity still further to $\sim$2000\lsun\ to match the tip of
the giant branch, we believe that the addition of carbon 
to the envelope will still cause envelope ejection.

There is some question about whether or not dredge-up can occur at the
low values of M$_{c,min}^{DU}$ indicated by our models (see
table~\ref{modpara}). While \nocite{latt89}Lattanzio (1989) found that
dredge-up can occur at a core mass above $0.605M_{\odot}$, the same
study also found a dependence of the minimum dredge-up mass on
metallicity, with lower metallicities giving lower mass
dredge-ups. \nocite{bs88c} Boothroyd and Sackmann (1988c) found that
if they increased the mixing length parameter $\alpha$ from 1 to 3,
they were able to cause a dredge-up in a model with Z=0.001, $\rm
M_c=0.566M_{\odot}$, and $\rm M=0.81M_{\odot}$, although it is unclear
if a mixing length this large is justified. Additionally, to match the
low luminosity end of the carbon star luminosity function of the LMC,
Groenewegen and deJong (1993) had to set $\rm
M^{DU}_{min}=0.58M_{\odot}$. From these studies it appears that our
values for M$_{c,min}^{DU}$ are not unreasonable.

Finally, we point out that each of the delayed models gives a very
natural explanation of the high carbon abundance of K648 and the lack
of carbon stars. Each model also predicts the observed mass of the
ionized gas to be a few times $10^{-2}M_{\odot}$. The C/O ratios of
each model range from 4 to 25, with model 2 giving the best fit, which
agrees reasonably well with the observed values of $4-11$. The He/H
ratios of the model stars also agree with the observed value of
0.09. The high N/O ratio inferred in the models may be an artifact of
our choice of initial O abundance and hence could be reduced with a
higher O abundance, which would also slightly reduce the C/O ratio.
Thus, our delayed models are consistent with several important
observed properties of the K648 system.

\subsection{Prompt Scenario}

An alternative scenario results if we apply our mass-loss formulation
to the secondary luminosity peak (SLP) which follows the helium shell
flash of the $\rm 1M_{\odot}$, Z=0.001 model of \nocite{bs88a}
Boothroyd and Sackmann (1988a, BS88a). The metallicity of the BS88a
model is a factor of $\sim$5 higher than M15, however, no models of
the appropriate metallicity exist and we attempted to use the closest
one in terms of Z, M$_{\rm c}$, and M. The SLP corresponds to the
region between point C and the vertical dashed line on figure 2
of Boothroyd and Sackmann, i.e. the same place
that Renzini (1989) and Renzini and Fuci-Pecci (1988) predict this
event to occur when the star expands. The SLP occurs when the excess
luminosity produced in the helium shell flash reaches the
surface. This peak can be seen in most models of low mass AGB stars
[Iben(1982), BS88a, VW93].

It should be noted that this is the
point where dredge-up can occur, although it does not necessarily
do so. This scenario does not require a dredge-up of carbon rich
material for envelope ejection. We define the prompt scenario as
ejection at the SLP without the dredge-up of carbon rich material.

\begin{table}
\begin{tabular}{cc}
Parameter&Value\\
Luminosity& $4000\ {\rm L}_{\odot}$\\
Radius&  $400\ {\rm R}_{\odot}$\\
Mass& $0.58$\\
Core Mass& $0.54\ {\rm M}_{\odot} $\\
Mass-Loss Rate& $3.2\times 10^{-5}\,\msolyr$\\
Time in Stage& 2000 yr\\
\end{tabular}
\caption{Adopted Prompt PN Ejection Parameters: 
The values in this table are estimated from figure~2 of
BS88a for a 1.0 M$_{\odot}$, Z=0.001 model. All values are appropriate
between point C and the vertical dashed line on this figure. The
radius and luminosity are the estimated lower limits. The mass and
core mass are taken from their listed values. The mass-loss rate is
calculated from our mass-loss prescription. The time in this stage is
estimated from the BS88a graph.}
\label{prompt}
\end{table}

The adopted parameters of this model are shown in Table
\ref{prompt}. The luminosity and radius are ``eyeballed'' lower limits
from the SLP of BS88a, while the mass and core mass are parameters
stated in their text. The mass-loss rate calculated from our
prescription [i.e. eq.~(1)] is $\sim 10^{-5}\msolyr$ (essentially the
Eddington limit), which will remove the $\rm 0.03M_{\odot}$ envelope
in a few thousand years. This model is similar to K648 in terms of
core mass and envelope mass.  Values for luminosity, radius, mass-loss
rate, and core mass for the prompt scenario are indicated with filled
diamonds in Fig.~\ref{fig:k648modx.eps} and observed quantities are
also compared to those predicted for this scenario in
Table~\ref{paramcomp}.

A carbon-rich nebula could be formed by the prompt mechanism if a
sufficient amount of helium and carbon-rich material is ejected during
the post-AGB phase and mixed with the ejected hydrogen-rich
envelope. The fast wind overtaking the slower wind will produce a
shock which would likely be Rayleigh-Taylor unstable, causing the
nebula to mix. Only $5-15\times$10$^{-5}$ M$_{\odot}$ of material with
mass fractions of $^{4}He$=0.75 and $^{12}$C=0.23 needs to be mixed
into the envelope to match the C/O ratio of K648. Carbon-rich material
can be ejected into the nebula during the post AGB phase. As a star
moves horizontally across the HR diagram from the AGB stage to CSPN
position, the mass-loss rate will decrease as the wind speed
increases, so the material ejected during this transition can be mixed
with the slower hydrogen rich envelope. And since the currently
observed mass-loss rate of the K648 central star is
$10^{-9}-10^{-10}$~\msolyr\ (Adams \etal\ 1984; Bianchi \etal\ 1995),
the nebula is no longer being polluted. Examination of the models of
Vassiliadis and Wood (1994) suggests that as the star moves from the
AGB phase to the CSPN phase, the mass loss rate drops from
10$^{-5}$~\msolyr\ to $\sim10^{-10}$~\msolyr, indicating that during
this transition the mass loss was higher in the past, and possibly
high enough to account for the carbon enrichments in K648.

In the prompt scenario, the envelope is ejected when the star is
burning helium and as a result the resulting CSPN will
be follow a helium burning track (Sch\"{o}nberner 1981, 1983;
Iben 1984).



Thus, in the prompt scenario, the evolved star ejects sufficient
carbon into a slower moving hydrogen-rich shell to produce the PN we
observe today.  Mixing is assumed to occur due to shock induced
instabilities. Since the prompt scenario postulates the removal of the
entire H-rich envelope during the He burning stage, we expect K648 to
follow a He burning track because the H-burning shell has been
extinguished during the thermal pulse. Ultimately, a white
dwarf of type DB will be produced. 

\section{Discussion}
One additional scenario is again a delayed one, but one in which the
CSPN is a helium burner. We have not as yet performed calculations
relevant to it. In this case, if a dredge-up occurs, it does so at the
SLP. The stellar envelope will be enriched in carbon and the added
opacity may allow an even greater expansion during the SLP, making it
more likely that the envelope will be ejected during this phase. The
resulting PN would be carbon rich and have a helium burning CSPN. We
feel that this is also a promising model, although, proper calculations
of this scenario need to be done.

Both the prompt and delayed scenarios can be made to match many of the
observed features of K648. With each mechanism the radius increases
dramatically: in the prompt because of the increase in luminosity of
the star after a thermal pulse and in the delayed because of an
increase in the opacity due to an infusion of carbon rich material.
In addition, both mechanisms produce $^{12}$C in sufficient amounts to
explain the observed C/O ratio.
 
The most serious difficulty with the prompt scenario is that it
can only explain the enhancement of the carbon and helium abundances
by essentially adhoc means, in this case assuming the central star
wind pollutes the rest of the nebula or by shocks and carbon-rich
pockets due to this wind. This may not be an unreasonable assumption,
since the mass of K648 is low compared to a ``typical'' PN
($\sim$0.1M$_{\odot}$). To test the prompt scenario would require a
detailed model following the star from the horizontal branch to the
central star phase with attention to the details of mass-loss to see
if the central star wind can truly enhance the carbon and helium
abundances of the PN plus multidimensional hydrodynamics to test the mixing
hypothesis. 

The difficulty with the delayed scenario is it predicts that the
CSPN should be a H-burner. The dynamical age of K648 favors a
He-burning CSPN which is more likely to occur in the prompt 
scenario as the envelope is ejected during a phase when helium burning
is dominant. Since we assume that for a given metallicity only one of
these scenarios will be operative, a strong observational test to
determine the correct scenario would be to search for white dwarfs in
M15. If they are found to be type DB, this would favor the prompt
scenario, and if they are type DA, the delayed scenario is more
likely. 

A point favoring the prompt scenario is that it naturally accounts for the
dynamical age. On the other hand, this scenario requires the
assumption of efficient mixing and there is some evidence
(cf. section~3.1) that signatures of the requisite shocks are not
actually observed. However, until detailed models are produced, both
remain as viable evolutionary scenarios for K648 and similar systems.

\newpage
\chapter{Conclusions}

The goals of this thesis were to study the origin of the abundance
patterns seen in PNe. We have shown that the abundances in PNe are the
result of both initial composition and nucleosynthesis. The
major conclusions of this work are as follows:
\begin{enumerate}
\item{Planetary Nebulae must be represented by a parameter space of at
least two dimensions in chemical space. One axis represents
metallicity axis and the other nucleosynthesis. Also, the bulge PNe
have chemical compositions that are indistinguishable from disk
PNe. Some of the apparent bulge PNe are of type I.}
\item{We have created a surface luminosity function based on stellar
mass, the mass of the hydrogen exhausted core, metallicity, and the
thermal pulse number. Using this function our synthetic TP-AGB models
closely approximate the behavior of more realistic TP-AGB models.}
\item{Our synthetic TP-AGB models incorporate the most realistic
available parameters and physics and produce results that match the
expected behavior. Low (M$\le$4$\msun$) and intermediate
(M$>$4$\msun$) mass stars respectively produce PNe rich in carbon and
nitrogen as expected. Intermediate mass models encounter hot-bottom
burning, preventing them from becoming carbon stars.}
\item{Helium is produced in low and intermediate mass stars via the
third and second dredge-ups, respectively.}
\item{There is good agreement between our synthetic TP-AGB models and
PN abundances. On each abundance ratio-abundance ratio plot we have
found generally good agreement.}
\item{In nitrogen-rich PNe (type I), we found that nitrogen is
produced via hot-bottom burning, a process which depletes
carbon. Helium is produced via the second dredge-up. Hot-bottom
burning shortens the TP-AGB lifetime, restricting the amount of
nitrogen that can be produced.}
\item{In carbon-rich PNe, we found that both carbon and helium are
produced via the third dredge-up.}
\item{The evidence for ON cycling is inconclusive. We have found that
the predicted amount of oxygen depletion via ON cycling is small and
difficult to detect observationally.}
\item{When we infer the progenitor masses of C-rich
(M$\gtrsim$1.7 $\msun$) and N-rich (M$\gtrsim$4.0 $\msun$) PNe, we find
that they lie in a narrow metallicity range ($\approx$0.2$dex$). }
\item{We have examined possible models for the formation of halo PNe
K648. To the best of our knowledge, we are the first group to actually
attempt to model the processes involved. We found three possible
models: 
\begin{itemize}
\item{A delayed model where carbon dredged into the envelope increases
the opacity, causing the envelope to expand and the mass-loss
rate to increase to levels where the envelope is rapidly ($\sim$1000
years) ejected.}
\item{A prompt model where the surface luminosity increase at the end
of the thermal pulse causes the envelope to expand, causing
accelerated mass-loss.}
\item{A combination model where the carbon dredged up and the
increased luminosity at the end of the thermal pulse cause envelope
expansion, also causing accelerated mass-loss.}
\end{itemize}}
\end{enumerate}

\section{Future Work}
There is other observational evidence to consider. The only halo PN
considered in detail was PN K648; we hope model other halo PNe. Also,
models will be made of the Magellanic Cloud PNe. The progenitors of
Magellanic Cloud PNe have a similar range as Galactic PNe, however,
the initial abundances are different. By comparing models of
Magellanic Cloud PNe to the abundance data, we can test our models as
a function of metallicity.

\newpage
\appendix
\chapter{Model Results}
\label{modresapp}

This appendix contains tables with the predicted PN abundance ratios
as a function of mass. The free parameters are the abundance of iron
relative to the solar abundance ([Fe/H]) and the mixing length
parameter ($\alpha$). The first column of each table indicates the
ZAMS mass of each model in solar units. The second column indicates
the white dwarf mass in solar units.

\begin{small}
\begin{table}
\begin{center}
\begin{tabular}{llllll}
Mass&M$_{\rm WD}$&O/H&He/H&N/O&C/O\\
  1.00&   0.555&  8.94&  0.112&  0.244&  0.329\\
  1.10&   0.568&  8.94&  0.111&  0.246&  0.327\\
  1.20&   0.575&  8.94&  0.110&  0.248&  0.324\\
  1.30&   0.588&  8.94&  0.109&  0.250&  0.322\\
  1.40&   0.602&  8.93&  0.108&  0.252&  0.320\\
  1.50&   0.617&  8.93&  0.107&  0.255&  0.318\\
  1.60&   0.626&  8.93&  0.106&  0.256&  0.316\\
  1.70&   0.640&  8.93&  0.105&  0.259&  0.388\\
  1.80&   0.648&  8.93&  0.107&  0.261&  0.634\\
  1.90&   0.654&  8.93&  0.108&  0.262&  0.916\\
  2.00&   0.659&  8.93&  0.111&  0.264&  1.352\\
  2.10&   0.666&  8.93&  0.114&  0.265&  1.816\\
  2.20&   0.676&  8.93&  0.117&  0.267&  2.249\\
  2.30&   0.680&  8.93&  0.116&  0.269&  2.173\\
  2.40&   0.690&  8.93&  0.118&  0.271&  2.399\\
  2.50&   0.699&  8.93&  0.118&  0.274&  2.432\\
  2.60&   0.710&  8.93&  0.118&  0.275&  2.559\\
  2.80&   0.726&  8.93&  0.117&  0.279&  2.523\\
  2.90&   0.735&  8.93&  0.116&  0.281&  2.500\\
  3.10&   0.752&  8.93&  0.116&  0.283&  2.500\\
  3.30&   0.773&  8.93&  0.116&  0.283&  2.588\\
  3.40&   0.782&  8.93&  0.116&  0.283&  2.624\\
  3.50&   0.786&  8.93&  0.116&  0.283&  2.570\\
  3.60&   0.805&  8.93&  0.114&  0.283&  2.388\\
  3.70&   0.822&  8.93&  0.111&  0.283&  2.032\\
  3.80&   0.844&  8.93&  0.110&  0.283&  1.892\\
  3.90&   0.869&  8.93&  0.109&  0.283&  1.778\\
  4.00&   0.885&  8.93&  0.107&  0.284&  1.603\\
  4.10&   0.854&  8.92&  0.104&  0.889&  0.519\\
  4.20&   0.860&  8.91&  0.102&  1.007&  0.245\\
  4.30&   0.873&  8.90&  0.101&  0.920&  0.247\\
  4.40&   0.891&  8.90&  0.100&  0.785&  0.219\\
  4.50&   0.900&  8.89&  0.102&  0.804&  0.284\\
  4.60&   0.906&  8.89&  0.104&  0.853&  0.229\\
  4.70&   0.913&  8.88&  0.106&  0.899&  0.247\\
  4.80&   0.919&  8.88&  0.108&  0.942&  0.249\\
  4.90&   0.927&  8.87&  0.111&  0.991&  0.324\\
  5.00&   0.933&  8.87&  0.112&  1.035&  0.286\\
  6.00&   0.991&  8.81&  0.130&  1.560&  0.412\\
  7.00&   1.037&  8.79&  0.142&  1.778&  0.498\\
  8.00&   1.079&  8.80&  0.150&  1.660&  0.459\\
\end{tabular}
\end{center}
\caption{[Fe/H]=0.0 $\alpha$=2.3 Galaxy}
\end{table}

\begin{table}
\begin{center}
\begin{tabular}{llllll}
Mass&M$_{\rm WD}$&O/H&He/H&N/O&C/O\\
  1.00&   0.553&  9.05&  0.117&  0.244&  0.329\\
  1.10&   0.554&  9.05&  0.116&  0.246&  0.327\\
  1.20&   0.567&  9.05&  0.115&  0.248&  0.324\\
  1.30&   0.574&  9.04&  0.114&  0.250&  0.322\\
  1.40&   0.587&  9.04&  0.113&  0.252&  0.320\\
  1.50&   0.601&  9.04&  0.112&  0.255&  0.318\\
  1.60&   0.610&  9.04&  0.111&  0.256&  0.316\\
  1.70&   0.625&  9.04&  0.110&  0.259&  0.313\\
  1.80&   0.634&  9.04&  0.110&  0.261&  0.353\\
  1.90&   0.645&  9.04&  0.111&  0.262&  0.540\\
  2.00&   0.652&  9.04&  0.112&  0.265&  0.729\\
  2.20&   0.668&  9.04&  0.118&  0.269&  1.416\\
  2.40&   0.686&  9.04&  0.122&  0.272&  1.875\\
  2.60&   0.701&  9.04&  0.122&  0.276&  1.928\\
  2.80&   0.721&  9.04&  0.122&  0.280&  2.023\\
  3.00&   0.737&  9.04&  0.121&  0.284&  1.991\\
  3.20&   0.757&  9.03&  0.121&  0.284&  2.046\\
  3.40&   0.772&  9.03&  0.121&  0.284&  2.089\\
  3.60&   0.785&  9.03&  0.121&  0.284&  2.084\\
  3.80&   0.805&  9.03&  0.123&  0.284&  2.239\\
  3.90&   0.816&  9.03&  0.124&  0.283&  2.360\\
  4.00&   0.821&  9.04&  0.122&  0.283&  2.218\\
  4.02&   0.825&  9.03&  0.122&  0.283&  2.158\\
  4.04&   0.824&  9.03&  0.121&  0.284&  2.061\\
  4.06&   0.830&  9.03&  0.121&  0.284&  2.046\\
  4.08&   0.832&  9.03&  0.121&  0.284&  2.046\\
  4.10&   0.831&  9.03&  0.119&  0.284&  1.950\\
  4.20&   0.851&  9.03&  0.119&  0.284&  1.849\\
  4.30&   0.883&  9.04&  0.118&  0.284&  1.754\\
  4.50&   0.886&  9.03&  0.114&  0.802&  0.843\\
  4.60&   0.864&  9.02&  0.110&  1.099&  0.260\\
  4.70&   0.871&  9.01&  0.108&  0.995&  0.248\\
  4.80&   0.886&  9.00&  0.106&  0.873&  0.221\\
  4.90&   0.895&  9.00&  0.108&  0.863&  0.249\\
  5.00&   0.901&  9.00&  0.110&  0.895&  0.239\\
  6.00&   0.961&  8.96&  0.128&  1.211&  0.352\\
  7.00&   1.010&  8.95&  0.140&  1.256&  0.359\\
  8.00&   1.062&  8.96&  0.149&  1.148&  0.321\\
\end{tabular}
\end{center}
\caption{[Fe/H]=0.1 $\alpha$=2.3 Galaxy}
\end{table}

\begin{table}
\begin{center}
\begin{tabular}{llllll}
Mass&M$_{\rm WD}$&O/H&He/H&N/O&C/O\\
  1.00&   0.543&  9.16&  0.124&  0.244&  0.329\\
  1.20&   0.557&  9.16&  0.122&  0.248&  0.324\\
  1.40&   0.578&  9.15&  0.120&  0.252&  0.320\\
  1.60&   0.599&  9.15&  0.118&  0.256&  0.316\\
  1.80&   0.630&  9.15&  0.116&  0.261&  0.311\\
  2.00&   0.647&  9.15&  0.117&  0.265&  0.532\\
  2.20&   0.666&  9.15&  0.125&  0.269&  1.156\\
  2.40&   0.677&  9.15&  0.127&  0.272&  1.340\\
  3.00&   0.735&  9.15&  0.130&  0.284&  1.702\\
  3.10&   0.740&  9.14&  0.128&  0.284&  1.629\\
  3.20&   0.750&  9.14&  0.128&  0.284&  1.648\\
  3.30&   0.761&  9.14&  0.129&  0.284&  1.722\\
  3.40&   0.765&  9.14&  0.128&  0.284&  1.694\\
  3.50&   0.775&  9.14&  0.129&  0.284&  1.730\\
  3.60&   0.785&  9.14&  0.130&  0.284&  1.795\\
  3.70&   0.790&  9.14&  0.130&  0.284&  1.816\\
  3.80&   0.800&  9.15&  0.131&  0.284&  1.879\\
  3.90&   0.811&  9.15&  0.132&  0.284&  1.972\\
  4.00&   0.815&  9.15&  0.132&  0.284&  1.995\\
  4.10&   0.826&  9.15&  0.133&  0.284&  2.056\\
  4.20&   0.838&  9.15&  0.134&  0.283&  2.163\\
  4.30&   0.844&  9.15&  0.136&  0.284&  2.275\\
  4.40&   0.858&  9.15&  0.138&  0.285&  2.415\\
  4.50&   0.863&  9.15&  0.139&  0.304&  2.455\\
  4.60&   0.879&  9.15&  0.140&  0.348&  2.547\\
  4.70&   0.883&  9.15&  0.140&  0.498&  2.355\\
  4.80&   0.868&  9.15&  0.134&  1.151&  1.300\\
  4.90&   0.910&  9.15&  0.134&  0.798&  1.626\\
  5.00&   0.934&  9.15&  0.133&  0.824&  1.503\\
  6.00&   0.926&  9.08&  0.124&  1.140&  0.238\\
  7.00&   0.979&  9.08&  0.136&  1.005&  0.289\\
  8.00&   1.044&  9.09&  0.146&  0.925&  0.235\\
\end{tabular}
\end{center}
\caption{[Fe/H]=0.2 $\alpha$=2.3 Galaxy}
\end{table}

\begin{table}
\begin{center}
\begin{tabular}{llllll}
Mass&M$_{\rm WD}$&O/H&He/H&N/O&C/O\\
  1.00&   0.563&  8.83&  0.107&  0.244&  0.329\\
  1.10&   0.575&  8.83&  0.106&  0.246&  0.327\\
  1.20&   0.583&  8.83&  0.105&  0.248&  0.324\\
  1.30&   0.596&  8.83&  0.105&  0.250&  0.322\\
  1.40&   0.611&  8.83&  0.104&  0.252&  0.320\\
  1.50&   0.626&  8.83&  0.103&  0.254&  0.318\\
  1.60&   0.634&  8.83&  0.102&  0.256&  0.426\\
  1.70&   0.645&  8.83&  0.104&  0.258&  0.824\\
  1.80&   0.651&  8.82&  0.106&  0.260&  1.233\\
  1.90&   0.657&  8.82&  0.107&  0.262&  1.581\\
  3.30&   0.782&  8.82&  0.112&  0.282&  3.177\\
  3.50&   0.826&  8.82&  0.107&  0.282&  2.415\\
  3.90&   0.857&  8.81&  0.097&  1.081&  0.291\\
  4.10&   0.891&  8.79&  0.096&  0.871&  0.321\\
  4.50&   0.926&  8.76&  0.104&  1.014&  0.358\\
  5.00&   0.960&  8.72&  0.114&  1.365&  0.502\\
  7.00&   1.058&  8.61&  0.142&  2.729&  0.622\\
  8.00&   1.094&  8.62&  0.149&  2.698&  0.547\\
\end{tabular}
\end{center}
\caption{[Fe/H]=-0.1 $\alpha$=2.3 Galaxy}
\end{table}

\begin{table}
\begin{center}
\begin{tabular}{llllll}
Mass&M$_{\rm WD}$&O/H&He/H&N/O&C/O\\
  1.00&   0.569&  8.73&  0.104&  0.244&  0.329\\
  1.10&   0.582&  8.73&  0.103&  0.246&  0.327\\
  1.20&   0.589&  8.73&  0.102&  0.248&  0.324\\
  1.30&   0.604&  8.72&  0.101&  0.250&  0.322\\
  1.40&   0.618&  8.72&  0.100&  0.252&  0.320\\
  1.50&   0.634&  8.72&  0.099&  0.255&  0.318\\
  1.60&   0.640&  8.72&  0.101&  0.256&  0.710\\
  1.70&   0.648&  8.72&  0.102&  0.258&  1.268\\
  1.80&   0.653&  8.72&  0.104&  0.259&  1.778\\
  1.90&   0.659&  8.72&  0.106&  0.261&  2.188\\
  2.00&   0.664&  8.72&  0.108&  0.262&  2.748\\
\end{tabular}
\end{center}
\caption{[Fe/H]=-0.2 $\alpha$=2.3 Galaxy}
\end{table}

\begin{table}
\begin{center}
\begin{tabular}{llllll}
Mass&M$_{\rm WD}$&O/H&He/H&N/O&C/O\\
  1.00&   0.594&  8.62&  0.101&  0.209&  0.283\\
  1.20&   0.610&  8.62&  0.099&  0.213&  0.279\\
  1.40&   0.633&  8.62&  0.097&  0.217&  0.392\\
  1.60&   0.650&  8.62&  0.100&  0.220&  1.563\\
  1.80&   0.657&  8.62&  0.104&  0.222&  2.904\\
  2.00&   0.670&  8.62&  0.109&  0.224&  4.581\\
  2.20&   0.688&  8.62&  0.111&  0.227&  5.284\\
  2.40&   0.705&  8.62&  0.110&  0.230&  5.433\\
  3.50&   0.856&  8.61&  0.093&  1.245&  1.074\\
\end{tabular}
\end{center}
\caption{[Fe/H]=-0.5 $\alpha$=2.3 Galaxy}
\end{table}

\begin{table}
\begin{center}
\begin{tabular}{llllll}
Mass&M$_{\rm WD}$&O/H&He/H&N/O&C/O\\
  1.00&   0.615&  8.31&  0.096&  0.188&  0.254\\
  1.20&   0.638&  8.31&  0.095&  0.191&  0.685\\
  1.40&   0.646&  8.31&  0.097&  0.193&  2.270\\
  1.60&   0.651&  8.31&  0.100&  0.195&  4.808\\
  1.80&   0.658&  8.31&  0.101&  0.196&  6.252\\
  2.00&   0.668&  8.32&  0.103&  0.198&  8.054\\
  2.20&   0.683&  8.32&  0.103&  0.200&  8.730\\
  2.40&   0.715&  8.32&  0.108&  0.201& 11.429\\
  3.00&   0.811&  8.32&  0.101&  0.210&  8.790\\
  3.50&   0.899&  8.27&  0.089&  3.412&  0.908\\
  3.60&   0.927&  8.18&  0.087&  3.614&  1.007\\
  3.70&   0.954&  8.09&  0.086&  3.917&  1.324\\
  3.80&   0.966&  8.05&  0.088&  4.375&  1.496\\
\end{tabular}
\end{center}
\caption{[Fe/H]=-1.0 $\alpha$=2.3 Galaxy}
\end{table}

\begin{table}
\begin{center}
\begin{tabular}{llllll}
Mass&M$_{\rm WD}$&O/H&He/H&N/O&C/O\\
  1.00&   0.549&  8.94&  0.112&  0.244&  0.329\\
  1.50&   0.591&  8.93&  0.107&  0.255&  0.318\\
  2.00&   0.645&  8.93&  0.104&  0.265&  0.512\\
  2.50&   0.672&  8.93&  0.109&  0.275&  1.330\\
  3.00&   0.709&  8.93&  0.111&  0.284&  1.816\\
  4.00&   0.828&  8.92&  0.103&  0.285&  1.033\\
  4.50&   0.878&  8.92&  0.100&  0.417&  0.316\\
  5.00&   0.908&  8.92&  0.111&  0.615&  0.285\\
  6.00&   0.958&  8.90&  0.127&  0.741&  0.270\\
  8.00&   1.073&  8.86&  0.149&  1.084&  0.425\\
\end{tabular}
\end{center}
\caption{[Fe/H]=0.2 $\alpha$=1.9 Galaxy}
\end{table}

\begin{table}
\begin{center}
\begin{tabular}{llllll}
Mass&M$_{\rm WD}$&O/H&He/H&N/O&C/O\\
  2.60&   0.725&  8.93&  0.122&  0.275&  3.020\\
  2.70&   0.732&  8.93&  0.121&  0.277&  2.938\\
  2.80&   0.743&  8.93&  0.121&  0.279&  2.944\\
  2.90&   0.754&  8.93&  0.120&  0.281&  2.917\\
  3.10&   0.772&  8.93&  0.119&  0.282&  2.871\\
  3.20&   0.783&  8.93&  0.119&  0.282&  2.891\\
  3.30&   0.794&  8.93&  0.119&  0.282&  2.965\\
  3.40&   0.804&  8.93&  0.119&  0.282&  2.965\\
  3.90&   0.899&  8.93&  0.111&  0.284&  2.023\\
  4.10&   0.876&  8.92&  0.105&  1.374&  0.265\\
  4.50&   0.925&  8.86&  0.103&  1.156&  0.284\\
  5.00&   0.962&  8.82&  0.115&  1.521&  0.417\\
  7.00&   1.059&  8.70&  0.146&  3.221&  0.624\\
  8.00&   1.090&  8.70&  0.154&  3.296&  0.533\\
\end{tabular}
\end{center}
\caption{[Fe/H]=0.2 $\alpha$=2.5 Galaxy}
\end{table}

\end{small}
\newpage
\chapter{Yields}

In this chapter we present the model yields for $\he$, $\ctw$, $\cth$,
$\nit$, and $\oxy$ for low and intermediate mass stars.

\begin{table}
\begin{tabular}{ccccccc}
Mass&M[H]&M[He]&M[C12]&M[N14]&M[O16]&M[C13]\\
1.00&-1.30E-02&1.30E-02&-3.53E-04&4.12E-04&-4.30E-05&3.11E-05\\
1.50&-1.79E-02&1.79E-02&-7.80E-04&9.10E-04&-8.66E-05&5.94E-05\\
2.00&-1.86E-02&1.71E-02&1.08E-04&1.47E-03&-1.79E-04&8.33E-05\\
2.50&-5.17E-02&3.82E-02&1.12E-02&2.02E-03&-6.44E-04&1.01E-04\\
3.00&-7.43E-02&5.15E-02&1.95E-02&2.67E-03&-1.00E-03&1.17E-04\\
4.00&-4.45E-02&2.95E-02&1.10E-02&3.87E-03&-7.66E-04&1.68E-04\\
4.50&-1.86E-02&1.72E-02&-5.99E-03&7.83E-03&-6.42E-04&1.07E-03\\
5.00&-1.01E-01&9.87E-02&-9.23E-03&1.42E-02&-2.11E-03&2.84E-04\\
6.00&-2.51E-01&2.50E-01&-1.18E-02&1.95E-02&-5.34E-03&1.84E-04\\
8.00&-5.36E-01&5.34E-01&-1.56E-02&3.13E-02&-1.27E-02&3.99E-04\\
\end{tabular}
\caption{[Fe/H]= 0.0 and $\alpha$= 1.90}
\end{table}

\begin{table}
\begin{tabular}{ccccccc}
Mass&M[H]&M[He]&M[C12]&M[N14]&M[O16]&M[C13]\\
1.00&-1.28E-02&1.28E-02&-3.46E-04&4.04E-04&-4.21E-05&3.05E-05\\
1.10&-1.44E-02&1.44E-02&-4.25E-04&4.96E-04&-5.07E-05&3.63E-05\\
1.20&-1.57E-02&1.57E-02&-5.06E-04&5.90E-04&-5.92E-05&4.19E-05\\
1.30&-1.66E-02&1.66E-02&-5.89E-04&6.87E-04&-6.78E-05&4.75E-05\\
1.40&-1.72E-02&1.72E-02&-6.73E-04&7.85E-04&-7.60E-05&5.27E-05\\
1.50&-1.74E-02&1.74E-02&-7.59E-04&8.85E-04&-8.42E-05&5.77E-05\\
1.60&-1.74E-02&1.74E-02&-8.51E-04&9.92E-04&-9.28E-05&6.29E-05\\
1.70&-1.93E-02&1.85E-02&-2.14E-04&1.09E-03&-1.28E-04&6.75E-05\\
1.80&-2.42E-02&2.16E-02&1.52E-03&1.19E-03&-2.03E-04&7.18E-05\\
1.90&-3.16E-02&2.61E-02&4.13E-03&1.29E-03&-3.10E-04&7.58E-05\\
2.00&-4.21E-02&3.26E-02&7.87E-03&1.38E-03&-4.58E-04&7.92E-05\\
2.10&-5.54E-02&4.12E-02&1.23E-02&1.46E-03&-6.29E-04&8.20E-05\\
2.20&-6.85E-02&4.97E-02&1.67E-02&1.54E-03&-7.98E-04&8.45E-05\\
2.30&-7.16E-02&5.17E-02&1.77E-02&1.66E-03&-8.46E-04&8.87E-05\\
2.40&-8.07E-02&5.75E-02&2.07E-02&1.77E-03&-9.64E-04&9.17E-05\\
2.50&-8.61E-02&6.09E-02&2.25E-02&1.88E-03&-1.04E-03&9.49E-05\\
2.60&-9.18E-02&6.45E-02&2.45E-02&2.00E-03&-1.12E-03&9.80E-05\\
2.80&-9.68E-02&6.71E-02&2.64E-02&2.26E-03&-1.21E-03&1.05E-04\\
2.90&-1.00E-01&6.91E-02&2.78E-02&2.39E-03&-1.27E-03&1.08E-04\\
3.10&-1.04E-01&7.07E-02&2.96E-02&2.64E-03&-1.35E-03&1.16E-04\\
3.30&-1.12E-01&7.53E-02&3.28E-02&2.83E-03&-1.48E-03&1.24E-04\\
3.40&-1.18E-01&7.91E-02&3.47E-02&2.93E-03&-1.55E-03&1.28E-04\\
3.50&-1.21E-01&8.06E-02&3.56E-02&3.04E-03&-1.59E-03&1.33E-04\\
3.60&-1.11E-01&7.40E-02&3.25E-02&3.17E-03&-1.48E-03&1.39E-04\\
3.70&-9.65E-02&6.42E-02&2.80E-02&3.32E-03&-1.31E-03&1.45E-04\\
3.80&-8.88E-02&5.90E-02&2.56E-02&3.45E-03&-1.22E-03&1.50E-04\\
3.90&-8.39E-02&5.56E-02&2.39E-02&3.56E-03&-1.16E-03&1.61E-04\\
4.00&-7.62E-02&5.05E-02&2.12E-02&3.69E-03&-1.07E-03&2.91E-04\\
4.10&-5.04E-02&3.34E-02&-1.54E-03&1.86E-02&-9.15E-04&2.21E-03\\
4.20&-3.75E-02&2.49E-02&-6.88E-03&2.24E-02&-1.31E-03&4.47E-04\\
4.30&-2.70E-02&1.79E-02&-7.67E-03&2.02E-02&-2.00E-03&3.17E-04\\
4.40&-1.58E-02&1.04E-02&-8.32E-03&1.70E-02&-2.11E-03&2.29E-04\\
4.50&-2.97E-02&2.45E-02&-8.52E-03&1.76E-02&-2.64E-03&2.27E-04\\
4.60&-4.64E-02&4.11E-02&-8.74E-03&1.86E-02&-3.40E-03&2.36E-04\\
4.70&-6.36E-02&5.79E-02&-8.90E-03&1.98E-02&-4.05E-03&2.48E-04\\
4.80&-8.06E-02&7.45E-02&-9.08E-03&2.08E-02&-4.61E-03&2.57E-04\\
4.90&-9.75E-02&9.11E-02&-9.20E-03&2.20E-02&-5.29E-03&2.71E-04\\
5.00&-1.14E-01&1.07E-01&-9.42E-03&2.31E-02&-5.83E-03&2.80E-04\\
6.00&-2.75E-01&2.64E-01&-1.12E-02&3.52E-02&-1.26E-02&4.14E-04\\
7.00&-4.18E-01&4.07E-01&-1.32E-02&4.16E-02&-1.68E-02&4.83E-04\\
8.00&-5.49E-01&5.41E-01&-1.56E-02&4.21E-02&-1.85E-02&4.75E-04\\
\end{tabular}
\caption{[Fe/H]= 0.0 and $\alpha$= 2.30}
\end{table}

\begin{table}
\begin{tabular}{ccccccc}
Mass&M[H]&M[He]&M[C12]&M[N14]&M[O16]&M[C13]\\
1.00&-1.24E-02&1.24E-02&-4.41E-04&5.15E-04&-5.37E-05&3.88E-05\\
1.10&-1.41E-02&1.41E-02&-5.46E-04&6.38E-04&-6.52E-05&4.67E-05\\
1.20&-1.52E-02&1.52E-02&-6.49E-04&7.57E-04&-7.60E-05&5.38E-05\\
1.30&-1.62E-02&1.62E-02&-7.56E-04&8.82E-04&-8.70E-05&6.09E-05\\
1.40&-1.67E-02&1.67E-02&-8.63E-04&1.01E-03&-9.75E-05&6.76E-05\\
1.50&-1.68E-02&1.68E-02&-9.72E-04&1.13E-03&-1.08E-04&7.39E-05\\
1.60&-1.68E-02&1.68E-02&-1.09E-03&1.27E-03&-1.19E-04&8.05E-05\\
1.70&-1.63E-02&1.63E-02&-1.20E-03&1.40E-03&-1.29E-04&8.65E-05\\
1.80&-1.65E-02&1.62E-02&-1.02E-03&1.54E-03&-1.54E-04&9.25E-05\\
1.90&-2.00E-02&1.81E-02&3.42E-04&1.68E-03&-2.36E-04&9.79E-05\\
2.00&-2.51E-02&2.10E-02&2.37E-03&1.80E-03&-3.48E-04&1.03E-04\\
2.20&-5.49E-02&4.01E-02&1.23E-02&2.00E-03&-8.48E-04&1.09E-04\\
2.40&-7.60E-02&5.36E-02&1.93E-02&2.24E-03&-1.21E-03&1.16E-04\\
2.60&-8.39E-02&5.83E-02&2.21E-02&2.55E-03&-1.37E-03&1.25E-04\\
2.80&-9.29E-02&6.35E-02&2.54E-02&2.86E-03&-1.55E-03&1.33E-04\\
3.00&-9.72E-02&6.56E-02&2.72E-02&3.21E-03&-1.67E-03&1.41E-04\\
3.20&-1.04E-01&6.91E-02&3.01E-02&3.46E-03&-1.83E-03&1.52E-04\\
3.40&-1.11E-01&7.34E-02&3.26E-02&3.72E-03&-1.97E-03&1.63E-04\\
3.60&-1.21E-01&7.94E-02&3.57E-02&3.97E-03&-2.14E-03&1.74E-04\\
3.80&-1.37E-01&8.99E-02&4.10E-02&4.18E-03&-2.40E-03&1.84E-04\\
3.90&-1.47E-01&9.64E-02&4.42E-02&4.29E-03&-2.55E-03&1.88E-04\\
4.00&-1.40E-01&9.18E-02&4.18E-02&4.47E-03&-2.45E-03&1.97E-04\\
4.02&-1.38E-01&9.07E-02&4.12E-02&4.50E-03&-2.42E-03&2.00E-04\\
4.04&-1.33E-01&8.71E-02&3.94E-02&4.56E-03&-2.34E-03&2.14E-04\\
4.06&-1.32E-01&8.65E-02&3.90E-02&4.58E-03&-2.32E-03&2.16E-04\\
4.08&-1.29E-01&8.45E-02&3.80E-02&4.63E-03&-2.28E-03&2.37E-04\\
4.10&-1.24E-01&8.12E-02&3.63E-02&4.68E-03&-2.20E-03&3.24E-04\\
4.20&-1.17E-01&7.66E-02&3.35E-02&4.86E-03&-2.09E-03&6.96E-04\\
4.30&-1.15E-01&7.52E-02&3.28E-02&4.97E-03&-2.03E-03&6.32E-04\\
4.50&-8.59E-02&5.62E-02&4.67E-03&2.18E-02&-1.69E-03&5.36E-03\\
4.60&-5.85E-02&3.83E-02&-1.02E-02&3.45E-02&-2.12E-03&5.93E-04\\
4.70&-4.25E-02&2.78E-02&-1.11E-02&3.09E-02&-3.19E-03&4.43E-04\\
4.80&-2.80E-02&1.83E-02&-1.19E-02&2.68E-02&-3.52E-03&3.42E-04\\
4.90&-3.80E-02&2.93E-02&-1.22E-02&2.66E-02&-4.05E-03&3.27E-04\\
5.00&-5.45E-02&4.57E-02&-1.25E-02&2.77E-02&-4.79E-03&3.35E-04\\
6.00&-2.13E-01&2.02E-01&-1.48E-02&3.87E-02&-1.19E-02&4.48E-04\\
7.00&-3.50E-01&3.42E-01&-1.76E-02&4.22E-02&-1.59E-02&4.71E-04\\
8.00&-4.79E-01&4.75E-01&-2.05E-02&4.22E-02&-1.69E-02&4.38E-04\\
\end{tabular}
\caption{[Fe/H]= 0.1 and $\alpha$= 2.30}
\end{table}

\begin{table}
\begin{tabular}{ccccccc}
Mass&M[H]&M[He]&M[C12]&M[N14]&M[O16]&M[C13]\\
1.00&-1.19E-02&1.19E-02&-5.65E-04&6.60E-04&-6.88E-05&4.97E-05\\
1.20&-1.45E-02&1.45E-02&-8.29E-04&9.67E-04&-9.71E-05&6.87E-05\\
1.40&-1.58E-02&1.58E-02&-1.10E-03&1.28E-03&-1.24E-04&8.61E-05\\
1.60&-1.57E-02&1.57E-02&-1.38E-03&1.61E-03&-1.51E-04&1.02E-04\\
1.80&-1.44E-02&1.44E-02&-1.68E-03&1.96E-03&-1.77E-04&1.17E-04\\
2.00&-1.90E-02&1.65E-02&3.85E-04&2.30E-03&-3.50E-04&1.31E-04\\
2.20&-5.18E-02&3.73E-02&1.14E-02&2.53E-03&-1.07E-03&1.38E-04\\
2.40&-6.45E-02&4.51E-02&1.57E-02&2.88E-03&-1.37E-03&1.49E-04\\
3.00&-9.48E-02&6.25E-02&2.67E-02&4.04E-03&-2.17E-03&1.77E-04\\
3.10&-9.51E-02&6.23E-02&2.71E-02&4.23E-03&-2.22E-03&1.85E-04\\
3.20&-9.87E-02&6.41E-02&2.86E-02&4.38E-03&-2.33E-03&1.92E-04\\
3.30&-1.03E-01&6.68E-02&3.04E-02&4.53E-03&-2.45E-03&1.98E-04\\
3.40&-1.06E-01&6.87E-02&3.12E-02&4.70E-03&-2.52E-03&2.06E-04\\
3.50&-1.13E-01&7.31E-02&3.34E-02&4.84E-03&-2.67E-03&2.12E-04\\
3.60&-1.21E-01&7.80E-02&3.59E-02&4.98E-03&-2.83E-03&2.18E-04\\
3.70&-1.25E-01&8.06E-02&3.71E-02&5.15E-03&-2.93E-03&2.26E-04\\
3.80&-1.34E-01&8.66E-02&4.01E-02&5.28E-03&-3.13E-03&2.32E-04\\
3.90&-1.44E-01&9.30E-02&4.33E-02&5.40E-03&-3.33E-03&2.37E-04\\
4.00&-1.51E-01&9.74E-02&4.55E-02&5.55E-03&-3.48E-03&2.45E-04\\
4.10&-1.63E-01&1.05E-01&4.93E-02&5.66E-03&-3.72E-03&2.58E-04\\
4.20&-1.76E-01&1.14E-01&5.36E-02&5.77E-03&-3.99E-03&3.04E-04\\
4.30&-1.87E-01&1.20E-01&5.66E-02&5.91E-03&-4.21E-03&6.67E-04\\
4.40&-2.02E-01&1.30E-01&6.10E-02&6.05E-03&-4.52E-03&1.49E-03\\
4.50&-2.15E-01&1.39E-01&6.14E-02&6.86E-03&-4.79E-03&4.94E-03\\
4.60&-2.34E-01&1.51E-01&6.31E-02&8.55E-03&-5.17E-03&8.43E-03\\
4.70&-2.34E-01&1.50E-01&5.36E-02&1.46E-02&-5.17E-03&1.29E-02\\
4.80&-2.02E-01&1.30E-01&1.98E-02&4.34E-02&-4.61E-03&1.13E-02\\
4.90&-2.02E-01&1.30E-01&3.08E-02&2.85E-02&-4.52E-03&1.33E-02\\
5.00&-1.95E-01&1.25E-01&2.72E-02&3.05E-02&-4.35E-03&1.28E-02\\
6.00&-1.28E-01&1.10E-01&-1.95E-02&5.16E-02&-1.20E-02&6.00E-04\\
7.00&-2.46E-01&2.38E-01&-2.33E-02&4.77E-02&-1.52E-02&5.12E-04\\
8.00&-3.73E-01&3.70E-01&-2.70E-02&4.78E-02&-1.61E-02&4.65E-04\\
\end{tabular}
\caption{[Fe/H]= 0.2 and $\alpha$= 2.30}
\end{table}

\begin{table}
\begin{tabular}{ccccccc}
Mass&M[H]&M[He]&M[C12]&M[N14]&M[O16]&M[C13]\\
1.00&-1.30E-02&1.30E-02&-2.71E-04&3.16E-04&-3.30E-05&2.38E-05\\
1.10&-1.47E-02&1.47E-02&-3.33E-04&3.88E-04&-3.97E-05&2.84E-05\\
1.20&-1.60E-02&1.60E-02&-3.97E-04&4.63E-04&-4.65E-05&3.29E-05\\
1.30&-1.71E-02&1.71E-02&-4.63E-04&5.40E-04&-5.32E-05&3.73E-05\\
1.40&-1.77E-02&1.77E-02&-5.29E-04&6.17E-04&-5.97E-05&4.14E-05\\
1.50&-1.79E-02&1.79E-02&-5.97E-04&6.96E-04&-6.62E-05&4.54E-05\\
1.60&-1.92E-02&1.88E-02&-2.98E-04&7.80E-04&-8.36E-05&4.95E-05\\
1.70&-2.38E-02&2.18E-02&1.23E-03&8.57E-04&-1.35E-04&5.30E-05\\
1.80&-3.03E-02&2.59E-02&3.44E-03&9.33E-04&-2.05E-04&5.64E-05\\
1.90&-3.91E-02&3.15E-02&6.50E-03&1.01E-03&-2.99E-04&5.93E-05\\
3.30&-1.13E-01&7.70E-02&3.30E-02&2.24E-03&-1.11E-03&9.84E-05\\
3.50&-8.64E-02&5.85E-02&2.47E-02&2.47E-03&-8.67E-04&1.08E-04\\
3.90&-3.45E-02&2.31E-02&-4.20E-03&1.76E-02&-8.22E-04&4.70E-04\\
4.10&-1.61E-02&1.08E-02&-5.67E-03&1.38E-02&-1.78E-03&2.05E-04\\
4.50&-7.09E-02&6.58E-02&-6.37E-03&1.62E-02&-3.90E-03&2.07E-04\\
5.00&-1.54E-01&1.47E-01&-7.01E-03&2.17E-02&-6.96E-03&2.68E-04\\
7.00&-4.65E-01&4.50E-01&-9.90E-03&4.34E-02&-1.72E-02&5.20E-04\\
8.00&-6.00E-01&5.86E-01&-1.17E-02&4.61E-02&-1.95E-02&5.50E-04\\
\end{tabular}
\caption{[Fe/H]=-0.1 and $\alpha$= 2.30}
\end{table}

\begin{table}
\begin{tabular}{ccccccc}
Mass&M[H]&M[He]&M[C12]&M[N14]&M[O16]&M[C13]\\
1.00&-1.32E-02&1.32E-02&-2.12E-04&2.48E-04&-2.58E-05&1.87E-05\\
1.10&-1.49E-02&1.49E-02&-2.61E-04&3.04E-04&-3.11E-05&2.23E-05\\
1.20&-1.63E-02&1.63E-02&-3.12E-04&3.64E-04&-3.65E-05&2.58E-05\\
1.30&-1.74E-02&1.74E-02&-3.64E-04&4.24E-04&-4.18E-05&2.93E-05\\
1.40&-1.80E-02&1.80E-02&-4.16E-04&4.85E-04&-4.70E-05&3.26E-05\\
1.50&-1.83E-02&1.83E-02&-4.70E-04&5.48E-04&-5.21E-05&3.57E-05\\
1.60&-2.20E-02&2.09E-02&6.07E-04&6.11E-04&-8.18E-05&3.89E-05\\
1.70&-2.82E-02&2.49E-02&2.58E-03&6.72E-04&-1.30E-04&4.17E-05\\
1.80&-3.54E-02&2.96E-02&5.01E-03&7.31E-04&-1.87E-04&4.43E-05\\
1.90&-4.65E-02&3.68E-02&8.77E-03&7.86E-04&-2.72E-04&4.65E-05\\
2.00&-5.64E-02&4.30E-02&1.22E-02&8.42E-04&-3.51E-04&4.87E-05\\
\end{tabular}
\caption{[Fe/H]=-0.2 and $\alpha$= 2.30}
\end{table}

\begin{table}
\begin{tabular}{ccccccc}
Mass&M[H]&M[He]&M[C12]&M[N14]&M[O16]&M[C13]\\
1.00&-1.27E-02&1.27E-02&-1.36E-04&1.59E-04&-1.93E-05&1.20E-05\\
1.20&-1.64E-02&1.64E-02&-2.07E-04&2.41E-04&-2.82E-05&1.72E-05\\
1.40&-1.88E-02&1.86E-02&-9.21E-05&3.23E-04&-3.94E-05&2.17E-05\\
1.60&-3.12E-02&2.73E-02&3.52E-03&4.03E-04&-1.06E-04&2.58E-05\\
1.80&-4.86E-02&3.89E-02&9.13E-03&4.82E-04&-2.03E-04&2.94E-05\\
2.00&-7.18E-02&5.40E-02&1.70E-02&5.54E-04&-3.32E-04&3.23E-05\\
2.20&-8.91E-02&6.51E-02&2.29E-02&6.36E-04&-4.27E-04&3.51E-05\\
2.40&-1.01E-01&7.29E-02&2.66E-02&7.32E-04&-4.89E-04&3.82E-05\\
3.50&-4.76E-02&3.32E-02&3.00E-03&1.14E-02&-3.17E-04&1.51E-03\\
\end{tabular}
\caption{[Fe/H]=-0.5 and $\alpha$= 2.30}
\end{table}

\begin{table}
\begin{tabular}{ccccccc}
Mass&M[H]&M[He]&M[C12]&M[N14]&M[O16]&M[C13]\\
1.00&-1.25E-02&1.25E-02&-5.80E-05&6.76E-05&-9.14E-06&5.10E-06\\
1.20&-1.66E-02&1.65E-02&4.49E-05&1.03E-04&-1.42E-05&7.35E-06\\
1.40&-2.61E-02&2.38E-02&2.20E-03&1.40E-04&-3.06E-05&9.48E-06\\
1.60&-3.79E-02&3.22E-02&5.49E-03&1.78E-04&-5.26E-05&1.15E-05\\
1.80&-5.43E-02&4.33E-02&1.07E-02&2.14E-04&-8.38E-05&1.31E-05\\
2.00&-7.10E-02&5.41E-02&1.65E-02&2.51E-04&-1.16E-04&1.46E-05\\
2.20&-8.28E-02&6.16E-02&2.07E-02&2.91E-04&-1.38E-04&1.60E-05\\
2.40&-1.12E-01&8.12E-02&2.98E-02&3.22E-04&-1.77E-04&1.68E-05\\
3.00&-1.05E-01&7.53E-02&2.93E-02&4.78E-04&-1.34E-04&2.09E-05\\
3.50&-4.68E-02&3.31E-02&4.19E-04&1.54E-02&-5.94E-04&3.10E-04\\
3.60&-3.72E-02&2.61E-02&1.53E-05&1.38E-02&-1.59E-03&2.29E-04\\
3.70&-2.87E-02&1.98E-02&-2.04E-04&1.23E-02&-2.57E-03&1.88E-04\\
3.80&-4.28E-02&3.39E-02&-2.64E-04&1.29E-02&-2.99E-03&1.90E-04\\
\end{tabular}
\caption{[Fe/H]=-1.0 and $\alpha$= 2.30}
\end{table}

\begin{table}
\begin{tabular}{ccccccc}
Mass&M[H]&M[He]&M[C12]&M[N14]&M[O16]&M[C13]\\
2.60&-1.06E-01&7.38E-02&2.91E-02&1.93E-03&-1.28E-03&9.52E-05\\
2.70&-1.06E-01&7.36E-02&2.93E-02&2.07E-03&-1.30E-03&9.92E-05\\
2.80&-1.10E-01&7.59E-02&3.08E-02&2.20E-03&-1.36E-03&1.02E-04\\
2.90&-1.14E-01&7.80E-02&3.22E-02&2.33E-03&-1.41E-03&1.05E-04\\
3.10&-1.18E-01&7.97E-02&3.40E-02&2.57E-03&-1.49E-03&1.13E-04\\
3.20&-1.21E-01&8.14E-02&3.54E-02&2.67E-03&-1.55E-03&1.17E-04\\
3.30&-1.26E-01&8.44E-02&3.73E-02&2.76E-03&-1.62E-03&1.21E-04\\
3.40&-1.32E-01&8.84E-02&3.93E-02&2.85E-03&-1.70E-03&1.25E-04\\
3.90&-9.50E-02&6.30E-02&2.70E-02&3.52E-03&-1.26E-03&7.19E-04\\
4.10&-6.15E-02&4.08E-02&-6.10E-03&3.01E-02&-1.39E-03&5.92E-04\\
4.50&-4.13E-02&3.21E-02&-8.38E-03&2.39E-02&-5.23E-03&2.90E-04\\
5.00&-1.28E-01&1.16E-01&-9.14E-03&3.13E-02&-9.14E-03&3.65E-04\\
7.00&-4.60E-01&4.32E-01&-1.24E-02&6.42E-02&-2.32E-02&7.42E-04\\
8.00&-5.97E-01&5.70E-01&-1.46E-02&6.94E-02&-2.69E-02&8.04E-04\\
\end{tabular}
\caption{[Fe/H]= 0.0 and $\alpha$= 2.50}
\end{table}

\begin{table}
\begin{tabular}{ccccccc}
Mass&M[H]&M[He]&M[C12]&M[N14]&M[O16]&M[C13]\\
1.00&-1.23E-02&1.23E-02&-4.40E-04&5.13E-04&-5.35E-05&3.87E-05\\
1.20&-1.51E-02&1.51E-02&-6.43E-04&7.50E-04&-7.52E-05&5.33E-05\\
1.40&-1.64E-02&1.64E-02&-8.51E-04&9.92E-04&-9.61E-05&6.66E-05\\
1.60&-1.65E-02&1.65E-02&-1.07E-03&1.25E-03&-1.17E-04&7.94E-05\\
1.65&-1.64E-02&1.64E-02&-1.13E-03&1.32E-03&-1.22E-04&8.24E-05\\
1.70&-1.68E-02&1.66E-02&-9.49E-04&1.38E-03&-1.39E-04&8.54E-05\\
1.75&-1.83E-02&1.74E-02&-4.13E-04&1.45E-03&-1.73E-04&8.82E-05\\
1.80&-2.22E-02&1.99E-02&9.49E-04&1.51E-03&-2.45E-04&9.07E-05\\
2.00&-3.94E-02&3.03E-02&7.16E-03&1.74E-03&-5.72E-04&9.97E-05\\
2.20&-7.56E-02&5.35E-02&1.93E-02&1.90E-03&-1.17E-03&1.05E-04\\
2.40&-9.28E-02&6.45E-02&2.50E-02&2.15E-03&-1.46E-03&1.12E-04\\
2.50&-9.89E-02&6.83E-02&2.70E-02&2.29E-03&-1.57E-03&1.16E-04\\
2.70&-1.03E-01&7.05E-02&2.86E-02&2.61E-03&-1.68E-03&1.25E-04\\
2.80&-1.07E-01&7.27E-02&3.00E-02&2.77E-03&-1.76E-03&1.29E-04\\
3.10&-1.15E-01&7.64E-02&3.33E-02&3.24E-03&-1.95E-03&1.42E-04\\
3.20&-1.18E-01&7.82E-02&3.48E-02&3.36E-03&-2.03E-03&1.48E-04\\
3.30&-1.20E-01&7.90E-02&3.55E-02&3.50E-03&-2.07E-03&1.54E-04\\
3.40&-1.26E-01&8.31E-02&3.76E-02&3.62E-03&-2.18E-03&1.59E-04\\
3.50&-1.33E-01&8.75E-02&3.98E-02&3.73E-03&-2.29E-03&1.64E-04\\
3.60&-1.37E-01&8.99E-02&4.11E-02&3.86E-03&-2.36E-03&1.70E-04\\
3.70&-1.45E-01&9.51E-02&4.37E-02&3.97E-03&-2.49E-03&1.74E-04\\
3.80&-1.54E-01&1.01E-01&4.67E-02&4.07E-03&-2.63E-03&1.82E-04\\
4.00&-1.57E-01&1.03E-01&4.71E-02&4.36E-03&-2.68E-03&6.22E-04\\
4.10&-1.41E-01&9.24E-02&3.82E-02&5.25E-03&-2.43E-03&3.74E-03\\
4.30&-1.28E-01&8.36E-02&3.02E-02&7.31E-03&-2.20E-03&5.93E-03\\
4.35&-1.22E-01&8.00E-02&2.45E-02&1.04E-02&-2.12E-03&7.32E-03\\
4.40&-1.16E-01&7.62E-02&1.84E-02&1.49E-02&-2.04E-03&7.74E-03\\
4.45&-1.11E-01&7.27E-02&1.19E-02&2.11E-02&-1.98E-03&7.14E-03\\
4.50&-9.80E-02&6.41E-02&-6.49E-03&4.36E-02&-2.03E-03&1.71E-03\\
5.00&-7.07E-02&5.60E-02&-1.22E-02&3.65E-02&-8.30E-03&4.22E-04\\
6.00&-2.42E-01&2.20E-01&-1.43E-02&5.46E-02&-1.75E-02&6.17E-04\\
7.00&-3.91E-01&3.66E-01&-1.67E-02&6.57E-02&-2.45E-02&7.43E-04\\
\end{tabular}
\caption{[Fe/H]= 0.1 and $\alpha$= 2.50}
\end{table}

\begin{table}
\begin{tabular}{ccccccc}
Mass&M[H]&M[He]&M[C12]&M[N14]&M[O16]&M[C13]\\
1.00&-1.18E-02&1.18E-02&-5.60E-04&6.53E-04&-6.81E-05&4.93E-05\\
1.20&-1.44E-02&1.44E-02&-8.21E-04&9.58E-04&-9.61E-05&6.81E-05\\
1.40&-1.56E-02&1.56E-02&-1.08E-03&1.27E-03&-1.22E-04&8.49E-05\\
1.60&-1.55E-02&1.55E-02&-1.37E-03&1.60E-03&-1.49E-04&1.01E-04\\
1.80&-1.64E-02&1.56E-02&-9.31E-04&1.93E-03&-2.20E-04&1.16E-04\\
2.00&-1.29E-02&1.29E-02&-2.16E-03&2.52E-03&-2.21E-04&1.43E-04\\
2.50&-9.38E-02&6.36E-02&2.57E-02&2.90E-03&-1.99E-03&1.47E-04\\
3.10&-1.13E-01&7.35E-02&3.30E-02&4.07E-03&-2.55E-03&1.79E-04\\
3.20&-1.15E-01&7.46E-02&3.42E-02&4.23E-03&-2.64E-03&1.86E-04\\
3.30&-1.19E-01&7.71E-02&3.58E-02&4.38E-03&-2.76E-03&1.93E-04\\
3.40&-1.23E-01&7.92E-02&3.68E-02&4.56E-03&-2.83E-03&2.00E-04\\
3.90&-1.63E-01&1.05E-01&4.97E-02&5.23E-03&-3.67E-03&2.88E-04\\
4.10&-1.82E-01&1.17E-01&5.45E-02&5.58E-03&-4.07E-03&1.76E-03\\
4.20&-1.95E-01&1.26E-01&5.63E-02&6.09E-03&-4.32E-03&3.89E-03\\
4.30&-2.05E-01&1.32E-01&5.24E-02&9.15E-03&-4.54E-03&8.94E-03\\
4.40&-2.20E-01&1.42E-01&5.13E-02&1.30E-02&-4.84E-03&1.19E-02\\
4.50&-2.33E-01&1.50E-01&4.43E-02&2.34E-02&-5.12E-03&1.47E-02\\
4.60&-2.51E-01&1.62E-01&4.15E-02&3.25E-02&-5.48E-03&1.57E-02\\
4.70&-2.51E-01&1.62E-01&2.82E-02&5.03E-02&-5.51E-03&1.37E-02\\
4.80&-2.23E-01&1.43E-01&-3.53E-03&8.72E-02&-5.18E-03&3.83E-03\\
4.90&-2.19E-01&1.41E-01&5.33E-03&7.18E-02&-4.92E-03&7.21E-03\\
6.00&-7.96E-02&7.96E-02&-9.03E-03&1.21E-02&-2.53E-03&4.20E-04\\
7.00&-2.91E-01&2.66E-01&-2.23E-02&7.42E-02&-2.56E-02&8.19E-04\\
\end{tabular}
\caption{[Fe/H]= 0.2 and $\alpha$= 2.50}
\end{table}

\begin{table}
\begin{tabular}{ccccccc}
Mass&M[H]&M[He]&M[C12]&M[N14]&M[O16]&M[C13]\\
1.00&-1.27E-02&1.27E-02&-2.64E-04&3.08E-04&-3.21E-05&2.32E-05\\
1.50&-1.84E-02&1.82E-02&-3.81E-04&6.88E-04&-7.15E-05&4.49E-05\\
2.00&-6.48E-02&4.82E-02&1.51E-02&1.03E-03&-5.44E-04&6.01E-05\\
2.50&-1.05E-01&7.43E-02&2.85E-02&1.43E-03&-9.43E-04&7.27E-05\\
3.00&-1.21E-01&8.35E-02&3.46E-02&1.94E-03&-1.13E-03&8.54E-05\\
3.50&-9.93E-02&6.71E-02&2.89E-02&2.41E-03&-9.57E-04&1.05E-04\\
4.00&-3.76E-02&2.51E-02&-5.13E-03&2.16E-02&-2.58E-03&3.12E-04\\
4.50&-8.18E-02&7.29E-02&-6.23E-03&2.23E-02&-6.30E-03&2.72E-04\\
5.00&-1.67E-01&1.55E-01&-6.84E-03&2.92E-02&-9.64E-03&3.45E-04\\
6.00&-3.38E-01&3.16E-01&-8.02E-03&4.62E-02&-1.60E-02&5.37E-04\\
8.00&-6.61E-01&6.22E-01&-1.05E-02&7.67E-02&-2.58E-02&9.17E-04\\
\end{tabular}
\caption{[Fe/H]=-0.1 and $\alpha$= 2.50}
\end{table}

\newpage

\end{document}